\documentclass[11pt,a4paper]{article}
\usepackage[dvipsnames,table]{xcolor}
\usepackage[normalem]{ulem}

\usepackage{tocloft}
\usepackage[]{todonotes}
\usepackage{verbatim}
\usepackage{fancyvrb}
\usepackage{mathrsfs}

\usepackage[shortlabels]{enumitem}
\usepackage{pgf}
\usepackage{tikz-cd}
\usetikzlibrary{shapes,arrows}
\usetikzlibrary{calc}
\usepackage{pgfplots}
\pgfplotsset{compat=1.12}
\usepackage{tikz-3dplot}
\usepackage{slashed}
\usepackage{multirow}
\usepackage{amsthm}
\usepackage{bm}
\usepackage{tikz}
\usetikzlibrary{fadings}
\usepackage{setspace,caption}
\usepackage{lipsum}
\usetikzlibrary{matrix,arrows,calc}
\usepackage{placeins}
\usepackage{amsmath}
\usepackage{amssymb}
\usepackage{amsfonts}
\usepackage{mathtools}
\usepackage{hhline}
\usepackage{sansmath}
\usepackage{tabularx}
\usepackage{float}

\usepackage{makecell}
\usepackage{subcaption}
\def\Re{{\rm Re\,}}
\def\Im{{\rm Im\,}}

\usepackage{comment}

\usepackage{jcapmod}
\usepackage{anyfontsize}
\usepackage{cleveref}

\newcommand{\nflux}{N_{\textrm{flux}}}

\theoremstyle{remark}

\theoremstyle{definition}

\renewcommand{\Re}{\textrm{Re}}
\renewcommand{\Im}{\textrm{Im}}
\newcommand{\hk}{\ensuremath{h^{1,1}}}
\newcommand{\hc}{\ensuremath{h^{1,2}}}

\usepackage{orcidlink}

\usepackage{fontawesome5}

\makeatletter
\newcommand{\github}[1]{%
   \href{#1}{{\normalsize \color{black}\faGithub}}%
}
\makeatother

\definecolor{airforceblue}{rgb}{0.36, 0.54, 0.66}

\definecolor{darkgreen}{rgb}{0.23, 0.56, 0.22}

\begin{document}
	\newcommand{\main}{.}
\begin{titlepage}

\setcounter{page}{1} \baselineskip=15.5pt \thispagestyle{empty}
\setcounter{tocdepth}{2}
\bigskip\

\vspace{.1cm}
\begin{center}
{\Large \bfseries Towards a foundation model for Type IIB flux vacua}
\end{center}

\vspace{0.55cm}

\begin{center}
\scalebox{0.95}[0.95]{{\fontsize{14}{30}\selectfont 
Sven Krippendorf \orcidlink{0000-0001-6374-6828},$^{a,b}$ Magdalena Larfors \orcidlink{0000-0003-1011-8777},$^{c}$ Zhimei Liu \orcidlink{0009-0007-7320-2251},$^{a}$
}}\\[0.5em]
\scalebox{0.95}[0.95]{{\fontsize{14}{30}\selectfont 
Andreas Schachner \orcidlink{0000-0002-7287-1476},$^{d,e}$ and Moritz Walden \orcidlink{0000-0001-9900-6485}$\,^{b}$\vspace{0.25cm}
}}

\end{center}

\begin{center}

\vspace{0.15 cm}
{\fontsize{11}{30}
\textsl{$^{a}$}Department of Applied Mathematics and Theoretical Physics, University of Cambridge, Wilberforce Road, Cambridge, CB3 0WA, United Kingdom\\[.1cm]
\textsl{$^{b}$}Cavendish Laboratory, University of Cambridge, JJ Thomson Avenue, Cambridge, CB3 0US, United Kingdom\\[.1cm]
\textsl{$^{c}$}Department of Physics and Astronomy \& Department of Mathematics, Uppsala~University, Lägerhyddsv. 1, SE-751 20 Uppsala, Sweden\\[.1cm]
\textsl{$^{d}$}Institute for Theoretical Physics, Heidelberg University, Philosophenweg 19, 69120~Heidelberg, Germany\\[.1cm]
\textsl{$^{e}$}Department of Physics, Cornell University, Ithaca, NY 14853 USA\\
}
\vspace{0.25cm}

\vspace{1.1cm}
\normalsize{\bf Abstract} \\[8mm]
\end{center}

\begin{center}

    \begin{minipage}[h]{15.0cm}

    Progress in characterising the Type IIB flux landscape requires scalable methods for constructing flux vacua. In this work, we investigate generative models as a framework for sampling vacua across multiple Calabi–Yau geometries and for directing searches towards specified physical properties. We compare transformers and diffusion models trained to generate quantised fluxes together with approximate values of the complex structure moduli and axio-dilaton. We exploit a staged training procedure comprising pre-training, and post-training via reinforcement learning, and assess how each stage improves the generated distributions and enables targeted sampling. We further examine whether structures learned from geometries with smaller $h^{1,2}$ can be transferred to geometries with larger $h^{1,2}$ through fine-tuning on specialised datasets. Our results provide evidence that generative models can learn distributions of flux vacua across distinct compactifications, improve searches for vacua with selected properties and transfer information between geometries. This establishes a concrete basis for developing a foundation model of the Type IIB flux landscape.
    
    \end{minipage}

\end{center}

\end{titlepage}
\tableofcontents

\hrulefill

\section{Introduction}\label{sec:intro}

The string theory landscape contains a vast range of possible vacuum solutions~\cite{Ashok:2003gk, Douglas:2003um}, with low-energy physics determined by the compactification geometry, background fluxes, and the stabilisation of moduli. Understanding the structure of the string landscape and the statistical properties of its vacua is a central task in connecting string theory to particle physics and cosmology~\cite{Bousso:2000xa,Susskind:2003kw,Denef:2004ze}. This requires both the construction of explicit vacua and methods for exploring their distributions. The challenge is therefore not merely to find individual solutions, but to develop systematic ways of identifying vacua with specified properties across different compactification geometries. 

Type IIB flux compactifications on Calabi-Yau (CY) orientifolds provide a well-studied setting for such investigations~\cite{Dasgupta:1999ss,Giddings:2001yu, Douglas:2006es}. Several complementary approaches have been developed to construct flux vacua and study their properties. Statistical studies based on the continuous flux approximation have provided estimates for the number and distribution of vacua~\cite{Ashok:2003gk, Denef:2004ze, Douglas:2003um, Eguchi:2005eh, Acharya:2006zw, Lu:2009aw, Cheng:2019mgz}, but do not directly construct the corresponding integer flux configurations. Analytical methods have enabled the explicit construction of vacua with particular physical properties, including exponentially small flux superpotentials at large complex structure and near conifold loci~\cite{Demirtas:2019sip, Alvarez-Garcia:2020pxd}, although such constructions often rely on special geometric regimes and carefully chosen fluxes. Systematic numerical searches have also been developed to explore flux vacua on specific Calabi-Yau geometries~\cite{Braun:2011av,Martinez-Pedrera:2012teo, Ebelt:2023clh,Krippendorf:2023idy,Plauschinn:2023hjw, Chauhan:2025rdj}. Advances in computational frameworks, such as \verb|CYTools| \cite{Demirtas:2022hqf} or \verb|JAXVacua| \cite{Dubey:2023dvu}, have enabled the construction of substantial ensembles of flux vacua and detailed studies of their distributions~\cite{Ebelt:2023clh,Plauschinn:2023hjw,Chauhan:2025rdj}. Complementary data-driven studies have used topological data analysis, in particular persistent homology, to probe the global structure of vacuum distributions and identify robust features in flux and moduli space~\cite{Cole:2018emh,Chauhan:2026gid}. In parallel, optimisation techniques based on genetic algorithms and reinforcement learning (RL) have been applied to search for vacua with desired properties and investigate structure in the flux landscape~\cite{Cole:2019enn,Krippendorf:2021uxu,Cole:2021nnt}. Nevertheless, finding vacua with prescribed properties can remain computationally expensive, particularly when the desired solutions are rarely encountered in numerical searches. More recently, generative models have offered an alternative by learning distributions of discrete flux configurations and enabling targeted sampling conditioned on physical quantities of interest~\cite{Krippendorf:2025mhp,Walden:2025cpf}. Related developments have demonstrated the use of transformer-based generative models to construct CY geometries through the generation of fine, regular, star triangulations of reflexive polytopes~\cite{Yip:2025hon,Arnal:2026zyo}. The success of these methods on individual geometries motivates their extension to a wider range of CY geometries.

Moving beyond individual geometries raises a question that successful generation on a fixed geometry cannot answer: can knowledge learned from flux vacua on one geometry facilitate the construction of vacua on another? Addressing this question requires targeted searches across different geometries, rather than restricting attention to those for which large datasets are already available. Since the flux lattices and moduli spaces depend on the geometry, a generative model intended for this purpose must do more than reproduce its training distribution. It should support searches subject to physical constraints and transfer useful information to geometries for which conventional methods have produced few known solutions. Establishing these capabilities would allow the computational effort invested in existing datasets to contribute to the construction of new compactifications and to systematic studies of their physical properties, without treating every geometry as an independent search problem.

This motivates the long-term goal of developing a \textit{foundation model} for Type IIB flux vacua: a model trained on a dataset of vacua from a broad range of CY geometries that can serve as a common starting point for different scientific investigations~\cite{bommasani2022opportunitiesrisksfoundationmodels, mccabe2024multiplephysicspretrainingphysical,zeni2024mattergengenerativemodelinorganic, Birk:2024knn,Golling:2024abg,bodnar2024foundationmodelearth}. In our context, its development would combine pre-training a generative model on existing dataset with RL techniques to improve the generation of valid solutions across those various geometries. The resulting model would provide a shared resource that researchers could use directly or adapt to new geometries and downstream tasks, without having to train separate models from scratch. 
The purpose of the present work is to investigate the capabilities underlying such a foundation model, particularly the ability to learn across geometries, perform targeted searches, and transfer knowledge to previously unseen geometries.

Our study considers twenty Calabi–Yau geometries with $h^{1,2}\in\{2,3,4,5\}$, using a dataset containing $100\,000$ distinct flux vacua per geometry, constructed from the dataset in~\cite{Ebelt:2023clh}. This provides a multi-geometry setting in which to investigate generation, conditioning, transfer learning, and post-training refinement. While these experiments demonstrate several key characteristics of foundation models, the extent to which the learned representations can be reused across a substantially broader range of geometries and physical tasks remains to be established. Our results therefore provide an initial demonstration of the potential for a broadly applicable foundation model of the Type IIB flux landscape.

We consider two complementary approaches to modelling the discrete flux distribution. The first is an encoder–decoder transformer~\cite{Vaswani:2017lxt}, which takes the CY geometry and any additional physical conditions as input, and generates a sequence of tokens representing the flux configuration and initial guesses for the complex-structure moduli and axio-dilaton. The second is based on a discrete diffusion model, which learns to generate samples by reversing a gradual corruption process. For discrete data, this process can be implemented through random transitions between categorical values~\cite{hoogeboom2021argmaxflowsmultinomialdiffusion}, or by progressively replacing individual entries with a special mask token~\cite{DBLP:conf/nips/AustinJHTB21,sahoo2024simpleeffectivemaskeddiffusion}. We adopt the latter approach, using a bidirectional transformer to reconstruct masked flux coordinates from the remaining visible coordinates. In contrast to the transformer's autoregressive generation, masked diffusion learns to predict flux components conditioned on arbitrary subsets of the other components, without imposing a fixed ordering during training. We compare these two approaches to investigate their ability to learn the distributions of flux vacua across different CY geometries.

\begin{figure}
    \centering
    \includegraphics[width=1\linewidth]{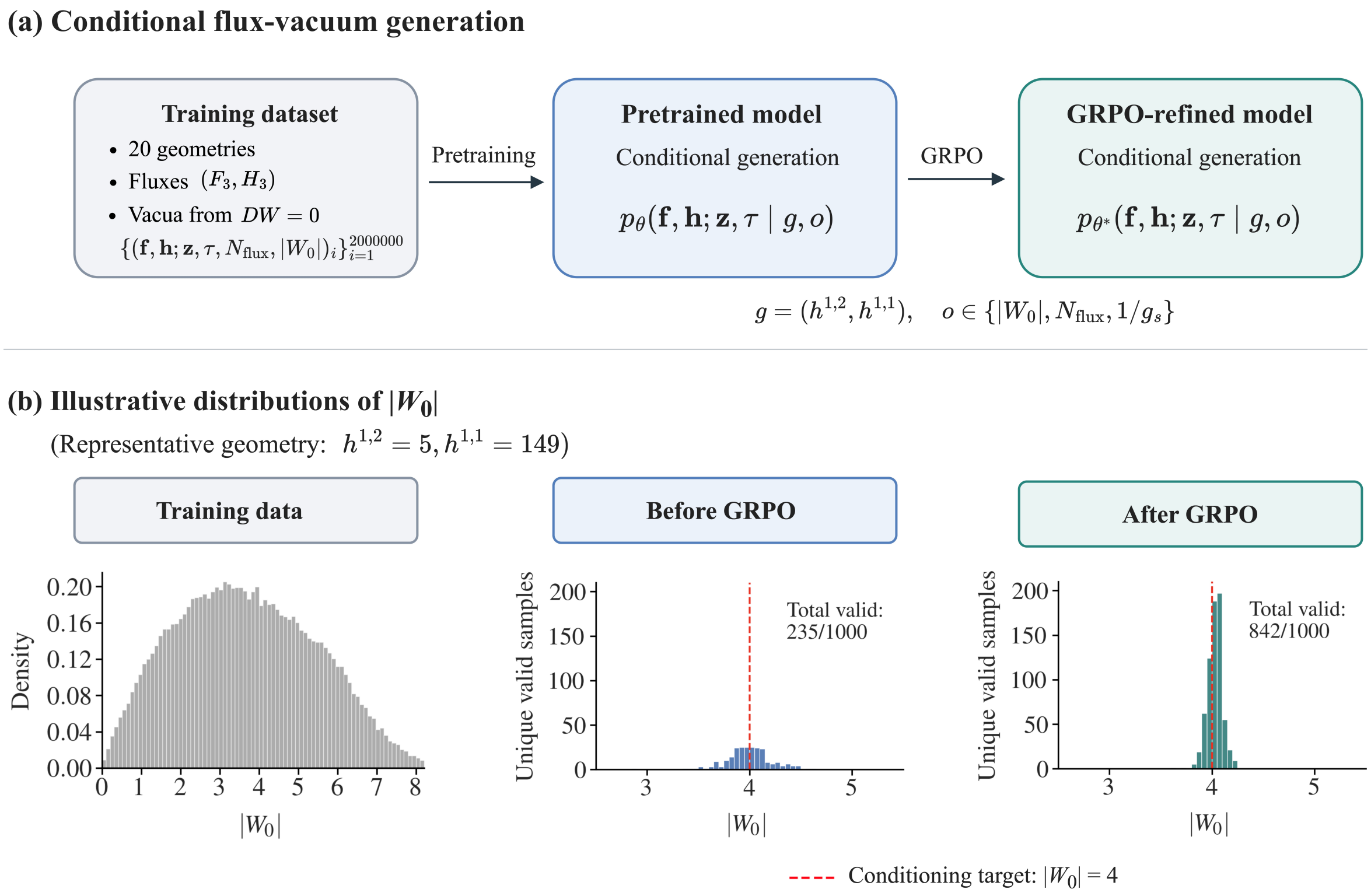}
    \caption{Overview of conditional flux generation through pre-training followed by post-training with GRPO. The geometry and flux vacuum shown in the dataset block are illustrative; the training datasets used in this work span 20 geometries. The lower panels show the training distribution of $|W_0|$ and conditional samples targeting $|W_0|=4$ before and after GRPO, illustrating improvements in conditional sampling accuracy and validity rate.}
    \label{fig:schematic_intro}
\end{figure}

An important part of our study concerns improvement beyond pre-training. We use RL techniques based on \textit{Group Relative Policy Optimisation} (GRPO)~\cite{shao2024deepseekmathpushinglimitsmathematical}, with an implementation adapted to each generative procedure, to refine the models using feedback from numerical validation. Fig.~\ref{fig:schematic_intro} summarises this workflow and illustrates how GRPO improves the sampled distribution more closely around the prompted target, which here is set to $|W_0|=4$. The post-training with GRPO connects the learned sampling distribution to the actual objective of producing valid solutions, rather than relying on the generative training loss alone. For both the transformer and the diffusion models, GRPO substantially improves sampling performance, particularly for the more complex flux and moduli spaces in our dataset. Moreover, we find that the validated solutions are absent from the training data, demonstrating that the improvement is not simply a consequence of memorising known data points. 

A central test of the potential of these generative approaches for a foundation model is whether pre-training facilitates transfer learning to new geometries using limited additional data on the corresponding geometries. To investigate this, we pre-train on geometries with $h^{1,2}\in \{2,3,4\}$ and subsequently fine-tune on five geometries with $h^{1,2}=5$, using only $1{,}000$ vacua per new geometry. For both generative approaches, the pre-trained models outperform counterparts trained from scratch on the same small target dataset, and subsequent GRPO further improves their performance. These experiments demonstrate transferability through limited-data adaptation to a larger flux and moduli space. They provide evidence that pre-training acquires information useful beyond the geometries on which it was performed, which is a key requirement for a future model intended to support progressively broader searches.

We further investigate whether sampling can be directed towards specified values of the flux-induced D3-brane charge $N_{\mathrm{flux}}$, the superpotential $|W_0|$, and the string coupling $g_s$. Both generative models learn distributions of flux vacua conditioned on these physical quantities, enabling targeted searches for vacua with prescribed properties, with GRPO improving accuracy towards the targets. Targeted generation also accesses low-$N_{\mathrm{flux}}$ regions that are sparsely represented in the training data.  

Taken together, these results establish several capabilities that make transformers and diffusion models promising starting points for a foundation model of Type IIB flux vacua: multi-geometry vacuum generation, targeted sampling, improvement through numerical feedback, and transfer learning with limited additional data. Whether these capabilities persist across a much broader list of geometries and substantially larger moduli spaces --- for example, extending to $h^{1,2} \sim 10$ and beyond --- remains to be established. The contribution of the present work is to provide the methodological framework and empirical evidence needed to undertake that next stage. In this sense, it is a step towards a foundation model: an investigation of the capabilities on which such a model would depend, and of the scientific value that would justify its construction.

The paper is organised as follows. In Sec.~\ref{sec:typeIIB}, we first review the background of constructing flux vacua in Type IIB compactifications. Sec.~\ref{sec:ml_section} introduces the dataset and the training procedures of our transformer and diffusion model. In Sec.~\ref{sec:experiments}, we present our experiments, including pre-training and post-training with GRPO, transfer learning to new CY geometries, conditional generation, scaling behaviour and computational performance. We conclude in Sec.~\ref{sec:con}. Additional architectural and training details are provided in the Appendix. Code and additional results, including conditional generation plots for all geometries considered in this work, are available in our GitHub repository \github{https://github.com/StringJAX/towards_type_IIB_foundation_model}.\footnote{\url{https://github.com/StringJAX/towards_type_IIB_foundation_model}}

\section{Type IIB flux compactification}\label{sec:typeIIB}

We consider compactifications of Type IIB superstring theory on CY threefold orientifolds $X$ with O3/O7-planes. Below the Kaluza-Klein scale, the light degrees of freedom are described by four-dimensional $\mathcal{N}=1$ supergravity, see \cite{Grana:2005jc,Douglas:2006es,Denef:2008wq,McAllister:2023vgy,McAllister:2025qwq} for reviews. Among the resulting chiral multiplets are the axio-dilaton, the complex structure moduli and the Kähler moduli. We denote the number of complex structure moduli surviving the orientifold projection by $\hc$ and write them as $z^a$, with $a=1,\dots,\hc$. Background three-form fluxes generate a tree-level potential for these fields and the axio-dilaton $\tau=c_0+\mathrm{i}s$, while the Kähler moduli do not enter the tree-level flux superpotential. We therefore restrict attention to this sector.

The complex structure moduli $z^a$ parametrise the complex geometry of $X$. Consequently, the holomorphic three-form $\Omega$ and its integrals over three-cycles depend on $z^a$. These periods couple directly to the RR and NSNS three-form fluxes $F_3$ and $H_3$. Choosing an oriented symplectic basis $(A_I, B^I)\in H_3(X, \mathbb{Z})$, with $I=0,\dots,\hc$, and working in units in which the flux quanta are integral, we introduce
\begin{align*}
    \bm{f}&=(f_I, f^I)^{\mathsf T}\,, \quad \text{with }\;   f_I = \int_{B^I}F_3\, , \; f^I = \int_{A_I} F_3\, , \\
    \bm{h}&=(h_I, h^I)^{\mathsf T}\,, \quad \text{with }\;   h_I = \int_{B^I}H_3\, , \; h^I = \int_{A_I}H_3\, . 
\end{align*}
Thus $\bm{f},\bm{h}\in\mathbb{Z}^{2\hc+2}$. The fluxes generate the Gukov-Vafa-Witten superpotential \cite{Gukov:1999ya}
\begin{equation} \label{eq:superpot}
    W = \int_X (F_3 - \tau H_3) \wedge \Omega = (\bm{f}-\tau \bm{h})^{\mathsf T} \cdot \Sigma \cdot \bm{\Pi} \,,\qquad \Sigma = \begin{pmatrix}
        \hphantom{-}{\bm{0}} & \hphantom{-}\mathbb{I} \\
        -\mathbb{I} & \hphantom{-}\bm{0} \\
    \end{pmatrix} \,,
\end{equation}
where the dependence on the complex structure moduli is carried by the period vector
\begin{equation}
    \bm{\Pi} =({\cal F}_I, X^I)^{\mathsf T} \; , \qquad  
    X^I = \int_{A_I} \Omega \,, \; 
    {\cal F}_I = \int_{B^I} \Omega \,. 
\end{equation}

Special geometry \cite{Strominger:1990pd,Candelas:1990pi} relates these periods to a holomorphic prepotential. On a patch with $X^0\neq0$, we introduce the affine coordinates $z^a=X^a/X^0$, $a=1,\dots,\hc$. After fixing the gauge to $X^0=1$, the period vector becomes
\begin{equation}
    \bm{\Pi}(\bm{z}) = \begin{pmatrix}
        2F - z^a F_a \\
        F_a \\
        1 \\
        z^a
    \end{pmatrix} \,, 
\end{equation}
where $F_a=\partial_{z^a}F$ and $F$ denotes the prepotential in this gauge. Near a large complex structure (LCS) point, mirror symmetry expresses $F$ in terms of topological data of the mirror threefold $\widetilde{X}$ \cite{Candelas:1990rm,Ceresole:1992su,Hosono:1993qy},
\begin{equation}\label{eq:prepotential}
    F(\bm{z}) = -\frac{1}{6} \widetilde{\kappa}_{a b c} z^a z^b z^c + \frac{1}{2}a_{a b}z^a z^b+b_a z^a + \tfrac{\mathrm{i}\zeta(3)\chi(\widetilde{X})}{16\pi^3} + F_{\rm inst}(\bm{z}) \, ,
\end{equation}
where $\widetilde{\kappa}_{abc}$ are the integer triple intersection numbers of $\widetilde{X}$. The coefficients $a_{ab}$ and $b_a$ are rational topological data in the chosen integral symplectic basis, following the conventions of \cite{Demirtas:2023als}. In particular, $a_{ab} = \widetilde{\kappa}_{abc}/2\, \rm{mod}\, \mathbb{Z}$,  $b_a$ is determined by the second Chern class of $\widetilde{X}$ and $\chi(\widetilde{X})$ is the Euler characteristic of $\widetilde{X}$. 

The remaining contribution arises from worldsheet instantons and is determined by the effective curves of $\widetilde{X}$ \cite{Hosono:1993qy,Hosono:1994av},
\begin{equation}\label{eq:Finst}
    F_{\rm inst}(\bm{z}) = -\frac{1}{(2\pi \mathrm{i})^3} \sum_{\tilde{\mathbf{q}} \in \mathcal{M}(\widetilde{X})} \mathscr{N}_{\tilde{\mathbf{q}}}\,\mathrm{Li}_3 \left( \mathrm{e}^{2\pi \mathrm{i}\,\tilde{\mathbf{q}}\cdot\bm{z}} \right)\; , \qquad \mathrm{Li}_3(x) = \sum_{\ell=1}^{\infty} \frac{x^\ell}{\ell^3} \,,
\end{equation}
where the sum is over non-zero effective curve classes $\tilde{\mathbf{q}}$ in the Mori cone $\mathcal{M}(\widetilde{X})$. The integers $\mathscr{N}_{\tilde{\mathbf{q}}}$ are the associated genus-zero Gopakumar-Vafa invariants \cite{Gopakumar:1998ii,Gopakumar:1998jq}. 

Having specified the geometry and the quantised fluxes, the vacuum problem reduces to a system of non-linear equations. For a fixed flux vector $\bm{F}=(\bm{f},\bm{h})^{\mathsf T}$, we seek values of $(\bm{z},\tau)$ that satisfy the complex structure and axio-dilaton $F$-term conditions
\begin{equation}
\begin{split}\label{eq:Fterms}
    D_\tau W &= \frac{1}{\bar{\tau}-\tau}(\bm{f}-\bar{\tau}\bm{h})^{\mathsf T} \cdot \Sigma \cdot \bm{\Pi}(\bm{z}) = 0\,,  \\
    D_{z^a} W &= (\bm{f}-\tau \bm{h})^{\mathsf T} \cdot \Sigma \cdot \left(\partial_{z^a} + \partial_{z^a} K(\bm{z}, \tau)\right)\bm{\Pi}(\bm{z}) = 0 \,,
\end{split}
\end{equation}
where the relevant part of the Kähler potential is
\begin{equation*}
    K(\bm{z}, \tau) = -\ln \left( -\mathrm{i}\, \bm{\Pi}^{\dagger}(\bm{z}) \cdot \Sigma \cdot \bm{\Pi}(\bm{z}) \right) - \ln \left( -\mathrm{i}\, (\tau - \overline{\tau} )\right)\, .
\end{equation*}

The fluxes are subject to the integrated D3-brane tadpole condition. For the non-trivial flux configurations considered here, and in the absence of anti-D3-branes, this gives
\begin{equation} \label{eq:tadpole}
       0 < N_{\rm flux} \leq Q_{D3} \,,
\end{equation}
where 
\begin{equation} \label{eq:nflux}
        N_{\rm flux}=\int_X H_3 \wedge F_3 = \bm{f}^{\mathsf T} \cdot \Sigma \cdot \bm{h} \,,
\end{equation}
and $Q_{D3}$ denotes the available D3-brane tadpole determined by the net localised D3-charge. It includes the relevant contributions from the orientifold planes and from curvature or gauge flux in the D7-brane sector. 

Control of the LCS expansion imposes additional conditions on the continuous fields. The saxions $\operatorname{Im}(\bm{z})$ must lie in the interior of the Kähler cone $\mathcal{K}(\widetilde{X})$ of $\widetilde{X}$. Equivalently,
$\tilde{\mathbf{q}}\cdot\operatorname{Im}(\bm{z})>0$ for every non-zero effective curve class of $\widetilde{X}$. The corresponding instanton weights $\mathrm{e}^{-2\pi\tilde{\mathbf{q}}\cdot\operatorname{Im}(\bm{z})}$ in \eqref{eq:Finst} must be sufficiently suppressed, and the truncated instanton expansion must converge to the required numerical accuracy. There is no known universal or analytic lower bound on each individual $\operatorname{Im}(z^a)$. The validity of the expansion must instead be assessed using the cone conditions and the convergence of the instanton series for the geometry under consideration. Finally, physical solutions require $s=\operatorname{Im}(\tau)>0$, with string coupling $g_s=1/s$.

The solutions of Eq.~\eqref{eq:Fterms} possess discrete redundancies, so distinct field and flux representatives need not correspond to distinct physical vacua. The first identification follows from an $\mathrm{SL}(2, \mathbb{Z})$ symmetry under which the axio-dilaton and fluxes transform as
\begin{equation}
    \tau \to \frac{\alpha \tau + \beta}{\gamma \tau + \delta} \; , \qquad 
    \begin{pmatrix}
        \bm{f} \\ \bm{h}
    \end{pmatrix} \to 
     \begin{pmatrix}
        \alpha & \beta \\ \gamma & \delta
    \end{pmatrix}
     \begin{pmatrix}
        \bm{f} \\ \bm{h}
    \end{pmatrix}
    \,, \; 
    \begin{pmatrix}
        \alpha & \beta \\ \gamma & \delta
    \end{pmatrix} \in \mathrm{SL}(2,\mathbb{Z}) \,.
\end{equation}
Under this combined transformation,
\begin{equation*}
    \bm{f}-\tau\bm{h}
    \to
    \frac{\bm{f}-\tau\bm{h}}{\gamma\tau+\delta}\; ,
    \qquad
    W
    \to
    \frac{W}{\gamma\tau+\delta}\,.
\end{equation*}
The Kähler potential undergoes the compensating Kähler transformation. The $F$-term equations are therefore covariant, while $N_{\rm flux}$ remains invariant.

A second set of identifications arises from monodromies of the period vector. For a given family of CY threefolds, these generate an integral subgroup
$\Gamma_{\rm mon}\subseteq\mathrm{Sp}(2\hc+2,\mathbb{Z})$. In particular, the discrete axionic shifts at large complex structure,
\begin{equation}
    z^a \longrightarrow z^a+n^a\,,
    \qquad n^a\in\mathbb{Z}\,,
\end{equation}
are realised as integral symplectic transformations of the periods. In an integral symplectic basis, the elementary shift $z^a\to z^a+1$ is represented by a matrix $T_a\in\Gamma_{\rm mon}$, such that
\begin{equation}
    \bm{\Pi}(\bm{z}+\bm{n})
    =
    M_{\bm{n}}\,\bm{\Pi}(\bm{z})\,,
    \qquad
    M_{\bm{n}}
    =
    \prod_a T_a^{\,n^a}
    \in\Gamma_{\rm mon}\,.
\end{equation}
The corresponding representative of a flux vacuum is obtained by transforming the fluxes simultaneously,
\begin{equation}\label{eq:flux_mon}
    \bm{f}\longrightarrow M_{\bm{n}}\bm{f}\,,
    \qquad
    \bm{h}\longrightarrow M_{\bm{n}}\bm{h}\,.
\end{equation}
Since $M_{\bm{n}}^{\mathsf T}\Sigma M_{\bm{n}}=\Sigma$, this joint action preserves the superpotential in \eqref{eq:superpot}, the Kähler potential, the F-term equations and $N_{\rm flux}$. The shifts may therefore be used to map the real parts of the complex structure moduli to a chosen LCS axion cell, while the same physical vacuum is represented by the monodromy-transformed flux choice in \eqref{eq:flux_mon}. More generally, the relevant identifications are generated by the geometric monodromy group $\Gamma_{\rm mon}$, which need not coincide with the complete integral symplectic group.

To avoid over-counting, the fields and fluxes must be mapped together. We map $\tau$ to the standard fundamental domain of $\mathrm{SL}(2,\mathbb{Z})$ and use the unit-shift monodromies at LCS to choose representatives for the complex structure axions. With the boundary convention adopted here, the representative region is
\begin{equation}\label{eq:FDdom}
    \mathcal{F} = \{ \Re(z^a) \in (-0.5, 0.5], c_0 \in (-0.5, 0.5], \tau\in\mathbb{H}, |\tau| \geq 1 \} \,.
\end{equation}
This region combines the axio-dilaton fundamental domain with a fundamental cell for the LCS shift monodromies. 

For later use, we define the reduced Kähler-covariant vacuum superpotential
\begin{equation}
    W_0(\bm{z}, \tau) = \sqrt{2/\pi} \left\langle e^{K(\bm{z},\tau)/2}\,  W(\bm{z}, \tau) \right\rangle \,,
\end{equation}
where the brackets denote evaluation at a solution of Eq. \eqref{eq:Fterms}. Under $\mathrm{SL}(2,\mathbb{Z})$,
\begin{equation*}
    W_0 \to \frac{|\gamma\tau+\delta|}{\gamma\tau+\delta}\,W_0\,.
\end{equation*}
Thus $W_0$ is covariant up to a phase, while its magnitude $|W_0|$ is invariant. 

In this work, we use \verb|JAXVacua| \cite{Dubey:2023dvu} to construct the period vector, evaluate the $F$-term equations and derived observables, and map each solution together with its flux vector to the representative region $\mathcal{F}$. Candidate solutions are obtained by employing standard optimisation routines such as \verb|scipy.optimize.root| to a real representation of Eq.~\eqref{eq:Fterms}. The solver therefore searches directly for zeros of the $F$-term equations rather than minima of their norm. For candidates produced by the generative models, the sampled values of $(\bm{z},\tau)$ provide the initial guesses, while the integer flux vector remains fixed during numerical minimisation. We retain the resulting solutions only if they satisfy the $F$-term equations to the prescribed tolerance and obey the constraints stated in Sec.~\ref{sec:conditions}.

\section{Generative models}\label{sec:ml_section}

The aim of this work is to take a step towards constructing a foundation model for Type IIB flux vacua. Such a model should learn statistical structures shared across different compactification geometries. It could then be used both to reproduce the distributions represented in existing datasets and to direct searches towards specified regions of the flux landscape.

Previous generative studies have demonstrated this approach for individual geometries~\cite{Krippendorf:2025mhp,Walden:2025cpf}. Here we instead seek to learn a common distribution across several geometries. Within the dataset considered in this work, each geometry is labelled by its Hodge pair~$(h^{1,2},h^{1,1})$. We model the conditional distribution
\begin{equation}
    \bm{F}, \bm{z}_0, \tau_0 \sim p(\bm{F},\bm{z}_0, \tau_0 \mid \hc, \hk)
\end{equation}
where $\bm{F}$ denotes the quantised flux vector and $(\bm{z}_0,\tau_0)$ is an approximate solution of the corresponding $F$-term equations. The distribution $p$ is
determined by multiple factors: the training distribution, the neural network architecture and the learning algorithm with their respective biases. It should not be interpreted as a canonical measure on the flux landscape.

The generated values $(\bm{z}_0,\tau_0)$ are treated as initial conditions for a numerical root finder rather than accepted directly as vacua. During this refinement, the flux vector $\bm{x}$ is held fixed while $\bm{z}$ and $\tau$ are varied until the $F$-term equations are satisfied to the required numerical precision. It is important to note that the generative models used here are inherently probabilistic, and there is no guarantee that such samples will always give rise to solutions. Therefore, careful evaluation of the samples is always required. We elaborate on this further in Sec. \ref{sec:conditions}.

The distribution can be refined further by conditioning on physical quantities such as $N_{\mathrm{flux}}$, $|W_0|$ or $1/g_s$. For example, conditioning on the flux-induced D3-brane charge gives
\begin{equation}
\bm{F}, \bm{z}_0, \tau_0 \sim p(\bm{F},\bm{z}_0, \tau_0 \mid h^{1,2},h^{1,1},N_{\rm flux})
\end{equation}
and targets a specified tadpole sector. Since $N_{\mathrm{flux}}$ is determined exactly by the generated flux vector, this condition can be verified before numerical refinement. By contrast, $|W_0|$ and $1/g_s$ are evaluated at the refined solution and must therefore be verified after solving the $F$-term equations. Conditional generation consequently provides a proposal mechanism for targeted searches, while the final physical properties remain subject to explicit numerical validation.

This is a mixed discrete-continuous generative problem. The flux vector $\bm{x}$ is integer valued, whereas the moduli $\bm{z}_0$ and $\tau_0$ are continuous. We compare models based on transformers \cite{Vaswani:2017lxt} and diffusion models \cite{DBLP:conf/nips/AustinJHTB21,sahoo2024simpleeffectivemaskeddiffusion} for learning these conditional distributions. We also perform a preliminary scaling study to determine how performance depends on the amount of training data and the model capacity.

Following pre-training, the resulting models can be adapted to particular geometries or physical targets. We investigate fine-tuning and GRPO as complementary post-training methods for improving generation in sparsely sampled regions and directing the model towards specified objectives. Their ability to generalise beyond the training distribution is assessed explicitly on held-out regions rather than inferred from their performance on the training data.

This work establishes a common data representation, generative objective and post-training framework for extending generative studies of individual compactifications towards a foundation model for Type IIB flux vacua.

\subsection{The dataset}\label{app:dataset}

\begin{table}[!t]
    \centering
    \begin{tabular}{ccc||ccc||ccc||ccc}
    $h^{1,2}$ & $h^{1,1}$ & $Q_{D3}$  & $h^{1,2}$ & $h^{1,1}$ & $Q_{D3}$  & $h^{1,2}$ & $h^{1,1}$ & $Q_{D3}$  & $h^{1,2}$ & $h^{1,1}$ & $Q_{D3}$ \\
    \hline \hline
    2 & 120 & 124   & 3 & 99 & 104  & 4 & 98 & 104  & 5 & 89 & 96  \\
    \hline
    2 & 128 & 132   & 3 & 107 & 112  & 4 & 118 & 124  & 5 & 149 & 156 \\
    \hline
    2 & 132 & 136  & 3 & 115 & 120  & 4 & 122 & 128  & 5 & 153 & 160  \\
    \hline
    2 & 144 & 148  & 3 & 119 & 124 & 4 & 130 & 136  & 5 & 165 & 172  \\
    \hline
    2 & 272 & 276  & 3 & 243 & 248  & 4 & 142 & 148  & 5 & 213 & 220  \\
    
    \end{tabular}
    \caption{Summary of the geometries making up the dataset used in this paper.} 
    \label{tab:dataset}
\end{table}

All models are trained on the flux vacuum dataset constructed in \cite{Ebelt:2023clh}. Before constructing the training sets, we convert the quadratic prepotential coefficients $a_{ab}$ to the convention of~\cite{Demirtas:2023als}. In this convention, the unit shifts $z^a\mapsto z^a+1$ are represented by integral symplectic monodromy matrices. The change of convention corresponds to a change of symplectic basis, so the period vector and the RR and NSNS flux vectors are transformed together.

For a subset of the original data, the transformed flux vectors are not integral. Since flux quantisation requires integer fluxes, these components are rounded to the nearest integers. This rounding changes the discrete flux background and therefore displaces the solution from its original position in moduli space. We use the transformed moduli as initial values and solve the $F$-term equations again at fixed rounded flux using \verb|JAXVacua| \cite{Dubey:2023dvu} and \verb|scipy.optimize.root|. The solver acts on the real representation of the $F$-term equations and uses the Jacobian computed by \verb|JAXVacua|. We retain only converged roots that satisfy the required numerical tolerance and physical consistency conditions.

Each certified solution is subsequently mapped to a common representative region. The axio-dilaton is reduced to the fundamental domain of $\mathrm{SL}(2,\mathbb{Z})$, while the complex structure axions are mapped to the chosen LCS axion cell \eqref{eq:FDdom} using the corresponding integral monodromies. In both cases, the fields and fluxes are transformed simultaneously. Equivalent representatives are then removed. This procedure yields $10^5$ distinct flux vacua for each geometry listed in Tab.~\ref{tab:dataset}.

A CY threefold is not uniquely specified by its Hodge numbers. In the present dataset, however, each Hodge pair $(h^{1,2},h^{1,1})$ is associated with a single geometry and can therefore be used as an unambiguous label. The remaining geometric information is implicit in the corresponding training samples. This choice is sufficient for the present comparison, but it is not a geometry-complete representation. A model trained on several geometries with the same Hodge pair, or intended to generalise to previously unseen geometries, must instead be conditioned on a richer description of the compactification. This could include explicit topological and period data or a learned embedding of the geometry which we leave for future investigations.

\begin{figure}[!t]
    \centering
    \includegraphics[]{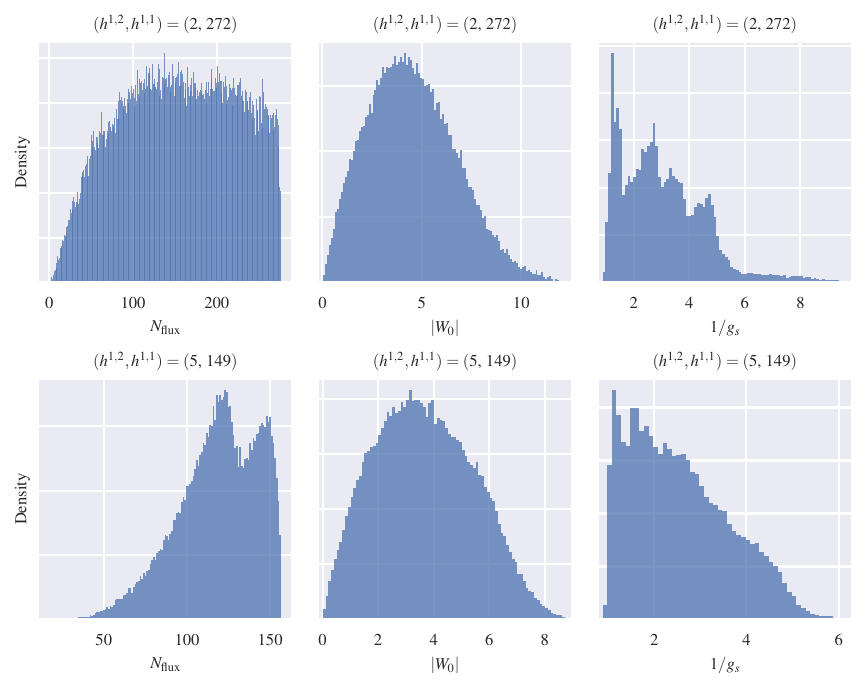}
    \caption{Distribution of training data for the geometries defined by $(\hc, \hk)= (2, 272)$ and $(\hc, \hk)=(5,149)$. Distributions for $\nflux$, $|W_0|$ and $1/g_s$ are shown. To increase readability, the plots exclude outliers.}
    \label{fig:train_data_2_272_5_149}
\end{figure}

Fig.~\ref{fig:train_data_2_272_5_149} shows the empirical distributions of $N_{\rm flux}$, $|W_0|$ and $1/g_s$ for two representative geometries. These distributions are strongly non-uniform and depend on both the underlying geometry and the sampling procedure used to construct the datasets. Consequently, the sparsely populated regions receive less statistical support during training, which must be taken into account when interpreting the model performance. We use the empirical ranges displayed in Fig.~\ref{fig:train_data_2_272_5_149} to define the target intervals employed in the subsequent refinement and evaluation of the models.

\subsection{Transformers}\label{sec:transformers}

Transformers \cite{Vaswani:2017lxt} have proven to be powerful tools in various scenarios and have previously been successfully applied to the Type IIB vacua setting \cite{Walden:2025cpf}. Roughly speaking, transformers map sequences to sequences, hence their original application for machine translation. In this paper, we exploit this property by taking an input sequence ($h^{1,2},h^{1,1},\dots)$ and mapping it to a solution of the $F$-term conditions as an output sequence. The $\dots$ include the remainder of the prompt which refers here to $\nflux, |W_0|, g_s$ or nothing. In the following, we sketch the training and sampling routine for transformers, leaving a more detailed discussion to Appendix \ref{app:transformer}.

Training a transformer is done in several steps. First, the training data is organised in pairs of input and output sequences. As previously mentioned, input sequences contain the respective Hodge pairs $(\hc, \hk)$ and potentially additional quantities. 
Output sequences consist of the flux vectors $\bm{F}$, complex structure moduli $\bm{z}$ and axio-dilaton~$\tau$. These sequences are then tokenised, i.e.~broken up into strings of tokens which are, in turn, mapped into an embedding space. Once embedded, the transformer is trained to learn dependencies between the different tokens of each string through the attention mechanism \cite{Vaswani:2017lxt}. Finally, when the model is fully trained, the model defines a probability distribution over complete sequences, from which new examples can be generated.

Drawing new sequences from the trained transformer is done through the following iterative procedure. Given an input sequence $(\hc, \hk, \dots)$, the first token $x_0$ of the output sequence is drawn from the distribution
\begin{equation}
    x_0 \sim p_{\bm \theta} ( x_0 \mid \hc, \hk, \dots) \,,
\end{equation}
where $\bm{\theta}$ denotes the optimised model parameters, by feeding the input sequence to the transformer which then assigns probabilities to each possible next token. Once $x_0$ has been drawn, it is again fed into the transformer, together with $(\hc, \hk)$, to draw the next token $x_1$. This results in repetitive draws from the distributions
\begin{equation}
    x_i \sim p_{\bm \theta}(x_i \mid x_{i-1}, \dots, x_0, \hc, \hk, \dots)
\end{equation}
until the desired output sequence is completed. Since the transformer is to map sequences to corresponding sequences this should lead to a sensible output sequence. In our case prompting the trained model with some Hodge pair ($h^{1,2},h^{1,1}$), allows us to sample a flux vector $\bm{F}$ and initial guesses $\bm{z_0}$ and $\tau_0$ for the complex structure moduli and axio-dilaton. After calling a numerical solver this should give rise to a solution for the respective $F$-term conditions \eqref{eq:Fterms}. If the prompt additionally contained a specific $\nflux, |W_0|$ or $1/g_s$ value, the solution is furthermore expected to approximately compute the prompted value.

\subsection{Diffusion models}\label{sec:diffusion}

We employ an absorbing-state discrete diffusion model~\cite{DBLP:conf/nips/AustinJHTB21,sahoo2024simpleeffectivemaskeddiffusion} to learn the distribution of integer flux configurations. In the forward diffusion process, individual flux coordinates are progressively replaced by a special \texttt{[MASK]} token, with the masking probability controlled by a continuous noise level. A bidirectional transformer is then trained to reconstruct the original flux values from the partially masked configurations, conditioned on the CY geometry and the noise level. Generation starts from a fully masked flux vector and proceeds by iteratively revealing the masked coordinates. Unlike the autoregressive transformer described above, the masked diffusion model does not impose a fixed left-to-right factorisation of the flux distribution, but instead learns to predict flux components conditioned on arbitrary subsets of the remaining coordinates. We summarise the formulation below and provide further architectural and training details in Appendix~\ref{app:discrete-diffusion}.

For a geometry defined by its Hodge pair $(h^{1,2},h^{1,1})$, we represent the RR and NS--NS fluxes by a sequence of categorical tokens $\bm{x}_0$, tokenised from the original flux vector $\bm{F}$. The forward diffusion process progressively replaces the entries of this sequence by the fixed \texttt{[MASK]} token with probability $t$, where $t \sim \text{Unif}[0,1]$. The forward transition kernel therefore is
\begin{equation}
    q_t(x_{t,i}\mid x_{0,i})
    =
    (1-t)\,\delta_{x_{t,i},x_{0,i}}
    +
    t\,
    \delta_{x_{t,i},\mathrm{MASK}}.
\end{equation}

A bidirectional transformer is trained to reverse this corruption. Given the corrupted sequence $\bm{x}_t$, the geometry $(h^{1,2}, h^{1,1})$, and the diffusion level $t$, it predicts
\begin{equation}
    p_\theta
    \left(
        x_{0,i}\mid\bm{x}_t,h^{1,2}, h^{1,1},t
    \right)
\end{equation}
for every masked coordinate $i$. Training minimises the categorical cross-entropy between these predictions and the original flux values, evaluated only at masked coordinates. Since the attention mechanism is bidirectional, each prediction can use all currently revealed flux entries, independently of whether they occur before or after the predicted coordinate.

Generation begins from a sequence in which all active flux coordinates are masked. At sampling iteration $s$, the model predicts the clean token distribution at every remaining masked coordinate. One coordinate $j_s$ is then selected and assigned a value according to
\begin{equation}
    x_{j_s}^{(s+1)}
    \sim
    p_\theta
    \left(
        x_{0,j_s}\mid
        \bm{x}^{(s)},h^{1,2}, h^{1,1},t_s
    \right),
\end{equation}
where $t_s$ is determined from the current fraction of masked coordinates. We consider sequential confidence-based sampling, in which the coordinate with the most confident coordinate prediction is revealed, and subsequently used for next token prediction. The procedure is repeated until the flux vector is complete. Although the coordinates are revealed sequentially, the model is not purely autoregressive: it is trained on arbitrary masking patterns, and the generation order is chosen only at sampling time.

Once a complete flux sequence $\bm{x}$ has been generated, a separate transformer-based initial-guess generator predicts the corresponding complex-structure moduli and axio-dilaton,
\begin{equation}
    \left(\bm{z}_0,\tau_0\right)
    =
    I_\phi(\bm{x},h^{1,2}, h^{1,1}).
\end{equation}
These predictions serve as initial guesses for the numerical solver, which keeps the flux vector fixed and refines \(\bm z\) and \(\tau\) until the $F$-term conditions are satisfied to the required numerical precision.

\subsection{Pre-training, GRPO, and fine-tuning}

Having introduced both transformers and diffusion models, we illustrate all training stages for both models in the following. 

\paragraph{Pre-training.}
 Training a transformer or a diffusion model usually starts with a pre-training stage. We sketch this stage here and refer the reader to Appendices \ref{app:transformer} and \ref{app:discrete-diffusion} for more details. 
 
To pre-train a model, some training data is collected as described in Sec.~\ref{app:dataset}, in our case solutions to the $F$-term conditions over several geometries, on which the generative models are then trained. During this pre-training stage, the models learn the training data distribution while being agnostic to the origin of the data. Once the distribution is learned, the models can be used to draw new samples from said distribution that are not contained in the training data, hence the models learn to interpolate.  Even more interesting is the question whether the models are able to extrapolate, i.e.~find samples that lie outside of the training data distribution. We elaborate on different methods to achieve this in the following paragraphs.

\paragraph{Post-training with GRPO.}
 In order to improve the performance of the pre-trained model, we add a Reinforcement Learning (RL) routine that involves \textit{Group Relative Policy Optimisation}~(GRPO)~\cite{shao2024deepseekmathpushinglimitsmathematical}. Roughly speaking, this routine consists of repetitively drawing samples from the model, rewarding those that give rise to solutions and penalising, or ignoring, those that do not. This leads to the models eventually learning the distinction between valid and invalid samples. An important subtlety of using such RL methods after pre-training is the possibility of sample collapse, a scenario in which the model learns to sample the same solutions over and over again. Since we are interested in diverse samples, we want to avoid such a scenario and modify the rewards such that new solutions are encouraged. A more detailed description of GRPO can be found in Appendices \ref{app:transformer_grpo} and \ref{app:diffusion-grpo}.

\paragraph{Fine-tuning.}
Once the models have been pre-trained, and potentially gone through post-training with GRPO, they can be fine-tuned to any specific task. This means that the pre-trained model will be trained again on a specialised dataset, instead of training a new model from scratch. This usually results in significantly shorter training times and also allows to extrapolate from a more general setting to a specialised one. This new dataset can for example include further conditions in the prompt to tune the model to specific solutions (e.g. small $|W_0|$). Another possibility is to fine-tune a pre-trained model on a small dataset of new geometries, hoping that the model is able to extrapolate from the larger dataset to the smaller. We provide evidence of this in Sec.~\ref{sec:fine-tuning}.

\subsection{Evaluation of models}\label{sec:conditions}

Generative models are often evaluated by comparing the distribution of their samples with that of the data. Such measures quantify statistical resemblance, but not whether an individual sample is correct. In our setting, correctness can be decided sample by sample. A sample is useful only if it describes a flux vacuum, and this can be verified by solving the $F$-term conditions at the sampled fluxes. Evaluation therefore reduces to counting the samples that pass a sequence of checks.

A sample $(\bm{F}, \bm{z}_0, \tau_0)$ is not itself a vacuum: The fluxes $\bm{F}=(\bm{f},\bm{h})$ are integers and are taken as given, whereas $\bm{z}_0$ and $\tau_0$ are at best approximate solutions. We therefore use them as initial guesses and apply the following routine:
\begin{enumerate}
    \item \emph{Tadpole:} Check that $\bm{F}$ obeys the tadpole bound~\eqref{eq:tadpole}.
    \item \emph{$F$-terms:} Keep $\bm{F}$ fixed and solve the $F$-term conditions $D_IW=0$ numerically, starting from $\bm{z}_0$ and $\tau_0$, to obtain $\bm{z}^*$ and $\tau^*$. A solution is kept only if the real and imaginary parts of all $F$-terms are below $10^{-10}$ in absolute value.
    \item \emph{Fundamental domain:} Map $(\bm{F}, \bm{z}^*, \tau^*)$ into the fundamental domain of $\Gamma_{\rm mon}$ and $\mathrm{SL}(2,\mathbb{Z})$. Configurations related by these transformations describe the same vacuum, so this step selects one representative per vacuum, and the uniqueness check below compares physically distinct vacua.
    \item \emph{Control:} Check that $\bm{z}^*$ lies in the LCS region, i.e., $\mathrm{Im}(\bm{z}^*)\in\mathcal{K}(\widetilde{X})$, cf.\ Sec.~\ref{sec:typeIIB}.
    \item \emph{Uniqueness:} Discard samples whose flux vector $\bm{F}$ coincides with that of a sample already accepted.\footnote{A single flux vector $\bm{F}$ can have several solutions of the $F$-term conditions, and can therefore give rise to several distinct vacua. We impose the stronger requirement that all accepted vacua have distinct flux vectors~$\bm{F}$.}
    \item \emph{Novelty:} Check whether the sample appears in the training data.
\end{enumerate}
A sample that passes all six checks is considered valid.

The checks probe different ways in which a generative model can fail. Steps 1--4 test whether a sample is a vacuum at all, i.e., whether the model has learned the physical constraints. Step~5 tests diversity: a model that returns the same vacuum repeatedly can succeed on most of its samples and still produce few distinct vacua. Step~6 tests memorisation, i.e., whether the model only reproduces its training data. None of the models produced a vacuum contained in the training data, so we do not report this check in the following.

\section{Experiments}\label{sec:experiments}

In this section, we demonstrate through experiments that transformers and diffusion models can generate type IIB flux vacua. The models reproduce
the distributions represented in their training data, and are capable of interpolating, i.e.~generating previously unseen, valid vacua across different CY geometries. An important question is whether their ability to generate valid solutions can be further improved through post-training with GRPO, and then not only in regions that are populated by the training data. In other words, we want to study not only improved interpolation, but also the possibility of extrapolation.

There is an important distinction to be made here. We are relying on reinforcement learning to improve a pre-trained model rather than learning a new distribution from scratch. The problem could in principle also be solved by ``pure" RL techniques, where the relevant physical information is fully encoded in the reward function. However, this would constitute a fundamentally different setting, in which the model learns the underlying physics through reinforcement learning rather than extrapolating from previously acquired knowledge. In our approach, the physical knowledge is acquired during pre-training, while GRPO is used to improve the model's ability to produce valid solutions. As a consequence, the rewards in our experiments are comparatively simple: they only distinguish valid from invalid output and contain essentially no physical information that the model could pick up on. Moreover, because the pre-trained model can already generate valid solutions, we avoid the difficulty of obtaining an initial reward signal that often arises in "pure" reinforcement learning.

A further requirement is the ability to transfer knowledge acquired during pre-training to new geometries for which only limited data are available. We therefore investigate whether fine-tuning provides an advantage over training a model from scratch on such datasets. We also investigate whether the models can be conditioned on physical quantities such as $N_{\mathrm{flux}}$, $|W_0|$, and $g_s$, enabling targeted searches for vacua with prescribed properties.
We also study how the performance of both approaches depends on model size and the amount of training data, to assess their potential for further scaling. Finally, we evaluate the sampling time of both the transformer and the diffusion model and compare with \texttt{JAXVacua} to understand advantages of our generative modelling approaches with traditional numerical sample generation.

Together, these experiments provide a systematic assessment of the generation, refinement, transfer learning, conditional sampling, and scaling capabilities that motivate the development of a future foundation model for type IIB flux vacua. Below, we evaluate each model through its capability of sampling valid solutions out of $1{,}000$ samples per prompt. The hyperparameters of both models are given in Appendix~\ref{app:transformer} and Appendix~\ref{sec:diffusion-hyperparameters}.

\subsection{Pre-training and post-training with GRPO}\label{sec:pre-train_grpo}

In this section, we investigate the capability of both models to find new valid solutions after both a pre-training and GRPO stage. First, the models are pre-trained on the aforementioned dataset (see Sec.~\ref{app:dataset}) containing all $20$ geometries. The models are prompted on the respective Hodge pair $(h^{1,2}, h^{1,1})$ and output a sequence of fluxes $\bm{F}=(\bm{f}, \bm{h})$ together with initial guesses $\bm{z}_0, \tau_0$ which are then used to solve the $F$-term conditions~\eqref{eq:Fterms} numerically. We are then interested in solutions that are valid according to Sec.~\ref{sec:conditions}. After both the pre-trained and the GRPO stage the models are evaluated on all geometries by drawing $1{,}000$ samples per geometry and evaluating them for each geometry with the same $h^{1,2}$. 

%----------------- transformer -------------------

\paragraph{Transformer.}
In Tab.~\ref{tab:transformer_pretrain_gpo_all_h12}, the performance of the pre-trained transformer is evaluated. The average validity, i.e., finding a solution for the prompted geometry, is given by
\begin{equation*}
    70.9\%\, ,\qquad 46.7\%\, ,\qquad 15.9\%\, ,\qquad 13.3\%
\end{equation*}
for $\hc \in \{2,3,4,5\}$, respectively. For all geometries with $h^{1,2}=2$, the performance is relatively high, while it gradually decreases as $h^{1,2}$ grows.

There also seems to be a large overlap in solutions between the different geometries with $h^{1,2}=2$, meaning that even if a solution was sampled by prompting the model with say $(h^{1,2}, h^{1,1})=(2, 120)$, the numerical solver still finds a solution for another geometry, e.g.~$(h^{1,2}, h^{1,1})=(2, 132)$. This suggests that for low $h^{1,2}$, it is significantly easier to find solutions and the setup is less sensitive to initial guesses. This effect dramatically reduces as $h^{1,2}$ grows and for $h^{1,2}=5$ almost no overlap is observed. 

\begin{table}[!t]
\renewcommand{\arraystretch}{1.3}
\centering
\resizebox{0.99\textwidth}{!}{
\begin{tabular}{c||c|ccccc||c|ccccc}
\multicolumn{1}{c}{\textbf{prompt}} & \multicolumn{6}{c}{\textbf{pre-trained}} & \multicolumn{6}{c}{\textbf{GRPO}} \\
$(h^{1,2}, h^{1,1})$ & $\#$valid &  $h^{1,1}=120$ & $128$ & $132$ & $144$ & $272$ & $\#$valid &  $h^{1,1}=120$ & $128$ & $132$ & $144$ & $272$ \\
    \hline
    $(2,120)$ &  849 & \cellcolor{yellow!25} 700 & 30 & 694 & 8 & 0 & 966 & \cellcolor{yellow!25} 959 & 33 & 952 & 31 & 0 \\ 
        $(2,128)$ & 772 & 28 & \cellcolor{yellow!25} 660 & 41 & 266 & 12 & 920 & 5 & \cellcolor{yellow!25} 877 & 65 & 355 & 17 \\ 
        $(2,132)$ &  756 & 409 & 16 & \cellcolor{yellow!25} 670 & 3 & 0 &  948 & 847 & 6 & \cellcolor{yellow!25} 942 & 7 & 0 \\ 
        $(2,144)$ & 822 & 45 & 193 & 94 & \cellcolor{yellow!25} 747 & 36 & 974 & 40 & 166 & 101 & \cellcolor{yellow!25} 962 & 88 \\ 
        $(2,272)$ & 834 & 5 & 6 & 12 & 257 & \cellcolor{yellow!25} 769 &  968 & 0 & 0 & 6 & 142 & \cellcolor{yellow!25} 965 \\ 
        \hline
    $(h^{1,2}, h^{1,1})$ & $\#$valid &  $h^{1,1}=99$ & $107$ & $115$ & $119$ & $243$ & $\#$valid &  $h^{1,1}=99$ & $107$ & $115$ & $119$ & $243$ \\
    \hline
    $(3,99)$ &  454 & \cellcolor{yellow!25} 420 & 1 & 28 & 24 & 0 &  960 & \cellcolor{yellow!25} 957 & 3 & 44 & 33 & 0 \\ 
        $(3,107)$ &  212 & 8 & \cellcolor{yellow!25} 146 & 28 & 15 & 26 & 968 & 1 & \cellcolor{yellow!25} 964 & 47 & 36 & 31 \\ 
        $(3,115)$ &  638 & 0 & 24 & \cellcolor{yellow!25} 574 & 3 & 147 &  985 & 0 & 22 & \cellcolor{yellow!25} 982 & 3 & 89 \\ 
        $(3,119)$ &  570 & 3 & 19 & 72 & \cellcolor{yellow!25} 528 & 6 & 946 & 0 & 27 & 86 & \cellcolor{yellow!25} 940 & 20 \\ 
        $(3,243)$ &  724 & 0 & 25 & 183 & 1 & \cellcolor{yellow!25} 669 & 986 & 0 & 0 & 1 & 1 & \cellcolor{yellow!25} 986 \\ 
        \hline
    $(h^{1,2}, h^{1,1})$ & $\#$valid & $h^{1,1}=98$ & $118$ & $122$ & $130$ & $142$ & $\#$valid & $h^{1,1}=98$ & $118$ & $122$ & $130$ & $142$ \\
    \hline
    $(4,98)$ &  250 & \cellcolor{yellow!25} 219 & 11 & 25 & 4 & 4 &  963 & \cellcolor{yellow!25} 960 & 3 & 6 & 19 & 10 \\ 
        $(4,118)$ &  134 & 0 & \cellcolor{yellow!25} 97 & 22 & 2 & 18 & 823 & 11 & \cellcolor{yellow!25} 816 & 26 & 3 & 9 \\ 
        $(4,122)$ &  204 & 2 & 6 & \cellcolor{yellow!25} 189 & 3 & 6 & 927 & 4 & 29 & \cellcolor{yellow!25} 925 & 0 & 11 \\ 
        $(4,130)$ &  223 & 0 & 9 & 18 & \cellcolor{yellow!25} 195 & 9 &  979 & 0 & 1 & 27 & \cellcolor{yellow!25} 978 & 7 \\ 
        $(4,142)$ &  138 & 0 & 13 & 23 & 8 & \cellcolor{yellow!25} 97 &  979 & 0 & 50 & 56 & 0 & \cellcolor{yellow!25} 975 \\ 
        \hline
    prompt &  $\#$valid &  $h^{1,1}=89$ & $149$ & $153$ & $165$ & $213$ &  $\#$valid &  $h^{1,1}=89$ & $149$ & $153$ & $165$ & $213$ \\
    \hline
    $(5,89)$ & 99 & \cellcolor{yellow!25} 96 & 2 & 0 & 0 & 1 & 951 & \cellcolor{yellow!25} 950 & 0 & 0 & 1 & 1 \\ 
        $(5,149)$ & 38 & 0 & \cellcolor{yellow!25} 38 & 0 & 0 & 0 & 739 & 0 & \cellcolor{yellow!25} 739 & 0 & 0 & 0 \\ 
        $(5,153)$ &  200 & 0 & 0 & \cellcolor{yellow!25} 200 & 0 & 0 &  916 & 0 & 0 & \cellcolor{yellow!25} 916 & 0 & 0 \\ 
        $(5,165)$ & 193 & 0 & 0 & 0 & \cellcolor{yellow!25} 193 & 0 & 944 & 0 & 0 & 1 & \cellcolor{yellow!25} 944 & 0 \\ 
        $(5,213)$ & 137 & 1 & 0 & 0 & 0 & \cellcolor{yellow!25} 136 & 969 & 0 & 0 & 0 & 0 & \cellcolor{yellow!25} 969 \\ 
        \hline
    \end{tabular}}
\caption{Evaluation of pre-trained versus GRPO-refined transformer. The first column indicates the prompted geometry. The column $\#$valid records the number of (unique) valid solutions (in any geometry with the same $h^{1,2}$) out of $1{,}000$ samples. In the remaining columns, the number of valid samples for each geometry is recorded where the yellow entry corresponds to the prompted geometry.} 
\label{tab:transformer_pretrain_gpo_all_h12}
\end{table}

Once pre-trained, the model is further improved via GRPO by prompting each geometry and rewarding valid solutions. In order to avoid sample collapse, i.e.~the model just learning to draw the same sample repetitively, each sample is collected and new samples, i.e.~not previously drawn, are further rewarded. The performance of the improved model is evaluated in Tab.~\ref{tab:transformer_pretrain_gpo_all_h12}. The average validity, i.e.~finding a solution for the prompted geometry, is given by
\begin{equation*}
    94.1\%\, ,\qquad 96.6\%\, ,\qquad 93.1\%\, ,\qquad 90.4\%
\end{equation*}
for $\hc \in \{2,3,4,5\}$, respectively. It can be observed that the performance for all geometries dramatically increased and the performance decrease with increasing $\hc$ is almost completely eradicated. For almost all geometries, more than $90\%$ of the samples drawn give rise to valid and unique solutions. Furthermore, no sample collapse can be observed ($\#$valid counts the number of unique solutions).

\paragraph{Diffusion model.}
With the same table setup as in Tab.~\ref{tab:transformer_pretrain_gpo_all_h12}, Tab.~\ref{tab:pretrain_all_h12_diffusion} presents the results for the pre-trained diffusion model. The average validity i.e.~finding a solution for the prompted geometry, is given by 
\begin{equation*}
    78.86\%\, ,\qquad 61.54\%\, ,\qquad 39.96\%\, ,\qquad 30.24\%
\end{equation*}
for $h^{1,2}\in \{2,3,4,5\}$, respectively. Again, we observe high validity for $\hc=2$, with performance decreasing as $\hc$ grows, albeit not as drastically as for the transformer.

Similarly, a large overlap in solutions between the different geometries with $h^{1,2}=2$ is observed, with a similar reduction of this trend as $h^{1,2}$ grows. This further supports the previously mentioned hypothesis, which states that for low $h^{1,2}$, it is significantly easier to find solutions, hence the setup is less sensitive to initial guesses.

\begin{table}[!t]
\renewcommand{\arraystretch}{1.3}
\centering
\resizebox{\textwidth}{!}{
\begin{tabular}{c||c|ccccc||c|ccccc}
\multicolumn{1}{c}{\textbf{prompt}} & \multicolumn{6}{c}{\textbf{pre-trained}} & \multicolumn{6}{c}{\textbf{GRPO}} \\
$(h^{1,2}, h^{1,1})$ & $\#$valid &  $h^{1,1}=120$ & $128$ & $132$ & $144$ & $272$ & $\#$valid &  $h^{1,1}=120$ & $128$ & $132$ & $144$ & $272$  \\
    \hline
    $(2,120)$ & 959 & \cellcolor{yellow!25}772  & 155 & 664 & 419 & 401  & 985  & \cellcolor{yellow!25}859  & 146 & 663 & 456 & 433  \\ 
        $(2,128)$ &  930 & 254 & \cellcolor{yellow!25}797  & 268 & 319 & 339  & 992 & 200 & \cellcolor{yellow!25}952  & 149 & 488  & 529  \\ 
        $(2,132)$ & 931 & 431 & 202 & \cellcolor{yellow!25}747 & 449  & 345   &  983  & 550 & 132 & \cellcolor{yellow!25}909 & 284 & 403  \\ 
        $(2,144)$ &  926 & 335 & 154 & 399 & \cellcolor{yellow!25}821  & 323  & 961 & 298 & 82 & 308 & \cellcolor{yellow!25}883 &  280  \\ 
        $(2,272)$ &  932 & 378 & 94 & 457 & 338 & \cellcolor{yellow!25}806  & 974  & 497 & 104 & 547  & 445 & \cellcolor{yellow!25}913  \\ 
        \hline
    $(h^{1,2}, h^{1,1})$ &  $\#$valid  & $h^{1,1}=99$ & $107$ & $115$ & $119$ & $243$ & $\#$valid  & $h^{1,1}=99$ & $107$ & $115$ & $119$ & $243$  \\
    \hline
    $(3,99)$ &  788 & \cellcolor{yellow!25}619  & 34 & 252 & 175 & 194  & 951 & \cellcolor{yellow!25}907  & 31 & 194 & 203 & 165  \\ 
        $(3,107)$ &  669 & 11 & \cellcolor{yellow!25}460  & 159 & 186 & 125 & 916 & 9 & \cellcolor{yellow!25}796  & 213 & 376 & 394  \\ 
        $(3,115)$ &  866 & 98 & 48 & \cellcolor{yellow!25}691  & 355 & 380  & 985 & 100 & 55 & \cellcolor{yellow!25}921  & 535 & 636  \\ 
        $(3,119)$ &  823 & 63 & 38 & 333 & \cellcolor{yellow!25}654  & 200   &  917 & 100 & 70 & 451 & \cellcolor{yellow!25}697  & 321  \\ 
        $(3,243)$ &  824 & 62 & 58 & 341 & 296 & \cellcolor{yellow!25}653   & 919 & 26 & 24 & 225 & 215 & \cellcolor{yellow!25}856   \\ 
        \hline
    $(h^{1,2}, h^{1,1})$ & $\#$valid & $h^{1,1}=98$ & $118$ & $122$ & $130$ & $142$ & $\#$valid & $h^{1,1}=98$ & $118$ & $122$ & $130$ & $142$  \\
    \hline
    $(4,98)$ & 552 & \cellcolor{yellow!25}483  & 33 & 66 & 6 & 21  & 748 & \cellcolor{yellow!25}717  & 13 & 48 & 4 & 24  \\ 
        $(4,118)$ &  357 & 4 & \cellcolor{yellow!25}301  & 35 & 11 & 42  & 886 & 1 & \cellcolor{yellow!25}878  & 10 & 2 & 35  \\ 
        $(4,122)$ &  523 & 2 & 49  & \cellcolor{yellow!25}470  & 2 & 39 & 987 & 8 & 30 & \cellcolor{yellow!25}986  & 4 & 34  \\ 
        $(4,130)$ &  513 & 5 &  31 & 45 & \cellcolor{yellow!25}453  & 30  & 953 & 3 & 80 & 53  & \cellcolor{yellow!25}948  &  31  \\ 
        $(4,142)$ &  355 & 5 & 27  & 46 & 19 & \cellcolor{yellow!25}291   & 984 & 0 & 91 & 45 & 3 & \cellcolor{yellow!25}983  \\ 
        \hline
    $(h^{1,2}, h^{1,1})$ & $\#$valid& $h^{1,1}=89$ & $149$ & $153$ & $165$ & $213$ & $\#$valid& $h^{1,1}=89$ & $149$ & $153$ & $165$ & $213$  \\
    \hline
    $(5,89)$ & 235 & \cellcolor{yellow!25}182  & 3 & 9 & 24 & 41 &  833 & \cellcolor{yellow!25}815  & 0 & 0 & 33 & 61  \\ 
        $(5,149)$ & 266 & 15 & \cellcolor{yellow!25}239  & 1 & 2 & 17  & 972 & 35 & \cellcolor{yellow!25}967  & 7 & 2 & 61  \\ 
        $(5,153)$ &  258 & 19 & 3  & \cellcolor{yellow!25}218  & 9 & 21 & 786 & 11 & 16 & \cellcolor{yellow!25}771  & 44 & 4 \\ 
        $(5,165)$ &  485 & 5 & 0 & 1 & \cellcolor{yellow!25}467 & 23  &  926 & 6 & 7 & 0 & \cellcolor{yellow!25}924  & 2  \\ 
        $(5,213)$ & 417 & 5 & 0 & 0 & 11 & \cellcolor{yellow!25}406   & 816 & 1 & 1 & 0 & 8 & \cellcolor{yellow!25}813 \\ 
        \hline
    \end{tabular}}
\caption{Evaluation of the diffusion model after pre-training and post-training with GRPO. The column $\#$valid records the number of (unique) valid solutions (in any geometry with the same $h^{1,2}$) out of $1{,}000$ samples. In the remaining columns, the number of valid samples for each geometry is recorded where the yellow entry corresponds to the prompted geometry.} 
\label{tab:pretrain_all_h12_diffusion}
\end{table}

We next refine the model with GRPO  (see Appendix~\ref{app:diffusion-grpo} for detailed implementation of GRPO on our diffusion model), using validity determined in Sec.~\ref{sec:conditions} as the reward. The resulting sampling performance is reported in Table~\ref{tab:pretrain_all_h12_diffusion}. The average validity, i.e.~finding a solution for the prompted geometry, is given by
\begin{equation*}
    90.32\%\, ,\qquad 83.54\%\, ,\qquad 90.24\%\, ,\qquad 85.80\% ,
\end{equation*}
for $h^{1,2} \in \{2,3,4,5\}$, respectively.  As we can see, performance substantially increases over all geometries, especially the originally hard geometries with $h^{1,2}=4,5$, showing the effectiveness of GRPO in diffusion model fine-tuning.
\newline
\paragraph{Comparison.}
Comparing both models, one can observe mixed performance at the pre-training stage, meaning that the models are able to find more solutions for geometries with lower $\hc$. We observe better performance of the diffusion model during the pre-training stage. However, this difference disappeared after the respective GRPO routines, leading to excellent performance over all geometries, where the transformer slightly outperforms the diffusion model.

\begin{figure}[!t]
\centering
\captionsetup[subfigure]{labelformat=empty}

% ---------- Transformer row ----------
\begin{minipage}[c]{0.02\textwidth}
    \centering
    \rotatebox{90}{\small\textbf{Transformer}}
\end{minipage}%
\begin{subfigure}[c]{0.32\textwidth}
    \centering
    \includegraphics{
        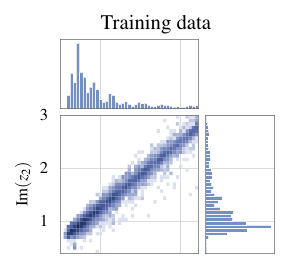
    }
\end{subfigure}%
\begin{subfigure}[c]{0.32\textwidth}
    \centering
    \includegraphics{
        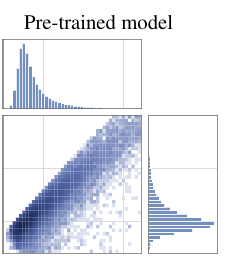
    }
\end{subfigure}%
\hspace{-0.03\textwidth}%
\begin{subfigure}[c]{0.32\textwidth}
    \centering
    \includegraphics{
        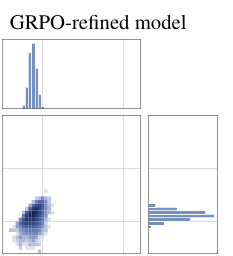
    }
\end{subfigure}

\vspace{0.2cm}

% ---------- Diffusion row ----------
\begin{minipage}[c]{0.02\textwidth}
    \centering
    \rotatebox{90}{\small\textbf{Diffusion model}}
\end{minipage}%
\begin{subfigure}[c]{0.32\textwidth}
    \centering
    \includegraphics{
        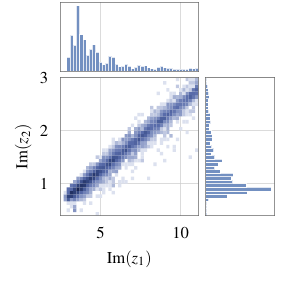
    }
\end{subfigure}%
\begin{subfigure}[c]{0.32\textwidth}
    \centering
    \includegraphics{
        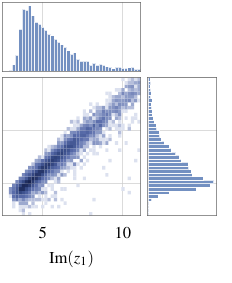
    }
\end{subfigure}%
\hspace{-0.03\textwidth}%
\begin{subfigure}[c]{0.32\textwidth}
    \centering
    \includegraphics{
        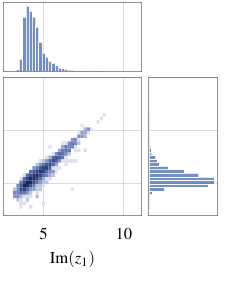
    }
\end{subfigure}
    \caption{Moduli distributions for the geometry $(\hc, \hk)=(2, 272)$ of the training data, and valid samples drawn from the pre-trained and GRPO-refined transformer (first row) and diffusion model (second row) respectively. To improve readability, the plots have been zoomed to the high density region. 
    }
    \label{fig:moduli_dists_2_272}
\end{figure}

\paragraph{Sample distributions.}
Since GRPO can in principle alter the distribution learned by the pre-trained model we investigate this further both for the transformer and diffusion model. In Fig.~\ref{fig:moduli_dists_2_272}, the distributions of the imaginary parts of the complex structure moduli are plotted for the training data, as well as valid samples drawn from the pre-trained and GRPO-refined model respectively. Those distributions are particularly interesting, since they demonstrate which parts of Kähler cone, which we know to become increasingly slim for higher $\hc$, are populated by solutions. The distributions of the training data and the valid samples from the pre-trained model are almost identical, verifying that the model indeed learns the training data distribution. The pre-training stage for the transformer seems to have the effect of slightly broadening the distribution.

The post-training stage with GRPO seems to reinforce the high density region of the previous distributions while outliers do not appear at all. Considering the choice of rewarding only valid solutions, this behaviour is to be expected since the model learns that most valid solutions appear around the peak of the distribution. If one aims to stay closer to the training data distribution, this can be controlled by increasing the magnitude of the KL divergence in the GRPO loss (see Appendix \ref{app:transformer_grpo}). If the goal is a more diverse distribution in terms of complex structure moduli or any other distributions, this can in principle be controlled through adding additional diversity rewards to the GRPO routine.

\subsection{Fine-tuning and extrapolation}\label{sec:fine-tuning}

\begin{figure}[!t]
    \centering
    \resizebox{0.75\textwidth}{!}{
  \begin{tikzpicture}
  \draw[fill, color=violet!75] (3.5,2) circle [radius=1.5];
    \node[] at (3.5,2.0) {\footnotesize \color{white} $\bm{\hc\leq4}$};
   \draw[rounded corners, fill, color=lightgray!20] (5.5, 2.5) rectangle (10.5, -2.5) node[pos=.5, align=left, color=black] { \textbf{Datasets} \\ \vspace{-0.7em} \\  \footnotesize \textbf{Pre-training (large):} \\ \footnotesize $\bullet$ all geometries with $\hc \leq 4$ \\ \footnotesize $\bullet$ total size: $1.5$M solutions \\ \vspace{-0.7em} \\ \footnotesize \textbf{Fine-tuning (small)} \\ \footnotesize $\bullet$ all geometries with $\hc=5$ \\ \footnotesize $\bullet$ total size: $5{,}000$ solutions};
   \draw[rounded corners, draw, color=lightgray, line width=1pt] (5.5, 2.5) rectangle (10.5, -2.5);
     \draw[->] (1.7,2) -- (1.2,2);
   \draw[rounded corners, fill, color=cyan!20] (-1,2.5) rectangle (1,1.5) node[pos=.5] {\footnotesize \color{black} pre-training};
   \draw[rounded corners, dotted, color=cyan, line width=1.5pt] (-1,2.5) rectangle (1,1.5);
     \draw[->, line width=1pt] (0,1.3) -- (0,0.8);
   \draw[rounded corners, fill, color=darkgreen!30] (-1.8,-.5) rectangle (1.8,.5) node[pos=.5] {\footnotesize \color{black} \textbf{pre-trained model}};
   \draw[rounded corners, draw, line width=1pt, color=darkgreen] (-1.8,-.5) rectangle (1.8,.5);
    \draw[->, line width=1pt] (0,-.7) -- (0,-1.2);
    \draw[rounded corners, fill, color=cyan!20] (-1, -1.5) rectangle (1, -2.5) node[pos=.5]  {\footnotesize \color{black} fine-tuning};
    \draw[rounded corners, dotted, line width=1.5pt, color=cyan] (-1, -1.5) rectangle (1, -2.5);
    \draw[->, line width=1pt] (0, -2.7) -- (0, -3.2);
    \draw[fill, color=violet!75] (2.5,-2) circle [radius=0.071];
    \node[] at (3.4,-2.) {\footnotesize $\hc=5$};
    \draw[->] (2.0, -2) -- (1.5, -2);
    \draw[rounded corners, fill, color=darkgreen!30] (-1.7, -3.5) rectangle (1.7,-4.5) node[pos=.5] {\footnotesize \color{black} \textbf{fine-tuned model}};
    \draw[rounded corners, draw, color=darkgreen, line width=1pt] (-1.7, -3.5) rectangle (1.7,-4.5);
     \draw[->, line width=1pt] (2.1,-4) -- (2.6,-4);
     \draw[rounded corners, fill, color=cyan!20] (2.9,-3.5) rectangle (4.3,-4.5) node[pos=.5] {\footnotesize \color{black} GRPO};
     \draw[rounded corners, dotted, line width=1.5pt, color=cyan] (2.9,-3.5) rectangle (4.3,-4.5);
     \draw[->, line width=1pt] (4.7,-4) -- (5.2,-4);
     \draw[rounded corners, fill, color=darkgreen] (5.5,-3.3) rectangle (10.5,-4.7) node[pos=.5] {\footnotesize \color{white} \textbf{improved fine-tuned model}};
     \node[] at (7.4, -7.0) {\includegraphics[width=0.4\linewidth]{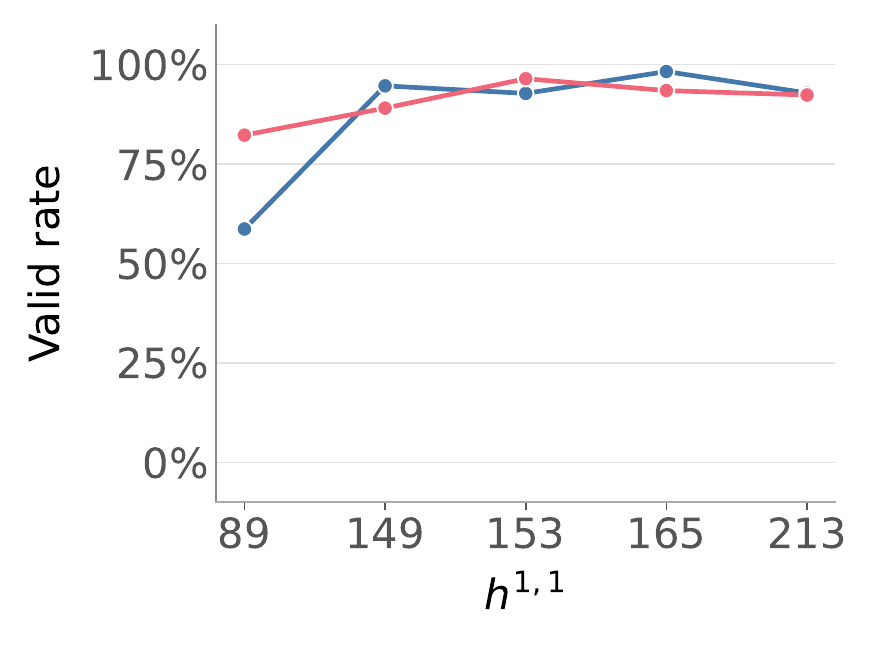}};
     \node[] at (7.7, -9.4) {\footnotesize \textbf{after GRPO}};
     \node[] at (-0.2, -7.0) {\includegraphics[width=0.4\linewidth]{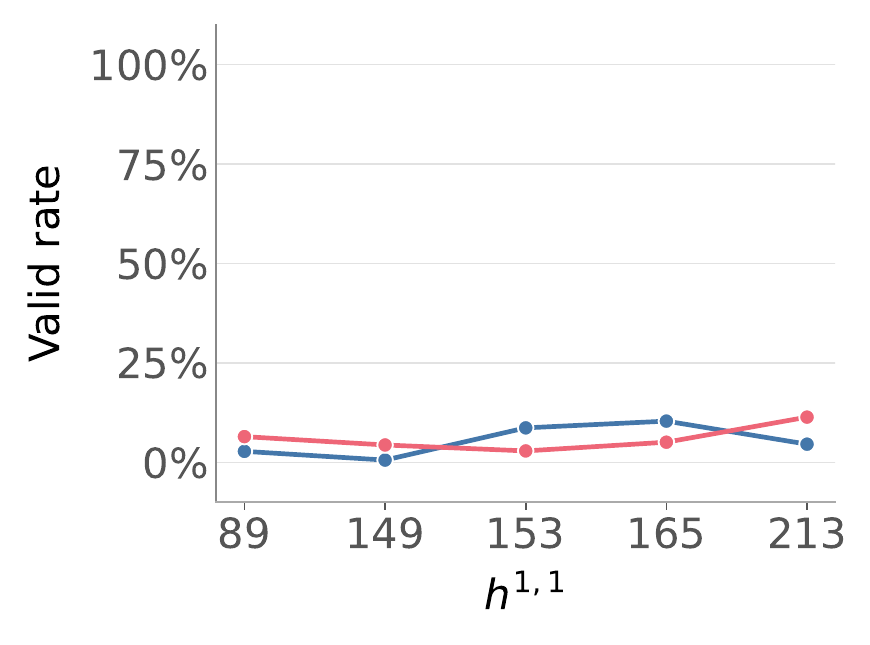}};
     \node[] at (0.4, -9.4) {\footnotesize \textbf{before GRPO}};
     \draw[->, line width=1pt] (3, -6.5) -- (4.2, -6.5);
     \node[] at (3.8, -10.1) {\includegraphics[width=0.35\linewidth]{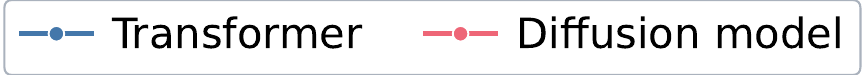}};
  \end{tikzpicture}}
    \caption{Schematic drawing of the fine-tuning routines used for the transformer and diffusion model. The large data set corresponds to all geometries with $\hc\leq4$, resulting in a total size of $1.5$M. The small dataset corresponds to $1{,}000$ randomly sampled solutions per geometry with $\hc=5$, resulting in a total size of $5{,}000$. The valid rates of both the transformer and diffusion model are plotted for each geometry with $\hc=5$ and $\hk\in\{ 89, 149, 153, 165, 213\}$ for the fine-tuned model both before and after the GRPO stage.}
    \label{fig:finetune_routnies}
\end{figure}

%------------------- transformer --------------------
In this section, we investigate how a transformer or diffusion model that has been pre-trained on a certain set of geometries can transfer its knowledge to new geometries. To this end, a model is pre-trained on a large dataset with all geometries with $h^{1,2}\in\{ 2, 3, 4\}$ ($100{,}000$ points per geometry) and later fine-tuned on a significantly smaller dataset with all geometries with $h^{1,2}=5$ ($1{,}000$ points per geometry). As a benchmark, another randomly initialised model is pre-trained on the same small dataset with $h^{1,2}=5$ and compared to the fine-tuned model.  Both models are then further improved by GRPO and again compared. We sketch this procedure in Fig.~\ref{fig:finetune_routnies}.

\paragraph{Transformer.}
The fine-tuned transformer is compared to the benchmark model in Tab.~\ref{tab:h12_5_small_dataset}. One can clearly see that the fine-tuned model outperforms the benchmark and has thus transferred knowledge from the pre-training stage. Both models can be improved with GRPO where the gains for the pre-trained and then fine-tuned model are significantly higher. This proves that transformer models are capable of extrapolation in this setting and can be pushed to excellent performance in regions where little numerically found solutions are available.
\begin{table}[!t]
\renewcommand{\arraystretch}{1.3}
\centering
\resizebox{0.99\textwidth}{!}{
\begin{tabular}{ccc||c|ccccc||c|ccccc}
\multicolumn{2}{c}{} & \multicolumn{1}{c}{\textbf{prompt}} &  \multicolumn{6}{c}{\textbf{pre-trained/fine-tuned}} & \multicolumn{6}{c}{\textbf{GRPO}} \\
&& $(\hc, \hk)$ & $\#$valid  & $h^{1,1}=89$ & $149$ & $153$ & $165$ & $213$   & $\#$valid  & $h^{1,1}=89$ & $149$ & $153$ & $165$ & $213$  \\
    \hhline{~--------------}
    \multirow{10}{*}{\rotatebox[origin=c]{90}{\textbf{Transformer}}} &
   \multirow{5}{*}{\rotatebox[origin=c]{90}{\textbf{benchmark}}} & $(5, 89)$ &  5 & \cellcolor{yellow!25} 4 & 0 & 0 & 0 & 1 & 16 & \cellcolor{yellow!25} 16 & 0 & 0 & 0 & 0  \\ 
        &&$(5,149)$ &  0 & 0 & \cellcolor{yellow!25} 0 & 0 & 0 & 0 & 0 & 0 & \cellcolor{yellow!25} 0 & 0 & 0 & 0  \\ 
        &&$(5, 153)$ &  9 & 0 & 0 & \cellcolor{yellow!25} 9 & 0 & 0   & 761 & 0 & 0 & \cellcolor{yellow!25} 761 & 0 & 0 \\ 
        &&$(5, 165)$ & 20 & 0 & 0 & 0 & \cellcolor{yellow!25} 19 & 1 & 779 & 0 & 0 & 0 & \cellcolor{yellow!25} 779 & 0  \\ 
        &&$(5, 213)$ &  6 & 0 & 0 & 0 & 0 & \cellcolor{yellow!25} 6 & 67 & 0 & 0 & 0 & 0 & \cellcolor{yellow!25} 67  \\ 
     \hhline{~--------------}
     &\multirow{5}{*}{\rotatebox[origin=c]{90}{\textbf{fine-tuned}}} &$(5, 89)$ & 31 &  \cellcolor{yellow!25} 28 & 1 & 2 & 0 & 0 &  589 & \cellcolor{yellow!25} 587 & 2 & 0 & 1 & 0  \\ 
        &&$(5, 149)$ &  6 & 0 & \cellcolor{yellow!25} 6 & 0 & 0 & 0   & 947 & 0 & \cellcolor{yellow!25} 947 & 0 & 0 & 0  \\ 
        &&$(5, 153)$ & 87 & 0 & 0 & \cellcolor{yellow!25} 87 & 0 & 0  & 928 & 0 & 0 & \cellcolor{yellow!25} 928 & 0 & 0 \\ 
        &&$(5, 165)$ & 104 & 0 & 0 & 0 & \cellcolor{yellow!25} 104 & 0 &  983 & 0 & 1 & 0 & \cellcolor{yellow!25} 983 & 0  \\ 
        &&$(5, 213)$ & 46 & 0 & 0 & 0 & 0 & \cellcolor{yellow!25} 46 & 929 & 0 & 0 & 0 & 0 & \cellcolor{yellow!25} 929 \\ 
         \hhline{~==============}  
&& $(\hc, \hk)$ & $\#$valid  & $h^{1,1}=89$ & $149$ & $153$ & $165$ & $213$ & $\#$valid  & $h^{1,1}=89$ & $149$ & $153$ & $165$ & $213$ \\
     \hhline{~--------------}
    \multirow{10}{*}{\rotatebox[origin=c]{90}{\textbf{Diffusion model}}} & \multirow{5}{*}{\rotatebox[origin=c]{90}{\textbf{benchmark}}} & $(5, 89)$ & 7 & \cellcolor{yellow!25}7  & 0 & 0 & 0 & 0   & 10 & \cellcolor{yellow!25} 10 & 0 & 0 & 0 & 0 \\ 
        &&$(5, 149)$ & 1 & 0 & \cellcolor{yellow!25} 1 & 0 & 0 & 0  & 2 & 0 & \cellcolor{yellow!25} 2 & 0 & 0 & 0  \\ 
        &&$(5, 153)$ & 3 & 0 & 0 & \cellcolor{yellow!25}3  & 0 & 0  & 8 & 0 & 0 & \cellcolor{yellow!25}8  & 0 & 0 \\ 
        &&$(5, 165)$ & 6 & 0 & 0 & 0 & \cellcolor{yellow!25}6  & 0   & 12 & 0 & 0 & 0 & \cellcolor{yellow!25}12  & 0  \\ 
        &&$(5, 213)$ & 2 & 0 & 0 & 0 & 0 & \cellcolor{yellow!25}2    & 2 & 0 & 0 & 0 & 0 & \cellcolor{yellow!25}2   \\ 
     \hhline{~--------------}
    &\multirow{5}{*}{\rotatebox[origin=c]{90}{\textbf{fine-tuned}}} &$(5, 89)$ & 65 & \cellcolor{yellow!25}65  & 0 & 0 & 0 & 0   & 823 & \cellcolor{yellow!25}789  & 0 & 0 & 29 & 5  \\ 
        &&$(5, 149)$ & 44 & 0 & \cellcolor{yellow!25}44  & 0 & 0 & 0  & 891 & 6 & \cellcolor{yellow!25}880  & 0 & 0 & 5  \\ 
        &&$(5, 153)$ & 29  & 0 & 0 & \cellcolor{yellow!25}29  & 0 & 0 & 965 & 0 & 8 & \cellcolor{yellow!25}956  & 1 & 0\\ 
        &&$(5, 165)$ & 51 & 0 & 0 & 0 & \cellcolor{yellow!25}51  & 0 & 935 & 5 & 4 & 0 & \cellcolor{yellow!25} 923 & 3\\ 
        &&$(5, 213)$ & 114 & 0 & 0 & 0 & 0 & \cellcolor{yellow!25} 114 & 924 & 0 & 0 & 0 & 5 & \cellcolor{yellow!25} 919 \\ 
         \hhline{~--------------}
    \end{tabular}}
\caption{Comparison of both the transformer and the diffusion model trained from scratch versus fine-tuned model. All models have been trained on all geometries with $h^{1,2}=5$ with $1{,}000$ points per geometry and then refined using GRPO. \textbf{benchmark:} Benchmark model pre-trained from scratch on the small data set. \textbf{fine-tuned:} On $\hc \leq 4$ pre-trained model that has been fine-tuned on the same dataset. For both cases, the pre-trained/fine-tuned and GRPO-refined model are shown.
The first column indicates the geometry that the transformer was prompted with. The column $\#$valid records the number of (unique) valid solutions (in any geometry with the same $h^{1,2}$) out of $1{,}000$ samples. In the remaining columns, the number of valid samples for each geometry is recorded where the yellow entry corresponds to the prompted geometry.}
\label{tab:h12_5_small_dataset}
\end{table}

%-------------------------- diffusion model -----------------------
\paragraph{Diffusion model.}
As shown in Tab.~\ref{tab:h12_5_small_dataset}, the pre-trained diffusion model clearly outperforms the benchmark diffusion model that only sees the small dataset during training. This pattern matches well with the transformer. This suggests that the diffusion model correctly learns from the pre-training stage, and transfers some knowledge, via the fine-tuning dataset, to the realm with higher $h^{1,2}$. We then apply GRPO with the same update setting to both models, and record the performance in Tab.~\ref{tab:h12_5_small_dataset}. The model with pre-training can receive significantly higher gains than the benchmark model, where the latter improves only marginally due to few reward signals for GRPO to improve.
\paragraph{Comparison.}
Both models proved capable of extrapolating to $\hc=5$ after having been pre-trained on a dataset with $h^{1,2}\leq 4$ and then fine-tuned on a significantly smaller dataset with $h^{1,2}=5$. The performance varies over geometries with different $\hk$ but is on average similar for both models with $5.42\%$ average validity for the transformer versus $6.06\%$ for the diffusion model. After the GRPO stage, both models could be pushed to excellent performance, where the transformer achieves an average validity of $87.48\%$ versus $90.76\%$ for the diffusion model.

\subsection{Conditioning on physical constraints}\label{sec:conditional-physical-constraints}

We now investigate whether the generative models can be used to search for flux vacua with particular physical properties. In addition to selecting a CY geometry, we consider conditioning on the flux-induced D3-brane charge $N_{\mathrm{flux}}$, the magnitude of the flux superpotential $|W_0|$, and the string coupling $g_s$. Unlike generating unconstrained vacua and subsequently filtering vacua with the desired properties, conditional generation aims to incorporate the target directly into the sampling process, hence resulting in more efficient conditional sampling. For each of these quantities, we train a separate conditional version of both the transformer and the masked diffusion model.

For a setting defined by a geometry with Hodge pair $(h^{1,2},h^{1,1})$ and a target value $\mathcal{O} \in \{N_{\text{flux}}, |W_0|, g_s \}$, the generative models are trained to draw from the distribution
\begin{equation}
    \bm{F},\bm{z}_0,\tau_0
    \sim p(\bm{F},\bm{z}_0,\tau_0\mid h^{1,2}, h^{1,1}, \mathcal{O}),
    \label{eq:conditional-sampling}
\end{equation}
where $\bm{F}$ denotes the flux vectors and $(\bm{z}_0,\tau_0)$ are initial guesses for the numerical solver. In the transformer, the (discretised) target value $\mathcal{O}$ is included in the input prompt and the fluxes and initial guesses are generated together. For the diffusion model, the target value $\mathcal{O}$ is embedded as an additional condition token in the denoising transformer along with the flux tokens, while the initial guesses are supplied by the flux-conditioned initial-guess generator. After pre-training, each conditional model is further refined using GRPO, with rewards favouring physically valid solutions and agreement with the requested value. The corresponding training procedures are detailed in Appendices~\ref{app:transformer_grpo} and~\ref{app:diffusion-grpo}.

We evaluate both approaches by sampling $1{,}000$ candidates for each target value and applying validity checks as described in Sec.~\ref{sec:conditions}. In the figures below, each two-dimensional histogram compares the prompted and sampled values of the conditioning quantity. Concentration near the diagonal $y=x$ line indicates accurate conditional generation. The rightmost panels show the fraction of unique valid solutions obtained at each target, together with the validity before deduplication for the GRPO-refined models. The distinction is important because increasing the probability of generating valid solutions may also increase the chance of sampling duplicated valid solutions after GRPO. 

To illustrate the conditional generation performance, we focus in this paper on two representative geometries, with Hodge pairs $(h^{1,2},h^{1,1})=(2,272)$ and $(5,149)$. These examples were chosen to display qualitatively different levels of difficulty across the geometries considered: the former exhibits comparatively strong conditional generation performance, while the latter provides a more challenging case in which the models show larger deviations from the requested targets and/or lower validity. Corresponding conditional generation plots for all remaining geometries are available in our GitHub repository \github{https://github.com/StringJAX/towards_type_IIB_foundation_model}.\footnote{\url{https://github.com/StringJAX/towards_type_IIB_foundation_model}}

\subsubsection{Conditioning on $N_{\rm flux}$}

\begin{figure}[t]
    \centering

    % Row 1: Transformer, h12 = 2
    \begin{minipage}[c]{0.02\textwidth}
        \centering
        \rotatebox{90}{\small\textbf{Transformer}}
    \end{minipage}%
    % \hfill
    \begin{minipage}[c]{0.98\textwidth}
        \centering
        \includegraphics{
            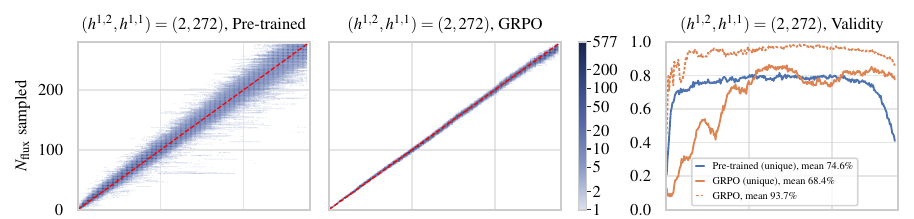
        }
    \end{minipage}

    % Row 2: Diffusion, h12 = 2
    \begin{minipage}[c]{0.02\textwidth}
        \centering
        \rotatebox{90}{\small\textbf{Diffusion model}}
    \end{minipage}%
    % \hfill
    \begin{minipage}[c]{0.98\textwidth}
        \centering
        \includegraphics{
            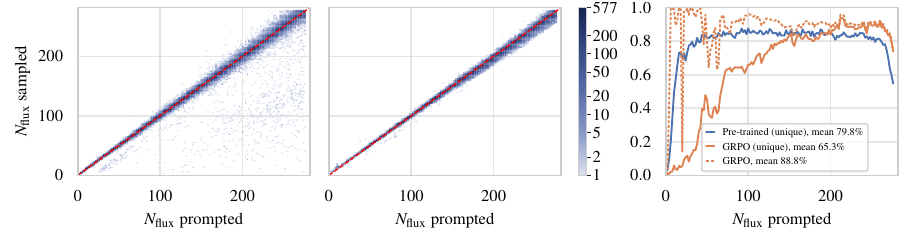
        }
    \end{minipage}

    \par\vspace{0.2cm}

    % Row 3: Transformer, h12 = 5
    \begin{minipage}[c]{0.02\textwidth}
        \centering
        \rotatebox{90}{\small\textbf{Transformer}}
    \end{minipage}%
    % \hfill
    \begin{minipage}[c]{0.98\textwidth}
        \centering
        \includegraphics{
            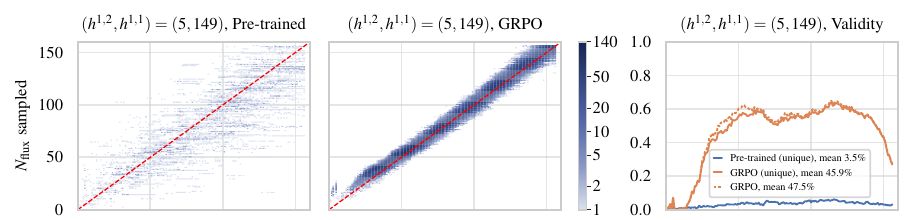
        }
    \end{minipage}

    % \par\vspace{0.2cm}

    % Row 4: Diffusion, h12 = 5
    \begin{minipage}[c]{0.02\textwidth}
        \centering
        \rotatebox{90}{\textbf{Diffusion model}}
    \end{minipage}%
    % \hfill
    \begin{minipage}[c]{0.98\textwidth}
        \centering
        \includegraphics{
            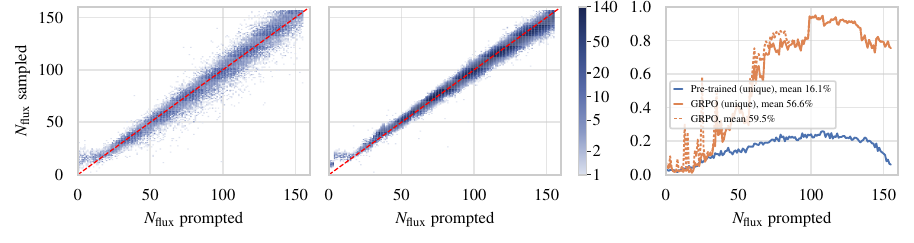
        }
    \end{minipage}
    \caption{ Conditional generation of flux vacua conditioned on $N_{\text{flux}}$ for the geometries $(\hc, \hk)=(2,272)$ (top two rows) and $(5,149)$ (bottom two rows). For each geometry, the upper row shows results for the transformer and the lower row for the diffusion model. The first two columns compare the prompted and sampled $N_{\mathrm{flux}}$ values for the pre-trained and GRPO-refined models, respectively, with the colour scale indicating the number of samples per bin on a logarithmic scale. The rightmost column shows the fraction of unique valid solutions for the pre-trained and GRPO-refined models (solid curves), together with the validity before deduplication for the GRPO-refined model (dashed curve), to illustrate possible sample collapse.    
    }
    \label{fig:nflux-cond-heatmap-combined}
\end{figure}

We first examine the performance of the generative models conditioned on the flux-induced D3-brane charge, $N_{\mathrm{flux}}=\bm{f}^{\mathsf T}\Sigma\bm{h}$. Unlike $|W_0|$ and $g_s$, this quantity is determined directly by the integer fluxes, without requiring a numerical solution for the moduli. We therefore expect this condition to be comparatively straightforward for the models to learn. Fig.~\ref{fig:nflux-cond-heatmap-combined} shows the results for integer targets $N_{\mathrm{flux}}\in [1,\ldots,Q_{\mathrm{D3}}]$.

% ------------ combined plot ------------

% ------------ transformer--------------
\paragraph{Transformer.} We pre-train the transformer with each geometry by appending the corresponding $N_{\rm flux}$ value to the input sequence and after pre-training the model is refined via GRPO, rewarding the proximity of the solution to the prompted value. In Fig.~\ref{fig:nflux-cond-heatmap-combined}, we evaluate the transformer by prompting each $N_{\rm flux} \in [1, 2, \dots, Q_{D3}]$ both for the pre-trained and GRPO-refined model, for the geometries defined by $(h^{1,2}, h^{1,1})=(2, 272)$ and $(5, 149)$. 
It can be immediately noticed that a strong correlation between prompted and sampled values is present both in the pre-trained and GRPO-refined model. However, GRPO significantly improves the performance for both geometries, while the model for $(\hc, \hk)=(5, 149)$ performs slightly worse overall, but still shows significant improvement. For $(\hc, \hk)=(2, 272)$, moderate sample collapse after GRPO can be observed over large ranges of $\nflux$, with the mean valid rate decreasing from 93.7\% to 68.4\%. However, it can be noted that this trend follows exactly the distribution of training data as seen in Fig.~\ref{fig:train_data_2_272_5_149}, i.e.~for lower $\nflux$ few solutions are available leading to worse performance in this regime. 
If one wants to prevent such sample collapse, additional diversity rewards could be imposed with the threat of moving further away from the training distribution, which in some cases might even be desirable.

% ------------ diffusion --------------

\paragraph{Diffusion model.}  The pre-trained diffusion model captures the target dependence on both geometries, although outliers and a finite spread around the diagonal remain. GRPO sharpens this dependence and reduces the outlier population. This improvement is accompanied by a trade-off in validity (after deduplication) over some target ranges similar as in the transformer case, most clearly for the $(\hc, \hk)=(2, 272)$ geometry, where the valid rate for unique solutions decreases from 79.8\% to 65.3\% after GRPO. We can also observe some sample collapse after deduplication for solutions generated after GRPO. For the more challenging higher-dimensional geometry $(\hc, \hk)=(5, 149)$, the conditional sampling performance is less accurate than the $(\hc, \hk)=(2, 272)$ case, and a small number of outliers are still present after refinement, although GRPO nevertheless improves the overall alignment between the prompted and sampled values. 

\paragraph{Comparison.} For the pre-trained diffusion model, it shows a stronger conditional learning of $N_{\text{flux}}$ than the pre-trained transformer, particularly at higher $h^{1,2}$ geometries such as $(h^{1,2}, h^{1,1})=(5, 149)$. After GRPO, both models significantly improve their conditional sampling performance and perform equally well. For $(h^{1,2},h^{1,1})=(2,272)$, however, both models also exhibit increased sample repetition after GRPO, as indicated by the gap between validity before and after deduplication.
    
% -------- extrapolation to low tadpole region for both models ------------

\begin{figure}[t]
    \centering
    \includegraphics[]{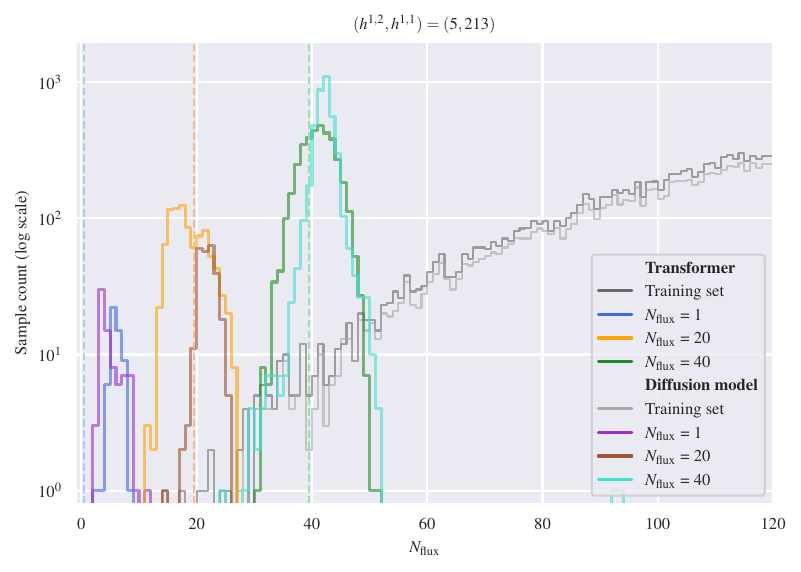}
    \caption{Distribution of $N_{\text{flux}}$ in the training set for the geometry $(\hc, \hk)=(5, 213)$, compared with samples generated under target conditions $N_{\text{flux}}\in\{1,20,40\}$ after deduplication from transformer and diffusion model. Vertical dotted lines correspond to the target value. Although the training data is sparsely populated for $N_{\text{flux}}<50$, the pre-trained models, subsequently refined with GRPO, generate substantial numbers of valid samples in this low-$N_{\text{flux}}$ region.}
\label{fig:extrapolation_small_tadpole}
\end{figure}

An interesting result we have observed is the extrapolation of vacua with small tadpole. As we can see in Fig.~\ref{fig:train_data_2_272_5_149}, on the geometry $(\hc, \hk)=(5, 149)$, vacua with $\nflux < 50$ are relatively rare in the dataset used for training. But both models still manage to learn that region to some extent, as shown in the GRPO-refined plot for the geometry $(\hc, \hk)=(5, 149)$ in Fig.~\ref{fig:nflux-cond-heatmap-combined}. This is made more explicit in Fig.~\ref{fig:extrapolation_small_tadpole}. Here, on the geometry $(h^{1,2}, h^{1,1})=(5, 213)$ the region $N_{\text{flux}}<50$ is sparsely populated. Nevertheless, both models are able to steer generation toward these comparatively under-represented regions and produce valid vacua around the prompted values.

\subsubsection{Conditioning on $|W_0|$}
As a second example, the model is conditioned on $|W_0|$ values, next to the corresponding Hodge pair. This case is of special interest if, for example, one strives to generate solutions with small $|W_0|$.

% ----------- combined plot ----------
\begin{figure}[t]
    \centering

    % Row 1: Transformer, h12 = 2
    \begin{minipage}[c]{0.02\textwidth}
        \centering
        \rotatebox{90}{\small\textbf{Transformer}}
    \end{minipage}%
    % \hfill
    \begin{minipage}[c]{0.98\textwidth}
        \centering
        \includegraphics{
            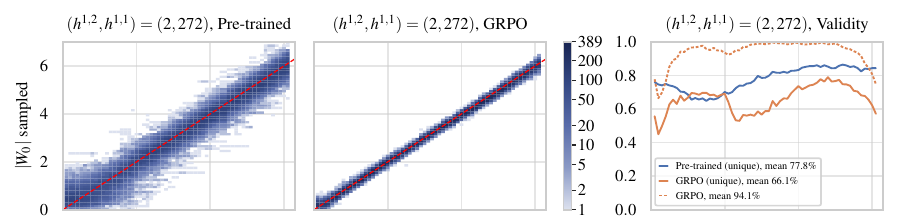
        }
    \end{minipage}

    % \par\vspace{0.2cm}

    % Row 2: Diffusion, h12 = 2
    \begin{minipage}[c]{0.02\textwidth}
        \centering
        \rotatebox{90}{\small\textbf{Diffusion model}}
    \end{minipage}%
    % \hfill
    \begin{minipage}[c]{0.98\textwidth}
        \centering
        \includegraphics{
            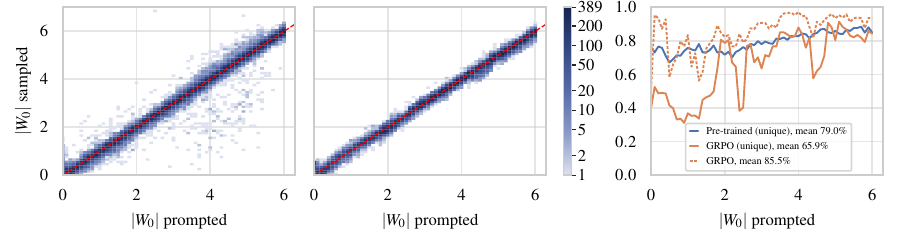
        }
    \end{minipage}

    \par\vspace{0.2cm}

    % Row 3: Transformer, h12 = 5
    \begin{minipage}[c]{0.02\textwidth}
        \centering
        \rotatebox{90}{\small\textbf{Transformer}}
    \end{minipage}%
    % \hfill
    \begin{minipage}[c]{0.98\textwidth}
        \centering
        \includegraphics{
            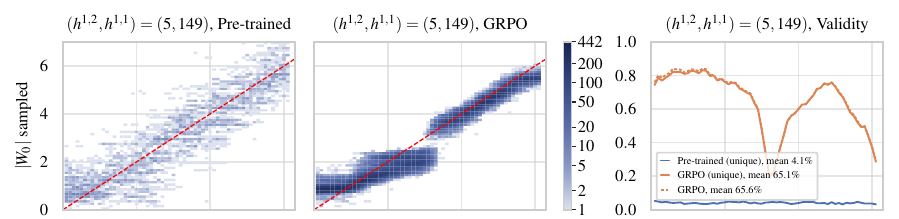
        }
    \end{minipage}

    % \par\vspace{0.2cm}

    % Row 4: Diffusion, h12 = 5
    \begin{minipage}[c]{0.02\textwidth}
        \centering
        \rotatebox{90}{\textbf{Diffusion model}}
    \end{minipage}%
    % \hfill
    \begin{minipage}[c]{0.98\textwidth}
        \centering
        \includegraphics{
            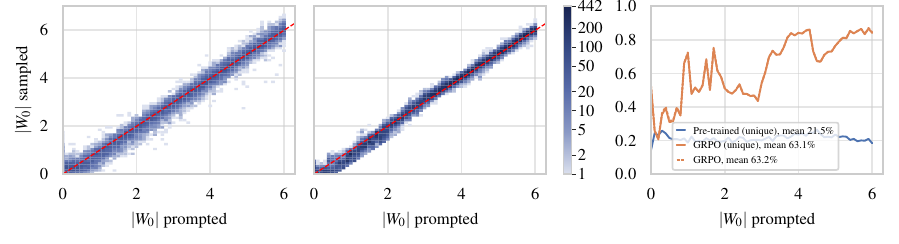
        }
    \end{minipage}
    \caption{Conditional generation of flux vacua conditioned on $|W_0|$ for the geometries $(\hc, \hk)=(2,272)$ and $(5,149)$. The arrangement of transformer and diffusion model results, as well as the pre-trained, GRPO-refined, and validity panels, follows Fig.~\ref{fig:nflux-cond-heatmap-combined}. }
    \label{fig:w0-cond-heatmap-combined}
\end{figure}

% ----------- transformer -------------
\paragraph{Transformer.}
 We pre-train the transformer by appending the corresponding discretised $|W_0|$ value to the input sequence and after pre-training the transformer is refined via GRPO, rewarding the proximity of the solution to the prompted value. In Fig.~\ref{fig:w0-cond-heatmap-combined}, we evaluate the transformer by prompting each $|W_0| \in [0.1,\dots,6.1]$ both for the pre-trained and GRPO-refined model, for the geometries defined by $(h^{1,2}, h^{1,1})=(2, 272)$ and $(5, 149)$.  Again, one immediately notes that the correlation between prompted and sampled values is successfully learned by the pre-trained model and significantly improved through GRPO. For $(\hc, \hk)=(2, 272)$, moderate sample collapse can be observed. For $(\hc, \hk)=(5, 149)$ the performance is overall worse with a significant dip around $|W_0|=3$. 

% -------- diffusion -----------
\paragraph{Diffusion model.}
Fig.~\ref{fig:w0-cond-heatmap-combined} shows the corresponding $|W_0|$ conditional generation results for the diffusion model on the same representative geometries. For both geometries, the pre-trained diffusion model already captures a clear correlation between the prompted and sampled values, although the generated distributions exhibit a noticeable spread around the diagonal. Post-training with GRPO sharpens this spread slightly and improves validity at the same time.

\paragraph{Comparison.}
As observed for $N_{\mathrm{flux}}$ in Fig.~\ref{fig:nflux-cond-heatmap-combined}, the pre-trained diffusion model exhibits closer agreement between prompted and sampled $|W_0|$ values than the pre-trained transformer. GRPO improves the conditional accuracy of both models, giving comparable performance in conditional sampling. For $(\hc,\hk)=(2,272)$, both models exhibit increased sample repetition after GRPO again, as indicated by the gap between validity before and after deduplication. The improved conditional accuracy is therefore accompanied by a decrease in the sampling of distinct valid solutions.

\subsubsection{Conditioning on $1/g_s$}

As a last example, both models are conditioned on the inverse of string coupling $1/g_s=\Im (\tau)$, in addition to the corresponding Hodge pair. This case is of special interest for achieving small string coupling. One can expect this to be the most challenging case since this removes $\Im(\tau)$ as a variable to be learned by the pre-trained model and to be optimised by GRPO.

% -------- combined plot ---------

\begin{figure}[t]
    \centering

    \begin{minipage}[c]{0.02\textwidth}
        \centering
        \rotatebox{90}{\small\textbf{Transformer}}
    \end{minipage}%
    \begin{minipage}[c]{0.98\textwidth}
        \centering
        \includegraphics{
            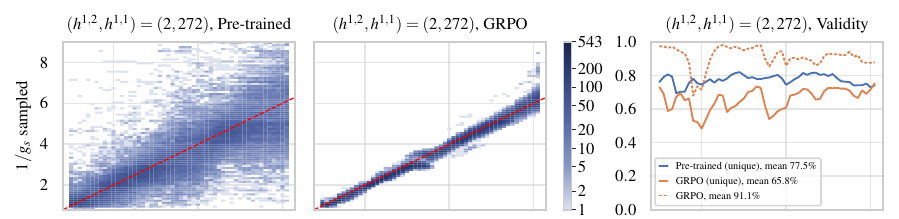
        }
    \end{minipage}

    \begin{minipage}[c]{0.02\textwidth}
        \centering
        \rotatebox{90}{\small\textbf{Diffusion model}}
    \end{minipage}%
    \begin{minipage}[c]{0.98\textwidth}
        \centering
        \includegraphics{
            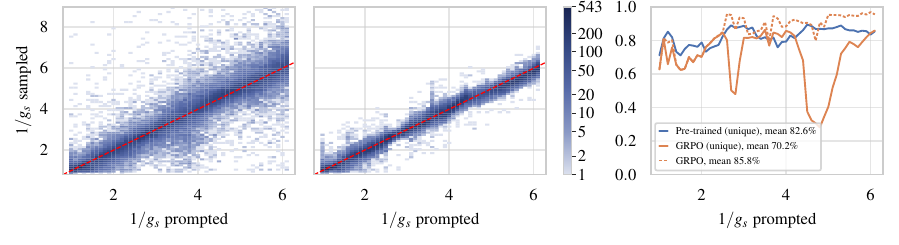
        }
    \end{minipage}

    \par\vspace{0.2cm}

    \begin{minipage}[c]{0.02\textwidth}
        \centering
        \rotatebox{90}{\small\textbf{Transformer}}
    \end{minipage}%
    \begin{minipage}[c]{0.98\textwidth}
        \centering
        \includegraphics{
            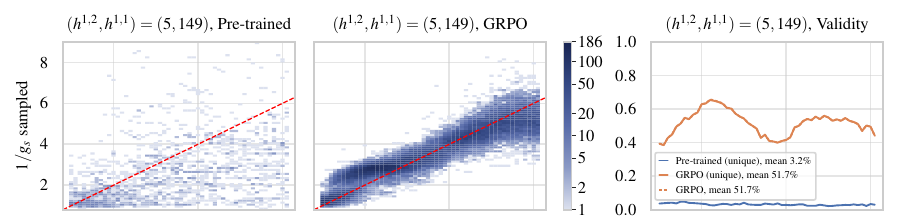
        }
    \end{minipage}

    \begin{minipage}[c]{0.02\textwidth}
        \centering
        \rotatebox{90}{\textbf{Diffusion model}}
    \end{minipage}%
    \begin{minipage}[c]{0.98\textwidth}
        \centering
        \includegraphics{
            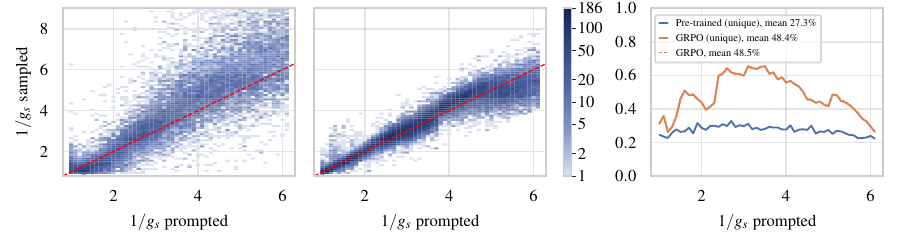
        }
    \end{minipage}
    \caption{Conditional generation of flux vacua conditioned on $1/g_s=\text{Im}(\tau)$ for the geometries $(\hc, \hk)=(2,272)$ and $(5,149)$. The arrangement of transformer and diffusion model results, as well as the pre-trained, GRPO-refined, and validity panels, follows Fig.~\ref{fig:nflux-cond-heatmap-combined}.}
    \label{fig:gs-cond-heatmap-combined}
\end{figure}

% ------- transformer ----------
\paragraph{Transformer.} We pre-train the transformer with each geometry by appending the corresponding discretised $1/g_s$ value to the input sequence and removing $\Im(\tau)$ from the output sequence. During evaluation, the numerical solver is then initialised with the prompt value instead of an $\Im(\tau)$ output by the transformer. After pre-training, the transformer is refined via GRPO, rewarding the proximity of the solution to the prompted value. In Fig.~\ref{fig:gs-cond-heatmap-combined}, we evaluate the transformer by prompting each $1/g_s \in [1.0,\dots, 6.1]$ both for the pre-trained and GRPO-refined model, for the geometries defined by $(h^{1,2}, h^{1,1})=(2, 272)$ and $(5, 149)$. One immediately notices worse performance of the pre-trained model, compared to the previous conditions, on both geometries. This performance is significantly improved by the post-training with GRPO and a large increase in validity is furthermore observed for $(\hc, \hk)=(5, 149)$. Again, moderate sample collapse is observed for $(\hc, \hk)=(2, 272)$. Rather remarkable is the fact that both models maintain high validity for larger $1/g_s$ values even though these are less populated in the training data (see Fig.~\ref{fig:train_data_2_272_5_149}). However, the spread of the distribution increases in this region (especially for $(\hc, \hk)=(5, 149)$), suggesting worse accuracy, i.e.~capability to sample the correct $1/g_s$ value, by the model.

% -------- diffusion -------------

\paragraph{Diffusion model.} Fig.~\ref{fig:gs-cond-heatmap-combined} shows the $1/g_s$ conditional generation results for the diffusion model. The pre-trained diffusion model captures the overall correlation between prompted and sampled values, although with less accuracy to the prompted values, while GRPO further concentrates the generated samples around the diagonal for both representative geometries. The effect on validity is again target dependent: for $(\hc,\hk)=(2, 272)$, GRPO produces comparable or reduced validity over parts of the target range, whereas for the higher-dimensional geometry it improves validity across most values of $1/g_s$.

\paragraph{Comparison.}
Compared with conditional sampling on $N_{\text{flux}}$ and $|W_0|$, the conditional sampling on $1/g_s$ is harder for both models, as shown by the wide spread around the prompted values in Fig.~\ref{fig:gs-cond-heatmap-combined} from the pre-trained models. GRPO significantly improves the conditional performance for both models, although at the cost of slight sample collapse.

\subsection{Scaling study}\label{sec:scaling_study}

The previous sections have shown that both a diffusion model and a transformer, in combination with GRPO, are powerful tools to sample flux vacua across several geometries and explore uncharted regions in the distribution of Type IIB flux vacua. The logical consequence of scaling up to more geometries and higher $h^{1,2}$ raises the question of how big of a model is needed to achieve adequate performance. The size of the dataset is also of relevance since it becomes increasingly more difficult to numerically generate solutions for higher $h^{1,2}$. 

Empirical studies of transformers have shown that generative losses often exhibit approximate power-law dependence on model size, dataset size, and training compute
\cite{Kaplan:2020trn,henighan2020scalinglawsautoregressivegenerative,hoffmann2022trainingcomputeoptimallargelanguage}. Related scaling behaviour has subsequently been observed in diffusion transformers and masked diffusion models \cite{peebles2023scalablediffusionmodelstransformers,li2024scalabilitydiffusionbasedtexttoimagegeneration,
ICLR2026_880d8999}.
Motivated by these observations, we investigate whether this trend can be reproduced by our models. To this end, we study how validation loss and sampling validity scale when model size and training set size are varied.

\begin{figure}[t]
    \centering

    \begin{subfigure}[c]{0.58\textwidth}
        \centering
        \includegraphics
            {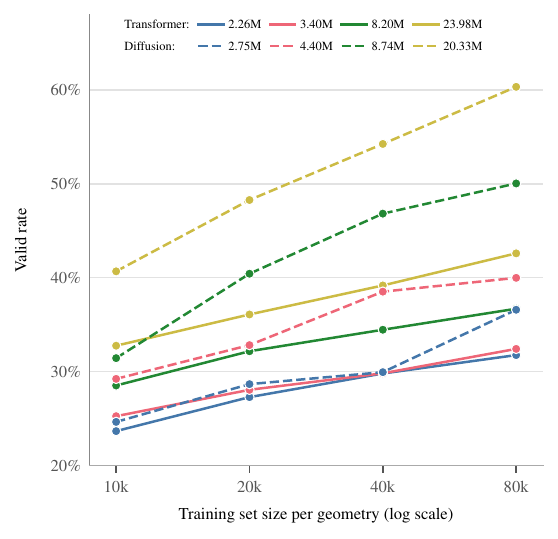}
        \caption{Valid rate for both models.}
        \label{fig:scaling-validity}
    \end{subfigure}%
    \hspace{0.02\textwidth}%
    \begin{subfigure}[c]{0.40\textwidth}
        \centering

        \includegraphics
            {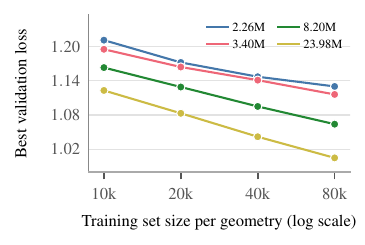}
        \caption{Transformer validation loss.}
        \label{fig:scaling-transformer-loss}

        \vspace{0.2cm}

        \includegraphics
            {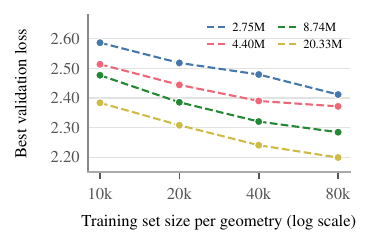}
        \caption{Diffusion model validation loss.}
        \label{fig:scaling-diffusion-loss}
    \end{subfigure}

    \caption{Scaling behaviour of the transformer and
    diffusion model with increasing model size and training-set size.
    \emph{Left:} Mean validity rate for both models, evaluated at the
    checkpoint with the lowest validation loss. \emph{Right:} Best
    validation loss for the transformer (top) and diffusion model
    (bottom). Each curve corresponds to a different model size.}
    \label{fig:scaling-transformer}
\end{figure}

In Fig.~\ref{fig:scaling-transformer}, the validation loss and mean (over all geometries) valid rate over various combinations of (pre-trained) model sizes and dataset sizes are shown. The model sizes of the diffusion model correspond to $2.75$M, $4.40$M, $8.74$M and $20.33$M parameters, while the transformer model sizes are given by $2.26$M, $3.40$M, $8.20$M and $23.98$M parameters respectively. Both models are trained on two distinct fixed training sets and also evaluated on two distinct validation sets respectively.

As expected, the validation loss decreases smoothly and monotonically with both model size and training-set size, consistent with a regular empirical scaling trend over the range studied. The improvement in validation loss translates directly into better sampling performance from each of the pre-trained model setting since the valid rate increases monotonically with both model size and training-set size, confirming that lower validation loss corresponds to higher-quality flux vacua generation. 

It can also be observed that a larger model trained on a small data set can outperform a smaller model trained on a large data set. Practically, this means that when the dataset is limited, investing in a larger model can partially compensate for the lack of data which offers a useful trade-off when generating additional training data is computationally expensive.

\subsection{Timing study}\label{sec:timing}

While we have demonstrated the capability of both the transformer and diffusion model to efficiently generate type IIB vacua over various geometries, their practical utility also depends on the computational cost of obtaining valid solutions. We compare the cost of targeted searches with the GRPO-refined conditional transformer, the GRPO-refined masked diffusion model and an ISD sampler implemented with \texttt{JAXVacua}~\cite{Dubey:2023dvu}. There are three separate experiments: we require either $N_{\rm flux}=40$, $1/g_s=4$ or $|W_0|=1.1$, without imposing a target on the other two observables. Each observable has its own trained pair of conditional models, as described in Sec.~\ref{sec:conditional-physical-constraints}. For each target, we accept roots in the windows $\pm20\%$, $\pm10\%$ or $\pm5\%$ of its value. The comparison covers all $20$ geometries introduced before.

Each method supplies integer fluxes and initial guesses for the moduli. For the \texttt{JAXVacua} comparison, we use a tuned imaginary self-dual proposal sampler referred to as $\mathrm{ISD}_+$ in \cite{Dubey:2023dvu}. It draws trial moduli and one half of the flux entries, completes the other half using the continuous ISD equations, then rounds the fluxes to integers. Concretely, for the comparison study, we uniformly sample the integer flux entries from the range $[-2,\ldots,2]$. The initial seeds for the moduli are sampled from ranges matched to the training data distribution.

For the $1/g_s$ experiment, both trained generative models set the initial $\operatorname{Im}(\tau)$ to the prompted value $4$, while the ISD sampler draws it within the requested window. The ISD sampler rejects draws that fail the cone or tadpole checks before full numerical minimisation. It tests $N_{\rm flux}$ exactly using the rounded fluxes. For $|W_0|$, it instead pre-selects using the continuous ISD flux at the starting point, so this provisional value can change upon rounding and solving. All three methods subsequently use the same solver, filtering steps and acceptance checks as described in Sec.~\ref{sec:conditions}.

We report the observed throughput as seconds per 100 accepted distinct vacua.
The time includes proposal generation, rejected draws, root finding and checks, but excludes model training, loading, compilation, a warm-up call and transfer between the generation and evaluation machines. For \texttt{JAXVacua}, the time per surviving candidate includes every draw discarded in pre-selection. Generation was timed on an NVIDIA Tesla T4 and numerical minimisation and certification on an Apple M4 Max. The single-seed experiment generated one candidate pool per learned model, geometry and target, reused across the three windows. The ISD sampler was rerun for each window because its pre-selection depends on the window. Stage~1 aimed for $8{,}192$ retained neural candidates or $4{,}096$ ISD candidates per window, subject to a $240$\,s generation cap. The subsequent evaluation stopped at $900$\,s if its candidate pool was not exhausted earlier.

\begin{figure}[t!]
    \centering
    \includegraphics[width=0.96\textwidth]{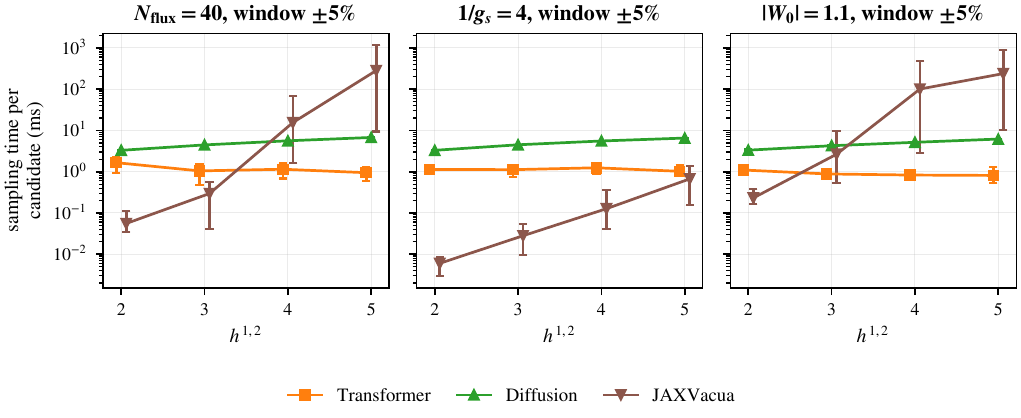}
    \caption{GPU time to produce one candidate as a function of $h^{1,2}$, averaged over five geometries at each value of $h^{1,2}$. The bars show their minimum and maximum. For \texttt{JAXVacua}, the time includes all draws rejected before a candidate is retained. The generative models are sampled once per target, so their values are the same in the three acceptance windows.}
    \label{fig:scaling}
\end{figure}

The cost of generating candidates depends differently on $h^{1,2}$ for the three methods. Fig.~\ref{fig:scaling} shows that the transformer requires about $1$\,ms per candidate throughout $h^{1,2}\in \{2,3,4,5\}$, while diffusion requires about $3$--$7$\,ms. By contrast, the cost for \texttt{JAXVacua} rises steeply with $h^{1,2}$, partly because fewer ISD draws survive pre-selection. We therefore expect the generative models to gain an advantage in candidate generation at larger $h^{1,2}$. This advantage is already visible at $h^{1,2}=5$ for the $N_{\rm flux}$ and $|W_0|$ searches, even though their sampling implementations have not been optimised specifically for computational efficiency. 

\begin{figure}[t!]
    \centering
    \includegraphics[width=\textwidth]{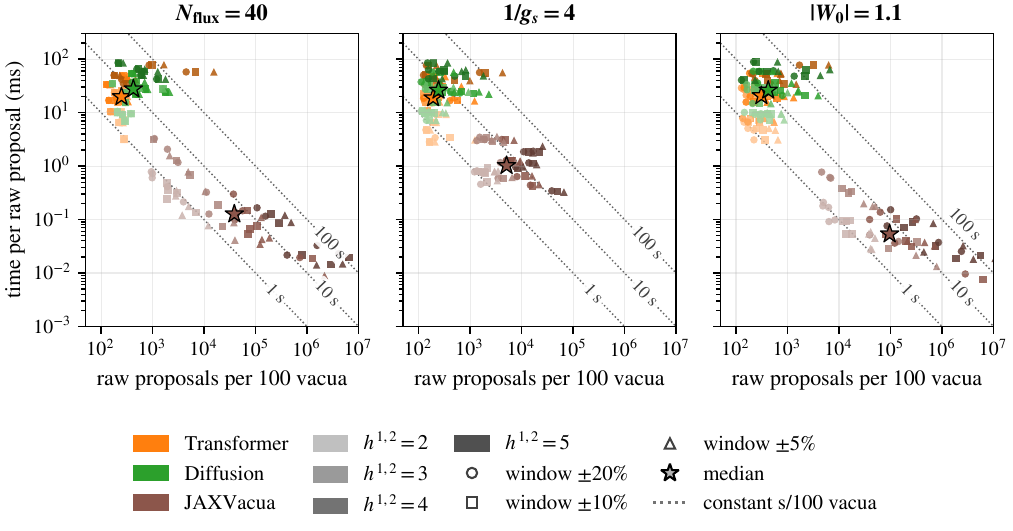}
    \caption{Cost of obtaining accepted vacua in the three conditional searches. The horizontal axis shows the number of raw proposals required per 100 accepted vacua, while the vertical axis shows the average elapsed time (in $\mathrm{ms}$) per proposal, including minimisation and checks. Symbols denote individual geometries and target windows with different shades showing $h^{1,2}$. Stars represent coordinate-wise medians for each target. The dotted diagonals indicate equal total cost in seconds per 100 accepted vacua, given by the product of the two axes.}
    \label{fig:cost}
\end{figure}

Fig.~\ref{fig:cost} shows why sampling efficiency alone does not determine the final running time. \texttt{JAXVacua} requires many more raw proposals, but most are rejected before numerical minimisation and are therefore inexpensive.\footnote{We stress that this rejection is currently only possible in the case of ISD vacua which we analyse here. It is left for the future to analyse the timing in non-ISD solutions as in~\cite{Krippendorf:2023idy}.} The generative models obtain accepted vacua from far fewer proposals, although each proposal takes longer on average. These opposing effects place the methods on similar equal-cost diagonals in several searches. In particular, the low cost per draw allows \texttt{JAXVacua} to remain competitive with the transformer for the $N_{\rm flux}$ and $|W_0|$ targets, whereas the transformer has a lower total cost for the $1/g_s$ target.

We provide more details in Appendix~\ref{app:C}, see in particular Fig.~\ref{fig:rates} for acceptance rates and Fig.~\ref{fig:distributions} for the vacuum distributions for the individual methods.

\section{Conclusions and outlook}\label{sec:con}

In this paper, we have demonstrated the effectiveness of two types of generative models, transformers and diffusion models, to generate Type IIB flux vacua. Our experiments cover twenty Calabi--Yau geometries with Hodge numbers $\hc \in \{2,3,4,5\}$. In contrast to previous work, where generative models were trained and evaluated on a single geometry~\cite{Krippendorf:2025mhp,Walden:2025cpf}, the present study probes how well such models capture the structure of the flux landscape across a family of compactification manifolds. Despite significant differences in architecture and training, we find that both transformers and diffusion models reliably learn the distribution of flux vacua. Moreover, both can be pushed to excellent performance in a post-training phase with GRPO.

Concretely, after post-training with GRPO, the models achieve an average validity rate of $90\%$, with some variance over the geometries. In more detail, out of $1{,}000$ samples drawn per geometry, $840$--$970$ samples give rise to valid $F$-flat solutions, in the sense of Sec.~\ref{sec:conditions}. Furthermore, we observe that in our experiments all sampled solutions are original, i.e., not contained in the training data. This attests to the excellent interpolation capabilities of both architectures, as they manage to draw samples that resemble, without simply copying, those in the training data. We stress that, since these generative models are approximate by nature, every sample must be checked a posteriori. In our case, this is done by solving the $F$-term conditions numerically with \verb|JAXVacua| \cite{Dubey:2023dvu}. In this sense, our method differs from ``exact'' (generative) approaches, where the output by construction satisfies all constraints. 

More remarkably, both the transformer and the diffusion model demonstrate significant extrapolation capabilities. When fine-tuned on a small dataset of new geometries, i.e., geometries not contained in the training data of the pre-trained model, they efficiently generate valid flux vacua on these new geometries which is arguably the most important result of this paper. Extrapolation of this type is a necessary prerequisite for a foundation model for the flux landscape: a single, pre-trained model that can be adapted, at small computational cost, to geometries which it has not previously encountered. A natural application of such a model would be to extrapolate from lower to higher $\hc$. Since computing the periods and solving the $F$-term conditions scale badly with the number of complex structure moduli, data generation at large $\hc$ is currently a bottleneck for these constructions.\footnote{The experiments featured in this work only probe a modest range of $\hc$. Establishing how performance and training cost scale with $\hc$ is an important question that we leave for future work.} In this context, it is particularly encouraging that fine-tuning to other geometries only seems to require modest number of training samples.

We have furthermore demonstrated that the models can be used for targeted searches, where the sampled vacua are required to obey additional constraints. Specifically, we have conditioned on target values for $\nflux$, $|W_0|$ and $1/g_s$. Such conditional searches are of particular interest for string model building and numerically controlled vacuum constructions. For example, the KKLT scenario \cite{Kachru:2003aw} requires solutions with $|W_0| \ll 1$, and $g_s \ll 1$. Since vacua with such properties are typically rare in the training data,\footnote{This rarity is a property of the sampling distribution used to construct the training set. Targeted flux choices, notably perturbatively flat vacua, can instead yield exponentially small $|W_0|$ at weak coupling \cite{Demirtas:2019sip,Demirtas:2021nlu}. Related constructions near conifold loci were developed in \cite{Demirtas:2020ffz,Alvarez-Garcia:2020pxd}.} it is significant that our GRPO routines are able to explore regions of the landscape that are absent in the training data. This adds a valuable component to the toolset for targeted searches. We expect that, combined with a careful study of the numerical precision needed to resolve extremely small values of $|W_0|$, these methods can be used to systematically chart phenomenologically interesting corners of the Type IIB flux landscape.

While the two architectures show no significant difference in the effectiveness of generating Type IIB flux vacua, there remain a few minor conceptual differences. Upon sampling, the transformer follows a fixed sequence, whereas the diffusion model allows for a more flexible generation order. The transformer offers the advantage of jointly generating both fluxes and initial guesses, resulting in a more compact system whereas the diffusion model provides greater flexibility in handling continuous inputs. In terms of computational cost, drawing samples from the diffusion model is slightly more expensive than from the transformer. In practice, however, the bottleneck of the full pipeline is the numerical solution and evaluation of the $F$-term conditions with \verb|JAXVacua|. As a consequence, the difference in sampling time between the two models is negligible (see Sec.~\ref{sec:timing}). Once pre-trained and refined, both models can thus be used to generate new solutions at a very fast pace, provided that suitable hardware is available. We find that either model serves as a good basis for a foundation and we will hence continue exploring both setups in future work.

It is highly encouraging to see the scaling behaviour of model generation with $h^{1,2}$ as this suggests that generative models can overcome one of the major bottlenecks for exploration of the flux landscape. We are excited to examine this behaviour further for geometries with $h^{1,2}>5.$ 

Taken together, these findings make a strong case for a foundation model for Type IIB flux vacua. Such a model would be pre-trained on a large and diverse set of geometries, and then fine-tuned and refined for specific geometries and targeted searches. There are several open questions in this direction, including the choice of architecture, the scaling of performance with $\hc$, and the extension beyond the large complex structure region of moduli space. We plan to present such models in future publications.

\section*{Acknowledgements}
We thank David Berman, Thomas Harvey, Xu He, Alessandro Mininno, Ayça Özcelikkale and Gary Shiu for useful discussion.
AS, ML and MW thank the ESI in Vienna, and the programme ``The unreasonable effectiveness of toric geometry" for hospitality during part of this work. 
SK and AS acknowledge the Aspen Center for Physics, which is supported by National Science Foundation grant PHY-2210452, for providing a productive environment where part of this work was performed.
The research of AS was supported by NSF grant PHY-2309456. ML is in part supported by a research grant from Olle Engkvists Stiftelse. Computational resources were, in part, provided by the National Academic Infrastructure for Supercomputing in Sweden (NAISS), funded by the Swedish Research Council. SK is partially supported by the STFC consolidated grants ST/X000664/1 and UKRI4224.

\section*{AI disclosure}
The authors used Codex, GPT-6.1 and Claude Opus 5.5 for assisted coding and text editing. All code and text have been subsequently edited and verified by the authors who take full responsibility for the published results.

\appendix

\section*{Appendix}

\addtocontents{toc}{\protect\setcounter{tocdepth}{1}}

\section{Transformer details}\label{app:transformer}

\subsection{Training data and tokenisation} \label{sec:transformer_data_and_tok}
For each geometry, $100{,}000$ distinct solutions were selected and split up during training into $90\%$ training and $10\%$ validation data. The data was tokenised as follows. Consider a solution to the $F$-term conditions \eqref{eq:Fterms} of the geometry defined by $(\hc, \hk)=(2,272)$
\begin{equation*}
    \bm{F} = (25, -2, -13, -3, -2, -2, 14, 2, 3, 1, -2, -1), \quad \bm{z} = (0.1+i 5.6, -0.5+i 1.4), \quad \tau = 0.4 + i 2.4 \,,
\end{equation*}
where $\bm{z}$ and $\tau$ have been rounded to the first decimal point and will be multiplied by $10$ in order to discretise them. A finer resolution for discretisation could also be considered but this scale was sufficient in our experiments.\footnote{During sampling, $\bm{z}$ and $\tau$ will be used as initial guesses for a numerical solver, correcting the remaining decimal points to numerical precision.} The corresponding training sequence is then given by
\begin{Verbatim}[fontsize=\fontsize{6.5pt}{7.5pt}\selectfont]
  + 2 + 272 <TAB> <BOS> + 25 - 2 - 13 - 3 - 2 - 2 + 14 + 2 + 3 + 1 - 2 - 1 <M_START> + 1 - 5 + 56 + 14 <T_START> + 4 + 24 <EOS>
  \end{Verbatim}
where each token is separated by a space. The first two numbers represent the Hodge pair with their respective signs, also given as tokens. \verb|<TAB>| marks the split point between input and output sequence which is needed for processing but is not used in training the transformer. The token \verb|<BOS>| marks the beginning of the output sequence which is needed during sampling and all the subsequent numbers and signs correspond to $\bm{F}$. \verb|<M_START>| indicates the beginning of $\bm{z}$ which is split by real and imaginary part. After that, $\tau$ is given after \verb|<T_START>| and the entire sequence is finished with the token \verb|<EOS>|. Both the \verb|<M_START>| and \verb|<T_START>| tokens are needed for the transformer to distinguish between fluxes, moduli and axio-dilaton. This allows to easily change dimensionalities and for clear sampling since each entry is clearly separated. This is not needed for the input sequence since the number of entries stays fixed. In some cases, the output sequence is padded with \verb|<PAD>| tokens to ensure a uniform shape of the training data. Those padding tokens are then masked out during training. If we prompt on an additional constraint (i.e.~$\nflux, |W_0|, 1/g_s)$, say $|W_0|=5.5$, this is added to the input sequence as \verb|+ 2 + 272 + 55 <TAB> ... |

All integers are mapped to base $1{,}000$, following \cite{charton2025int2intframeworkmathematicstransformers}, meaning that larger numbers are split accordingly. E.g.~$3~456~789$ would be tokenised as \verb|+ 3 456 789|, where the lack of a sign in-between the digits encodes the fact that this is one consecutive number. This allows for arbitrarily large flux integers with the cost of arbitrarily long sequences. In practice, the length of output sequences is restricted. In our case the maximum output sequence length is chosen to be $200$ tokens, which leaves enough room for large numbers.
After the data is tokenised, it is mapped to IDs $\in [0, 1,  \dots,s-1]$, where $s$ corresponds to the vocabulary size, i.e.~the number of possible tokens. Since we have five special tokens, and two signs, the first seven IDs are assigned to those tokens and the remaining IDs correspond to the integer tokens, in an ordered fashion. After each token has been mapped to its respective ID, it is then one-hot encoded, according to $k\to\left( \delta_{0k}, \delta_{1k}, \dots, \delta_{s-1 \, k} \right)^T$, which is the final representation that is fed into the transformer.

\subsection{Architecture}
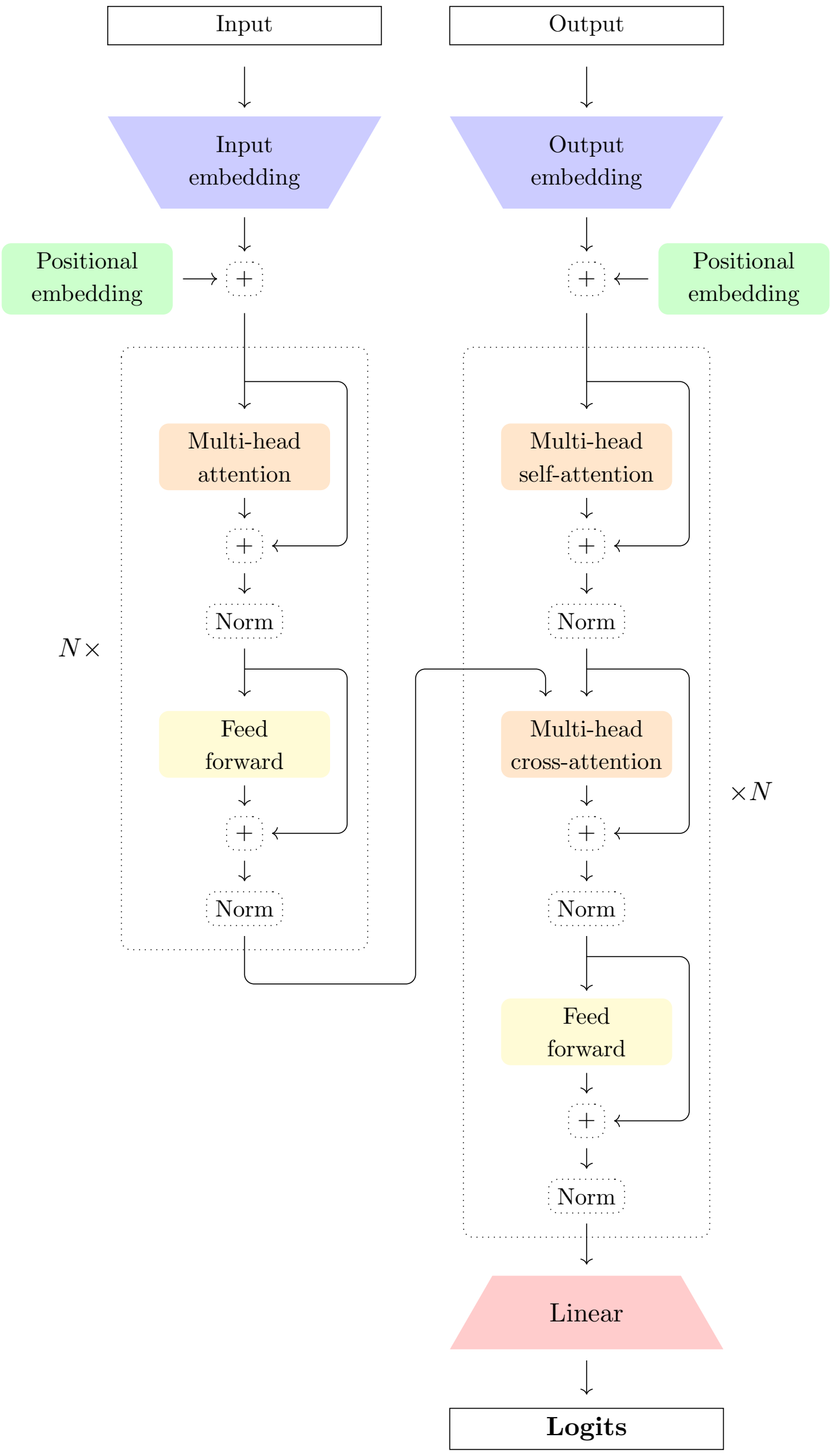
\begin{figure}[!t]
    \centering
    \resizebox{!}{0.95\textwidth}{
    \hspace{-3em}
  \begin{tikzpicture}
    % embeddings
    \node[draw, minimum width=4cm] at (0,2) {\small Input};
    \node[draw, minimum width=4cm] at (5,2) {\small Output};
    \draw[->] (5,1.4) -- (5, 0.8);
    \draw[->] (0,1.4) -- (0, 0.8);
    \node[trapezium, fill=blue!20, minimum width=4cm, align=center, shape border rotate=180] at (0,0) {\small Input \\ \small embedding};
    \node[trapezium, fill=blue!20, minimum width=4cm, align=center, shape border rotate=180] at (5,0) {\small Output \\ \small embedding};
    % positional embeddings
    \draw[->] (0, -0.8) -- (0, -1.3);
    \draw[->] (5, -0.8) -- (5, -1.3);
    \node[draw, rounded corners, dotted] at (0, -1.7) {\small +};
    \node[draw, rounded corners, dotted] at (5, -1.7) {\small +};
    \node[rectangle, rounded corners, fill=green!20, align=center, minimum width = 2.5cm] at (-2.3, -1.7) {\small Positional \\ \small embedding};
    \node[rectangle, rounded corners, fill=green!20, align=center, minimum width=2.5cm] at (7.3, -1.7) {\small Positional \\ \small embedding};
    \draw[->] (-0.9,-1.7) -- (-0.4,-1.7);
    \draw[->] (5.9,-1.7) -- (5.4,-1.7);
    \draw[->] (0, -2.2) -- (0, -3.6);
    \draw[->] (5,-2.2) -- (5,-3.6);
    % encoder block
    \draw[->, rounded corners] (0, -3.2) -- (1.5, -3.2) -- (1.5, -5.6) -- (0.4, -5.6);
    \draw[rounded corners, dotted] (-1.8,-2.7) rectangle (1.8, -11.5);
     \node[] at (-2.4, -7.1) {$N\times$};
    \node[rectangle, rounded corners, fill=orange!20, align=center, minimum width=2.5cm] at (0,-4.3) {\small Multi-head \\ \small attention};
    \draw[->] (0, -4.9) -- (0, -5.2);
    \node[draw, rounded corners, dotted] at (0, -5.6) {\small +};
    \draw[->] (0, -6) -- (0, -6.3);
    \node[draw, rounded corners, dotted] at (0, -6.7) {\small Norm};
    \draw[->] (0, -7.1) -- (0, -7.8);
    \draw[->, rounded corners] (0, -7.4) -- (1.5, -7.4) -- (1.5, -9.8) -- (0.4, -9.8);
    \node[rectangle, rounded corners, fill=yellow!20, align=center, minimum width=2.5cm] at (0, -8.5) {\small Feed \\ \small forward};
     \draw[->] (0, -9.1) -- (0, -9.4);
    \node[draw, rounded corners, dotted] at (0, -9.8) {\small +};
    \draw[->] (0, -10.2) -- (0, -10.5);
    \node[draw, rounded corners, dotted] at (0, -10.9) {\small Norm};
    \draw[->, rounded corners] (0, -11.3) -- (0, -12) -- (2.5, -12) -- (2.5, -7.4) -- ( 4.4, -7.4) -- ( 4.4, -7.8);

    % decoder block
     \draw[->, rounded corners] (5, -3.2) -- (6.5, -3.2) -- (6.5, -5.6) -- (5.4, -5.6);
    \draw[rounded corners, dotted] (3.2,-2.7) rectangle (6.8, -15.7);
    \node[] at (7.4, -9.2) {$\times N$};
    \node[rectangle, rounded corners, fill=orange!20, align=center, minimum width=2.5cm] at (5,-4.3) {\small Multi-head \\ \small self-attention};
    \draw[->] (5, -4.9) -- (5, -5.2);
    \node[draw, rounded corners, dotted] at (5, -5.6) {\small +};
    \draw[->] (5, -6) -- (5, -6.3);
    \node[draw, rounded corners, dotted] at (5, -6.7) {\small Norm};
    \draw[->] (5, -7.1) -- (5, -7.8);
    \draw[->, rounded corners] (5, -7.4) -- (6.5, -7.4) -- (6.5, -9.8) -- (5.4, -9.8);
    \node[rectangle, rounded corners, fill=orange!20, align=center, minimum width=2.5cm] at (5, -8.5) {\small Multi-head \\ \small cross-attention};
    \draw[->] (5, -9.1) -- (5, -9.4);
    \node[draw, rounded corners, dotted] at (5, -9.8) {\small +};
    \draw[->] (5, -10.2) -- (5, -10.5);
    \node[draw, rounded corners, dotted] at (5, -10.9) {\small Norm};
    \draw[->] (5, -11.3) -- (5, -12.1);
    \draw[->, rounded corners] (5, -11.6) -- (6.5, -11.6) -- (6.5, -14) -- (5.4, -14);
    \node[rectangle, rounded corners, fill=yellow!20, align=center, minimum width=2.5cm] at (5, -12.7) {\small Feed \\ \small forward};
     \draw[->] (5, -13.3) -- (5, -13.6);
    \node[draw, rounded corners, dotted] at (5, -14) {\small +};
    \draw[->] (5, -14.4) -- (5, -14.7);
    \node[draw, rounded corners, dotted] at (5, -15.1) {\small Norm};
    \draw[->] (5, -15.5) -- (5, -16.1);
    \node[trapezium, fill=red!20, minimum width=4cm] at (5, -16.8) {Linear};
    \draw[->] (5, -17.5) -- (5, -18);
    \node[draw, minimum width=4cm] at (5, -18.5) {\bf Logits};

  \end{tikzpicture}}
\caption{Schematic drawing of the encoder-decoder model used in the transformer experiments.}
\label{fig:transformer-sketch}
\end{figure}

The transformer is a simple encoder-decoder model given in Fig.~\ref{fig:transformer-sketch}. Its adaptation to type IIB flux vacua is through the tokenisation of input and output data,  as described in  Sec. \ref{sec:transformer_data_and_tok}. Via adequate hyperparameter choices, specified at the end of this section, we ensure that the model performs optimally for the problem at hand. In the following, we describe each part of the architecture of the encoder-decoder model in more detail.

\paragraph{Input.} The  input corresponds to the tokenised Hodge pairs (and, in the conditional case, the tokenised $\nflux, |W_0|$ or $1/g_s$ value). Each token is represented by a one-hot encoded vector $\bm{x}\in \mathbb{R}^s$, where $s$ is the vocabulary size.

\paragraph{Output.} The output corresponds to (potentially incomplete) tokenised sequences of flux vectors $\bm{F}$ together with complex structure moduli $\bm{z}$ and axio-dilaton $\tau$. Again, each token is represented by a one-hot encoded vector $\bm{x}\in \mathbb{R}^s$, where $s$ is the vocabulary size.

\paragraph{Embedding.}
The first part of the architecture is given by an embedding that simply consists of a linear transformation of the one-hot encoded token $\bm{x}\in \mathbb{R}^s$ into an embedding space. More precisely,  let $A\in \mathbb{R}^{d_{\rm emb}\times s}$, where $d_{\rm emb}$ is the embedding dimension, which is a hyperparameter of the architecture. Then the embedding is given by $\bm{x} \to A\bm{x}$.

\paragraph{Positional embedding.}
The positional embedding is important to carry on information about the position of a token in a sequence. It corresponds to a linear transformation of the one-hot encoded sequence position of a token into the embedding space. More precisely, let $\bm{x}_{0:l-1}$ be some sequence with length $l$ containing one-hot encoded tokens. Then the position of the token $\bm{x}_k$, with $k\leq l$, is encoded as $\bm{k} = \left( \delta_{0k}, \delta_{1k}, \dots, \delta_{l_{\rm max}-1 \, k} \right)^T$, where $l_{\rm max}$ is a hyperparameter which sets the maximum allowed sequence length. Since in our case input and output sequences generally differ in length, $l_{\rm max}$ is chosen differently for input and output sequences, respectively. Let $A\in \mathbb{R}^{d_{\rm emb} \times l_{\rm max}}$, then the positional embedding of the token $\bm{x}_k$ is given by $A\bm{k}$. Since this results in a vector in the same embedding space, the positional embedding can then simply be added to the embedding of the token itself. 

\paragraph{Multi-head attention.}
As the name suggests, multi-head attention consists of several concatenated attention heads. This is the most crucial part of the transformer architecture since it allows tokens to attend to other tokens in the sequence. More precisely, the model learns e.g.~that a sign is followed by a number, or on a bigger scale that a flux vector $\bm{F}$ is followed by the initial guesses $\bm{z}_0$ and $\tau_0$. Each attention head performs the following calculation
\begin{equation*}
    {\rm Attention} \left( Q, K, V \right) = {\rm softmax} \left( \frac{Q K^T}{\sqrt{d_{\rm head}}}\right) V \,.
\end{equation*}
Here, $Q, K, V$ correspond to query, key and value and are simply obtained through linear transformation of the embedded sequence, and $d_{\rm head}$ is the head size. Let $\bm{x}_{0:l-1}\in \mathbb{R}^{l\times d_{\rm emb}}$ be a sequence of embedded tokens, then
\begin{equation}
    Q = \bm{x}_{0:l-1} A_Q\,, \quad K=\bm{x}_{0:l-1} A_K  \,, \quad V = \bm{x}_{0:l-1} A_V \,,
\end{equation}
where $A_Q, A_K, A_V \in \mathbb{R}^{d_{\rm emb}\times d_{\rm head}}$. Here, softmax is applied row-wise which after multiplication with $V$ results in the output Attention$(Q, K, V)\in \mathbb{R}^{l\times d_{\rm head}}$. In multi-head attention the input is projected onto several distinct attention heads with $d_{\rm emb} = n \, d_{\rm heads}$, where $n$ is the number of heads. Afterwards, the head outputs are concatenated retaining the original dimension $d_{\rm emb}$. 

In the multi-head attention blocks in the decoder, the term $Q K^T$ is masked to ensure that each token in a sequence can only attend to previous tokens, maintaining an autoregressive structure that is needed during sampling. In the self-attention block, the decoder only attends to tokens contained in the output sequence while the cross-attention block also attends to the input tokens in the form of the encoder block output.

In the encoder, attention is not masked, meaning that each token can both attend to previous and following tokens.

\paragraph{Feed forward.}
The feed forward block consists of a linear transformation, followed by ReLU followed by another linear transformation. More precisely, let $\bm{x}\in \mathbb{R}^{d_{\rm embed}}$ be the embedded token, $A\in \mathbb{R}^{4d_{\rm emb}\times d_{\rm emb}}, \bm{a} \in \mathbb{R}^{4d_{\rm emb}}, B \in \mathbb{R}^{d_{\rm emb}\times 4 d_{\rm emb}}, \bm{b} \in \mathbb{R}^{d_{\rm emb}}$, then the map on $\bm{x}$ is given by
\begin{equation*}
    \bm{x} \to B \, {\rm ReLU} (A\bm{x} + \bm{a}) +  \bm{b} \,.
\end{equation*}

\paragraph{Norm.}
Each multi-head attention and feed forward block is followed by layer normalisation which is applied to each embedded token. Let $\bm{x}_{0:l-1}$ be a sequence of embedded tokens, then each token $\bm{x}_i$ is normalised over its embedding dimension by
\begin{equation*}
    \bm{x}_{i,j} = \gamma_{ij}\frac{\bm{x}_{i,j}-\mu_i}{\sqrt{\sigma_i^2+\epsilon}} + \beta_{ij} \,,
\end{equation*}
where
\begin{equation}
    \mu_i = \frac{1}{d_{\rm emb}} \sum_{j=1}^{d_{\rm emb}} \bm{x}_{i,j} \,, \quad \sigma_i = \frac{1}{d_{\rm emb}} \sum_{j=1}^{d_{\rm emb}}\left( \bm{x}_{i,j} - \mu_i \right)^2 \,,
\end{equation}
with $\gamma_{ij}, \beta_{ij}$ being learnable parameters and $\epsilon=10^{-8}$ being a constant preventing division by zero. Such layer normalisation generally helps to decrease training time \cite{ba2016layernormalization}.

\paragraph{Linear.}
This is the last layer consisting of a single linear map, to map the output back to the vocabulary size. Let $\bm{x}\in\mathbb{R}^{d_{\rm emb}}$ be an embedded token and $A\in \mathbb{R}^{s\times d_{\rm emb}}$, where $s$ is the vocabulary size, then $\bm{x}$ is mapped according to $x\to A \bm{x}$.

\paragraph{Dropout.}
The embeddings, multi-head attention and feed forward blocks are followed by a dropout layer, that randomly drops input parameters, i.e.~sets them to zero, according to some fixed probability. This procedure regularises the model and helps to prevent over-fitting to the training data \cite{JMLR:v15:srivastava14a}.

\paragraph{Logits.} The final output of the transformer consists of the unnormalised log probabilities (logits) for each possible next token. This corresponds to a vector $\bm{p}\in\mathbb{R}^s$, where each entry corresponds to the unnormalised log probability of the respective token to appear next in the sequence. E.g.~if the sequence consists of a completed flux vector $\bm{F}$, then the next token to appear is $\verb|<M_START>|$, indicating the beginning of the complex structure moduli. This means that the transformer will be trained to maximise the respective entry $p_i$ in $\bm{p}$, where $i$ is the ID of the token $\verb|<M_START>|$.

\subsection{Objective}
Denote by $\bm{x}_{0:l-1}$ the output sequences in the respective one-hot representation, where $\bm{x}_0$ corresponds to the \verb|<BOS>| token. Denote by $\bm{p}_{1:l-1}$ the transformer prediction corresponding to logits for each token in the sequence. During training, we then optimise the model by minimising the cross-entropy loss between predicted token probabilities and one-hot encoded output sequences 
\begin{equation} \label{eq:cross-entropy-loss}
    \mathcal L_{\mathrm{transformer}}(\bm{\theta})
    =
     -\mathbb{E}_{ \bm{x}_{0:l-1} \sim p_{\text{data}}} \left[ \frac{1}{l-1} \sum_{i=1}^{l-1} \log \left( \frac{e^{\bm{p}_i\cdot \bm{x}_i}}{\sum_{j=1}^s e^{\bm{p}_{ij}}} \right) \right] \,.
\end{equation}
Note that padding tokens are masked, i.e.~their loss contribution is set to zero. To make this more precise, denote by $\bm{x}_k$ the one-hot encoded $\verb|<M_START>|$ token that appears after the flux vector $\bm{F}$. Then the only term that remains after the dot product $\bm{p}_i\cdot \bm{x}_i$ is $\bm{p}_k$ and minimising $\mathcal L_{\mathrm{transformer}}(\bm{\theta})$ implies maximising $\bm{p}_k$. This means that the model learns to predict a $\verb|<M_START>|$ token after each flux vector $\bm{F}$. In this way, the transformer learns to maximise the probabilities of each token in a sequence, learning the distribution over solutions to the F-term conditions \eqref{eq:Fterms}.

\subsection{Sampling}
Samples of tokens corresponding to $\bm{F},\bm{z}_0,$ and $\tau_0$, are drawn from a categorical distribution, treating the transformer output as logits. 

To illustrate this further, consider the generation of a single token. Let $p_i$, $i=0,\dots,s-1$ be the transformer output. The logits are rescaled as $p_i\to p_i/T$, where $T$ is the sampling temperature. From this, we calculate the probability of each token through the softmax function
\begin{equation}
    \pi_i = \frac{e^{p_i/T}}{\sum_{j=0}^{s-1} e^{p_j/T}} \,.
\end{equation}
Clearly, this is a number between 0 and 1, and so Cat$(\bm{\pi})$ defines a categorical probability distribution from which the next token is drawn. In our experiments, each evaluation contains $1{,}000$ samples per prompt that have been drawn using $0.6$ as the sampling temperature.

\subsection{GRPO}\label{app:transformer_grpo}

After pre-training, the models were refined using Group Relative Policy Optimisation (GRPO) \cite{shao2024deepseekmathpushinglimitsmathematical}. During this routine, the model is prompted with a sampled or fixed set of prompts, i.e.~either a fixed set of Hodge pairs or a sampled set of Hodge pairs together with $\nflux, |W_0|$ or $1/g_s$. For each prompt, a group of samples is drawn for which rewards are computed and used to steer the model towards more favourable samples. The objective that is maximised after each draw is given by
\begin{equation*}
    \mathcal{J}_{\text{GRPO}} (\theta) = \mathbb{E}_{q\sim P(Q), \, \{ o_i \}_{i=1}^G \sim \pi_{\theta_{\text{old}}}(O|q)} \, \left[ L \left(q, \{ o_i \}_{i=1}^G \right) \right] \,,
\end{equation*}
where
\begin{align*}
    &L \left(q, \{ o_i \}_{i=1}^G \right) = \\
     &\frac{1}{G} \sum_{i=1}^G \frac{1}{|o_i|} \sum_{t=1}^{|o_i|} \Bigg[ \textrm{min} \left[ \frac{\pi_{\theta}(o_{i,t}|q,o_{i,<t})}{\pi_{\theta_{\text{old}}}(o_{i,t}|q,o_{i,<t})}\hat{A}_{i,t}, \textrm{clip} \left( \frac{\pi_{\theta}(o_{i,t}|q,o_{i,<t})}{\pi_{\theta_{\text{old}}}(o_{i,t}|q,o_{i,<t})}, 1- \epsilon, 1+\epsilon \right)\hat{A}_{i,t} \right] \\
      & \hspace{8em}- \beta \mathbb{D}_{\text{KL}} \left[ \pi_{\theta} | \pi_{\text{ref}} \right] \Bigg] \,.
\end{align*}
The different parts of the objective are as follows. $q$ are the prompts for the model that are either drawn from some distribution $P(Q)$ or can be fixed if the set of possible prompts is small enough. $\{ o_i \}_{i=1}^G$ is the group of outputs per prompt $q$. Each $o_i$ corresponds to a sequence with $|o_i|$ tokens. $G$ controls the group size, i.e.~the number of samples to draw per prompt. $\pi_{\theta_{\text{old}}}$ corresponds to the transformer state when the samples were generated, i.e.~before the update step(s), while $\pi_{\theta}$ is the model to be optimised.\footnote{This differs slightly from the notation used in Sec.~\ref{sec:ml_section} where we used $p(\dots|\dots)$ to denote the probabilities output by the transformer. Using $\pi$ for the policy model is rather common in the RL literature so we will not deviate from this.} $\pi_{\text{ref}}$ corresponds to the model parameters before GRPO which is needed to compute the Kullback–Leibler (KL) divergence which serves as a regulator. $\beta$ is the corresponding hyperparameter that controls the strength of regulation. $\hat{A}_{i,t}$ are the advantages of token $t$ in sequence $i$ which are computed as $\hat{A}_{i,t}=\frac{r_i - \textrm{mean}(\mathbf{r})}{\textrm{std}(\mathbf{r})}$, i.e.~normalised rewards, where $\bm{r}=(r_1,\dots,r_G)$ includes all rewards in one group $\{ o_i \}_{i=1}^G$. Note that each token $t$ is assigned the same advantage in one sequence. Lastly, $\epsilon$ is a hyperparameter that controls the clipping of the ratio of probabilities, regularising the gradient updates per token.

The input to $\pi_{\theta}$ consists of the prompt $q$ together with the already sampled tokens $o_{i,<t}$, determining the probability of the next token through $\pi_{\theta}(o_{i,t}|q,o_{i,<t})$. 

Furthermore, the KL divergence is given by
\begin{equation*}
    \mathbb{D}_{\text{KL}} \left[ \pi_{\theta} | \pi_{\text{ref}} \right] = \frac{\pi_{\text{ref}}(o_{i,t}|q,o_{i,<t})}{\pi_{\theta}(o_{i,t}|q,o_{i,<t})} - \ln \frac{\pi_{\text{ref}}(o_{i,t}|q,o_{i,<t})}{\pi_{\theta}(o_{i,t}|q,o_{i,<t})} - 1\,,
\end{equation*}
which penalises updates that move too far away from the original distribution learned by the pre-trained model.

 During each epoch, a fixed number of samples is then drawn for each prompt and the GRPO objective is maximised for a fixed number of iterations. During these iterations, the old model parameters $\pi_{\bm{\theta}_{\text{old}}}$ are kept fixed. 

Let us illustrate this procedure through an example. Consider a model being prompted with a specific geometry, say $(\hc, \hk)=(2, 272)$. Suppose we chose $G=2$ and we end up with a valid sample $o_1$, i.e.~a solution to the F-term conditions of $(\hc, \hk)=(2, 272)$ that obeys all constraints in Sec. \ref{sec:conditions}, and an invalid sample $o_2$, i.e.~one that is either an invalid sequence, not a solution to the F-term conditions or fails any other constraint in Sec. \ref{sec:conditions}. This means we assign a positive reward, say $10$, to $o_1$ and a reward of zero to $o_2$. The respective advantages are then given by $A_1 \approx 0.71$ and $A_2 \approx -0.71$. Since $A_1$ is positive, maximising the objective implies increasing the term $\pi_{\theta}(o_{1,t}|q,o_{1,<t})/\pi_{\theta_{\text{old}}}(o_{1,t}|q,o_{1,<t})$ which can only be achieved by increasing the probability of each token of the sampled sequence $\pi_{\theta}(o_{1,t}|q,o_{1,<t})$. The transformer thus learns that such sequences are favourable. For the sample $o_2$ with reward zero, the corresponding advantage $\hat{A}_{2,t}$ is negative, which requires the term $\pi_{\theta}(o_{2,t}|q,o_{2,<t})/\pi_{\theta_{\text{old}}}(o_{2,t}|q,o_{2,<t})$ to be decreased in order to maximise the overall objective. Through this, the transformer learns that such sequences are undesirable, eventually converging to sampling mostly favourable sequences.

\subsection{Hyperparameters}

In Tab.~\ref{tab:hyperparameters_transformer_architecture}, we list the hyperparameters for the transformer architecture used in Sec.~\ref{sec:pre-train_grpo}, \ref{sec:fine-tuning} and \ref{sec:conditional-physical-constraints}.

For the scaling study in Sec.~\ref{sec:scaling_study}, the architectural parameters were as in Tab.~\ref{tab:hyperparameters_transformer_architecture} with the embedding dimension and number of attention heads being varied as shown in Tab.~\ref{tab:hyperparameters_transformer_architecture_scaling}.

The hyperparameters for the training setting are given in Tab.~\ref{tab:train_parameters_transformer}. All models have been trained to convergence of the validation loss.

\begin{table}[!h]
    \centering
    \small
    \begin{tabular}{cc}
        \hline
        \textbf{Architectural parameter} & \textbf{Value} \\
        \hline \hline
        Number layers (N) & 4 \\
        Embedding dimension & 256 \\
        Positional embeddings in & 50 \\
        Positional embeddings out & 200 \\
        Number attention heads & 8 \\
        Dropout & 0.2 \\
        \hline
        \textbf{Total number of parameters} & $\bm{8.20}$\textbf{M} \\
        \hline
    \end{tabular}
    \caption{Hyperparameters for the transformer architecture used in all runs (except scaling study).}
    \label{tab:hyperparameters_transformer_architecture}
\end{table}
\begin{table}[!h]
    \centering
    \small
    \begin{tabular}{ccccc}
        \hline
        \textbf{Architectural parameter} & \textbf{Value (\#1)} & \textbf{Value (\#2)} & \textbf{Value (\#3)} & \textbf{Value (\#4)}\\
        \hline \hline
        Embedding dimension & 128 & 160 & 256 & 512 \\
        Number attention heads & 8 & 8 & 8 & 16 \\
        \hline
        \textbf{Total number of parameters} & $\bm{2.26}$\textbf{M} & $\bm{3.40}$\textbf{M} & $\bm{8.20}$\textbf{M} & $\bm{23.98}$\textbf{M}\\
        \hline
    \end{tabular}
    \caption{Variation of embedding dimension and number of attention heads used in Sec.~\ref{sec:scaling_study}.}
    \label{tab:hyperparameters_transformer_architecture_scaling}
\end{table}

\begin{table}[!h]
    \centering
    \small
    \begin{tabular}{ccc}
        \hline
        \textbf{Training parameter} & \textbf{pre-training} & \textbf{fine-tuning/benchmark} \\
        \hline \hline
        Optimiser & Adam & Adam \\
        $(\beta_1,\beta_2, \epsilon, \overline{\epsilon})$ (Adam)
            & $(0.9,0.98, 10^{-9}, 0)$ & $(0.9,0.98, 10^{-9}, 0)$ \\
        Training batch size & $1024$ & $32$ \\
        Decay schedule & Cosine with linear warmup & Cosine linear warmup \\
        Initial learning rate & $0$ & $0$ \\
        Peak learning rate & $10^{-5}$ & $10^{-5}$ \\
        Warmup steps & $40{,}000$ & $40{,}000$ \\
        Decay steps & $3{,}474{,}000$ & $100{,}000$ \\
        Epochs & $2{,}000$ & $1{,}000$ \\
       
        \hline
    \end{tabular}
    \caption{Hyperparameters for the transformer used in pre-training (all runs) and fine-tuning/benchmark.}
    \label{tab:train_parameters_transformer}
\end{table}

The hyperparameters for the GRPO runs are given in Tab.~\ref{tab:grpo_parameters_transformer}. During training, samples corresponding to valid solutions, were given a reward. Additionally, previously drawn samples were saved and samples were additionally rewarded if they have not been previously drawn. The rewards for the conditional runs were given by
\begin{equation*}
    \frac{R_0}{|v_{\rm prompted} - v_{\rm sampled}| + 1} \,,
\end{equation*}
where $v_{\rm prompted}$ and $v_{\rm sampled}$ correspond to the prompted and sampled value respectively. For the $N_{\rm flux}$ run $v\in\{ 1, \dots, Q_{D3} \}$, while $v\in \{1, \dots, 61\}$ for $|W_0|$ and $v\in\{ 10, \dots, 61 \}$ for $1/g_s$. Note that in the $|W_0|$ and $1/g_s$ cases, $v$ corresponds to the discretised values. $R_0$ is chosen to be $5$ for valid and $10$ for valid and new solutions. All runs were trained until convergence of the mean (over prompts) reward.

\begin{table}[!h]
    \centering
    \resizebox{\textwidth}{!}{
    \begin{tabular}{cccc}
        \hline
        \textbf{Training parameter} & \textbf{Unconditional} & \textbf{fine-tuning/benchmark} & \textbf{Conditional} \\
        \hline \hline
        Loaded checkpoint (Epoch) & $2{,}000$ & $400$/$1{,}000$ & $2{,}000$ \\
        Optimiser & Adam & Adam & Adam \\
        Learning rate & $10^{-5}$ & $10^{-5}$ & $10^{-5}$\\
        $(\beta_1,\beta_2, \epsilon, \overline{\epsilon})$ (Adam)
            & $(0.9,0.999, 10^{-8}, 0)$ & $(0.9,0.999, 10^{-8}, 0)$ & $(0.9,0.999, 10^{-8}, 0)$ \\
        Prompts per epoch & 20 (all geometries) & 5 (all $\hc=5$ geometries) & 40 (two conditions per geometry) \\
        Samples per prompt & 200 & 100 & 100 \\
        GRPO iterations & 4 & 4 & 2 \\
        Sampling temperature & 0.6 & 0.6 & 0.6 \\
        $\epsilon$ (GRPO) & 0.2 & 0.2 & 0.2 \\
        $\beta$ (GRPO) & 0.004 & 0.004 & 0.004 \\
        Rewards & $5$ ($+5$ if new) & $5$ ($+5$ if new) & see text \\
        Epochs & 200 & 200 & 400 \\
        \hline
    \end{tabular}}
    \caption{Hyperparameters used in the GRPO runs for the transformer.}
    \label{tab:grpo_parameters_transformer}
\end{table}

\subsection{Scaling over $\hc$}

\begin{figure}
    \centering
    \includegraphics[]{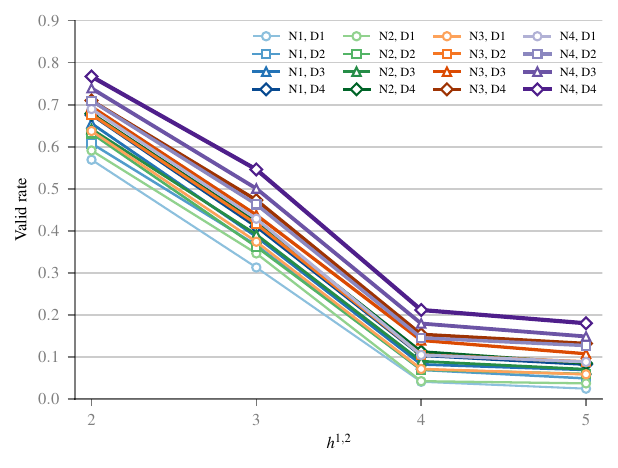}
    \caption{Average valid rate scaling across geometries with same $h^{1,2}$ for the transformer. The model sizes in the legends correspond to $N1=2.26$M, $N2=3.40$M, $N3=8.2$M, $N4=23.98$M; The dataset sizes correspond to $D1=10$k, $D2=20$k, $D3=40$k, $D4=80$k per geometry in the training set.}
    \label{fig:per_h12_scaling_validity_transformer}
\end{figure}
In this section, we analyse the sensitivity of the model capacity to increasing model size with respect to different $\hc$. Although we have shown in Sec. \ref{sec:pre-train_grpo} that the difference in performance over various $\hc$ is eliminated after the GRPO stage, one still requires a minimal rate of valid samples after the pre-training stage to serve as a reward signal during refinement. In Fig.~\ref{fig:per_h12_scaling_validity_transformer}, the valid rate over $\hc \in \{ 2, 3, 4, 5 \}$ is shown for various combinations of model and data set sizes. Interestingly, the decrease in valid rate as $\hc$ increases does not change significantly while scaling up model and data set size. Instead, all valid rates experience a constant shift. While this is in principle still sufficient to generate a reward signal, this requires further care as one scales up $\hc$. The decrease in performance seems to become less steep for $\hc\geq4$, although a more thorough investigation would require more examples with $\hc>5$ which we will leave for future work.

\section{Diffusion model details}
\label{app:discrete-diffusion}

\subsection{Diffusion models with a discrete absorbing state}

\paragraph{Flux representation.}

For a compactification geometry $g$ with $h^{1,2}$ complex-structure
moduli, we collect the RR and NS--NS flux integers into the vector
\begin{equation}
    \bm{F}
    =
    \left(
        f^{0},\ldots,f^{2h^{1,2}+1},
        h^{0},\ldots,h^{2h^{1,2}+1}
    \right)
    \in \mathbb{Z}^{L_g},
    \qquad
    L_g=4\left(h^{1,2}+1\right).
    \label{eq:flux-vector}
\end{equation}
Each integer is shifted by a fixed offset and represented by a single
categorical token. The vocabulary contains the complete range of flux
values occurring in the dataset, together with the special symbols
\texttt{[MASK]} and \texttt{[PAD]}. These special symbols are used only
as model inputs and are excluded from the output vocabulary.

To train a single model across geometries with different values of
$h^{1,2}$, all flux vectors are padded to a common length
$L_{\max}=24$, corresponding to $h^{1,2}_{\max}=5$. A padding mask
excludes coordinates $i>L_g$ from self-attention and from the training
loss.

\begin{figure}
    \centering
    \includegraphics[width=1\linewidth]{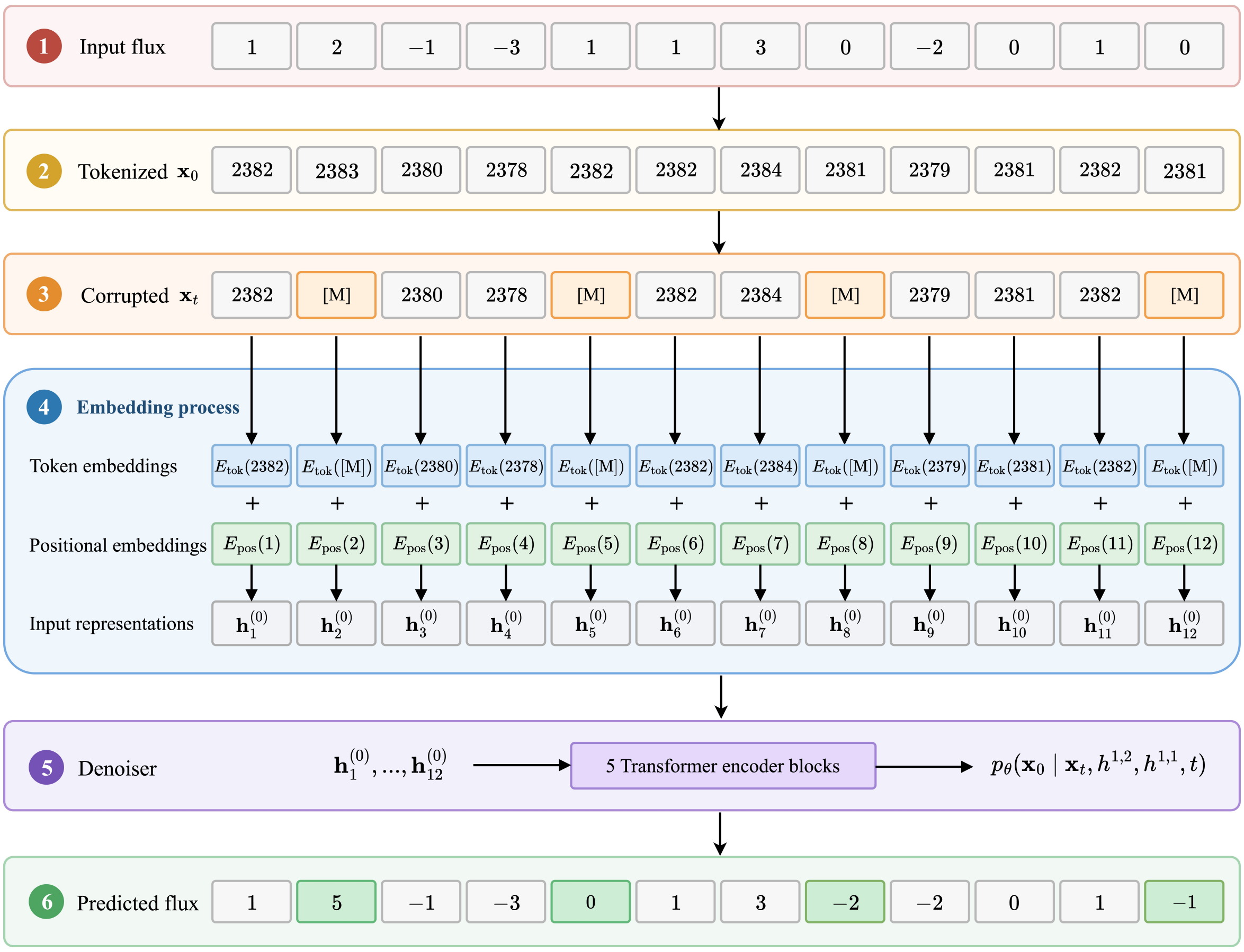}
    \caption{ Schematic illustration  for reconstructing a flux vector example from the training set on the geometry $(\hc, \hk)=(2, 120)$ at training time of the masked diffusion model. A clean integer flux vector from the training set is first tokenised by shifting each flux value by a fixed offset of $2381$. During the forward diffusion process, each token is independently replaced by \texttt{[MASK]} (denoted by \texttt{[M]} in this figure) with probability $t \sim \text{Unif}[0,1]$. The corrupted tokens are then mapped to learned token embeddings, which are combined with positional embeddings and processed by a bidirectional transformer conditioned on the Hodge pair $(h^{1,2}, h^{1,1})$ and diffusion time. The denoiser predicts categorical distributions over the original flux values at masked coordinates. The final row illustrates the predicted flux values, which need not coincide with the original configuration.}
    \label{fig:diffusion_workflow}
\end{figure}

\paragraph{Forward masking process.}

We use an absorbing-state discrete diffusion process
\cite{DBLP:conf/nips/AustinJHTB21,
sahoo2024simpleeffectivemaskeddiffusion}. Given a clean flux vector $\bm{x}_0 \sim p_{\text{data}}$ and masking probability $t\sim\operatorname{Unif}[0,1]$, the marginal corruption distribution for an active coordinate is
\begin{equation}
    q_t(x_{t,i}=k\mid x_{0,i})
    =
    (1-t)\,\delta_{k,x_{0,i}}
    +
    t\,\delta_{k,\mathrm{MASK}} .
    \label{eq:forward-masking}
\end{equation}
where $t=0$ corresponds to the clean flux vector and $t=1$
corresponds to a fully masked vector. The joint corruption distribution factorises over the active coordinates,
\begin{equation}
    q_t(\bm{x}_t\mid\bm{x}_0)
    =
    \prod_{i=1}^{L_g}
    q_t(x_{t,i}\mid x_{0,i}),
\end{equation}
while coordinates with $i>L_g$ remain deterministically equal to
\texttt{[PAD]}. Although the diffusion time $t$ is continuous, the states $\bm{x}_t$ remain discrete sequences of flux and special tokens.

\paragraph{Denoising transformer.}

The reverse process is parametrised by a bidirectional transformer that predicts the original flux value at each masked position. Following the standard embedding construction for transformer encoders, we form the input representation of coordinate \(i\) by adding learned token and positional embeddings \cite{Vaswani:2017lxt,Devlin:2018mgb,chang2022maskgitmaskedgenerativeimage},
\begin{equation*}
    \bm{h}_i^{(0)}
    =
    E_{\mathrm{tok}}(x_{t,i})
    +
    E_{\mathrm{pos}}(i),
    \qquad
    \bm{h}_i^{(0)}\in\mathbb{R}^{d}.
    \label{eq:flux-token-embedding}
\end{equation*}
The learned positional embeddings distinguish individual flux coordinates in the fixed ordering of Eq.~\eqref{eq:flux-vector}. In particular, they provide coordinate-specific representations even when all flux tokens are masked at the beginning of generation.

In addition to the partially masked sequence, the denoiser is conditioned on the Calabi--Yau geometry and the diffusion time. We describe each geometry by its Hodge pair
\begin{equation*}
    g
    =
    \left(
        h^{1,1},h^{1,2}
    \right),
    \qquad
    \bm{c}_{g,t}
    =
    \operatorname{MLP}_{g}(\widehat{g})
    +
    \operatorname{MLP}_{t}\!\left(
        \operatorname{PE}(t)
    \right) \in \mathbb{R}^d,
    \label{eq:geometry-time-conditioning}
\end{equation*}
where $\widehat{g}$ denotes standardisation using the training-set statistics and $\operatorname{PE}(t)$ is a sinusoidal timestep
embedding. The conditioning vector $\bm c_{g,t}$ is supplied to each transformer block through adaLN-Zero
\cite{peebles2023scalablediffusionmodelstransformers}. The scale, shift, and residual-gating parameters of the adaptive normalisation layers are
obtained from $\bm c_{g,t}$, with the residual gates initialised to zero.

The denoiser consists of five transformer encoder blocks with bidirectional self-attention. The first three blocks are shared with the initial-guess generator described below, after which the denoiser applies two additional encoder blocks. All blocks have hidden dimension $d=256$, eight attention heads, and feed-forward dimension $4d$. Padding
coordinates are excluded by the attention mask.

After the final encoder block, layer normalisation and a linear output projection produce logits over the $K$ flux values, where $K$ is the vocab size,
\begin{equation*}
    \bm{\ell}_i
    =
    W_{\mathrm{out}}
    \operatorname{LN}\!\left(\bm h_i^{(5)}\right)
    +
    \bm b_{\mathrm{out}},
    \qquad
    p_\theta(x_{0,i}=k\mid\bm x_t,g, t)
    =
    \operatorname{softmax}(\bm\ell_i)_k .
    \label{eq:clean-token-prediction}
\end{equation*}
This is an $\bm x_0$-prediction parametrisation: rather than predicting the noise, as commonly done in continuous Gaussian diffusion models, the network directly predicts a categorical distribution over the original flux values at each coordinate. This is the same parametrisation commonly used in masked diffusion language models
\cite{sahoo2024simpleeffectivemaskeddiffusion,
yang2026d3lmdiscretednadiffusion}.

\paragraph{Denoising objective.}
For a corrupted flux vector $\bm{x}_t$, let
\begin{equation*}
    \mathcal M_t
    =
    \left\{
        i\leq L_g:
        x_{t,i}=\mathrm{MASK}
    \right\}
    \label{eq:masked-coordinate-set}
\end{equation*}
denote the masked active coordinates. The model is trained to reconstruct the original flux tokens at these positions, using the time-weighted masked cross-entropy objective,
\begin{equation}
    \mathcal L_{\mathrm{diffusion}}(\theta)
    =
    - \mathbb{E}_{(\bm{x}_0, g) \sim p_{\text{data}}} \mathbb{E}_{t \sim \text{Unif}[0,1]}
    \mathbb E_{\bm x_t \sim q_t(\cdot \mid \bm{x}_0, g)}
    \left[
        \frac{1}{t L_g}
        \sum_{i = 1}^{L_g} \mathbf{1}[x_{t,i}=\mathrm{MASK}]
        \log
        p_\theta
        \left(
            x_{0,i}\mid\bm x_t, g, t
        \right)
    \right].
    \label{eq:denoising-loss}
\end{equation}
Here $p_{\text{data}}$ is the empirical distribution of clean flux vectors and their geometry labels from the dataset, while $q_t$ is the forward masking distribution at timestep $t$. The loss is evaluated only at masked positions, while the factor $1/(tL_g)$ normalises by the expected number of masked active coordinates under the independent corruption process.
Masked-token cross-entropy objectives of this form are commonly used for bidirectional masked generation
\cite{chang2022maskgitmaskedgenerativeimage,sahoo2024simpleeffectivemaskeddiffusion,
yang2026d3lmdiscretednadiffusion}.

\paragraph{Sequential masked decoding.}

Generation starts from a sequence in which every active coordinate is
equal to \texttt{[MASK]}. At each iteration, the denoiser predicts the
clean-token distribution at every remaining masked coordinate, after
which one coordinate is selected and revealed. The diffusion level
provided to the network is obtained by matching the current fraction of
masked coordinates to the nearest value of the training schedule
$m_t$.

Let $\mathcal M_s$ denote the masked coordinates at decoding iteration
$s$. Given a sampling temperature $\tau_{\mathrm{samp}}>0$, the
categorical distribution and confidence at coordinate $i$ are
\begin{equation}
    p_i^{(s)}(k)
    =
    \operatorname{softmax}
    \left(
        \frac{\bm\ell_i^{(s)}}{\tau_{\mathrm{samp}}}
    \right)_k,
    \qquad
    c_i^{(s)}
    =
    \max_k p_i^{(s)}(k).
    \label{eq:sampling-probability}
\end{equation}
Temperature-scaled categorical sampling is standard in iterative masked
decoding \cite{yang2026d3lmdiscretednadiffusion}.

We compare two coordinate-selection strategies:
\begin{align*}
    & j_s
    \sim
    \operatorname{Uniform}(\mathcal M_s)
    \quad\text{for random sampling},
    \\
    &j_s
    =
    \underset{i\in\mathcal M_s}{\arg\max}\;
    c_i^{(s)}
    \quad\text{for confidence sampling}.
    \label{eq:coordinate-selection}
\end{align*}
Random selection provides a model-independent decoding-order baseline, while confidence selection follows the confidence-based ordering used in iterative masked generation
\cite{chang2022maskgitmaskedgenerativeimage,yang2026d3lmdiscretednadiffusion}. We observed that confidence sampling outperforms random sampling in valid rate for flux generations, hence we choose confidence sampling throughout the work.

After selecting $j_s$, its flux value is sampled according to
\begin{equation*}
    x_{j_s}^{(s+1)}
    \sim
    \operatorname{Cat}
    \left(
        p_{j_s}^{(s)}
    \right),
    \qquad
    \mathcal M_{s+1}
    =
    \mathcal M_s\setminus\{j_s\}.
    \label{eq:sequential-unmasking}
\end{equation*}
All other coordinates remain unchanged, and revealed coordinates are not remasked. The procedure terminates after $L_g$ iterations.

Although one coordinate is revealed per iteration, the denoising model is not trained autoregressively. It is trained on arbitrary masking patterns and uses bidirectional attention over all currently revealed coordinates. A fixed left-to-right selection rule would therefore define a particular decoding order, but it would not turn the model into a causal transformer or replace the masked denoising objective by an autoregressive likelihood factorisation.

\paragraph{Initial-guess generator.}

To obtain initial guesses for the complex-structure moduli and axio-dilaton, we introduce a regression network that takes a complete flux vector as input. The network reuses the first three transformer encoder blocks of the denoiser, evaluated at $t=0$, whose parameters are kept frozen. This design reuses the flux representations learned during denoising, avoiding the need to train a separate flux encoder from scratch. The choice of three blocks keeps the initial-guess generator lightweight; we have not systematically optimised this depth.

The encoded flux representations from the pre-trained three transformer blocks are passed to two cross-attention blocks with learned queries, following the output-query approach of~\cite{jaegle2022perceiveriogeneralarchitecture}. These queries are trainable vectors that attend to the encoded flux coordinates, which serve as keys and values in the cross-attention mechanism. The model contains six queries, corresponding to $z^1,\ldots,z^5$ and $\tau$, of which only $h^{1,2}+1$ are active for a given geometry. The queries do not attend to one another. Each output representation is passed through layer normalisation and a linear layer to predict the real and imaginary components of the corresponding modulus.

To facilitate regression, we transform the target values according
to
\begin{equation}
    z^a \mapsto
    \left(
        \operatorname{Re}z^a,\,
        \operatorname{arsinh}(\operatorname{Im}z^a)
    \right),
    \qquad
    \tau \mapsto
    \left(
        \operatorname{Re}\tau,\,
        \log(\operatorname{Im}\tau)
    \right).
    \label{eq:moduli-transformations}
\end{equation}
The transformed values are standardised component-wise using the mean and standard deviation of the training data for each geometry. We denote the concatenated standardised targets by $\widetilde{\bm y}$ and the corresponding network predictions by $\widetilde{\bm y}^{\mathrm{pred}}$.

The initial-guess generator is trained using the Huber loss
with transition parameter $\delta=1$,
\begin{equation}
    \mathcal L_{\mathrm{init}}
    =
    \frac{1}{2(h^{1,2}+1)}
    \sum_{j=1}^{2(h^{1,2}+1)}
    \operatorname{Huber}_{1}
    \left(
        \widetilde y_j^{\mathrm{pred}}
        -
        \widetilde y_j
    \right).
    \label{eq:initialiser-loss}
\end{equation}

Training proceeds in two stages. First, the denoising transformer is trained using Eq.~\eqref{eq:denoising-loss}. Its parameters are then frozen, and only the learned queries, cross-attention blocks, and regression layers are trained using Eq.~\eqref{eq:initialiser-loss}.

\paragraph{Conditional modelling.}

To generate flux vacua with prescribed physical properties, we train
three independent conditional denoisers, one for each physical quantity
$\mathcal O\in\{N_{\mathrm{flux}},g_s,|W_0|\}$. Each model retains the
architecture of the base denoiser but receives an additional input
token encoding the target value of the corresponding quantity. Unlike the geometry and diffusion-time conditions, which are incorporated through adaLN-Zero, the observable condition is supplied through bidirectional self-attention.

For a target value $o$ of the physical quantity $\mathcal O$, we first
apply the transformation
\begin{equation}
    \widehat u
    =
    \frac{f_{\mathcal O}(o)-\mu_{g,\mathcal O}}
         {\sigma_{g,\mathcal O}},
    \qquad
    f_{\mathcal O}(o)
    =
    \begin{cases}
        o, & \mathcal O=N_{\mathrm{flux}},\\
        \log o, & \mathcal O=g_s,\\
        \log(o+10^{-8}), & \mathcal O=|W_0|,
    \end{cases}
    \label{eq:observable-standardisation}
\end{equation}
where $\mu_{g,\mathcal O}$ and $\sigma_{g,\mathcal O}$ denote the
mean and standard deviation of the transformed quantity in the
training data for geometry $g$.

Following the Fourier feature encoding of continuous inputs~\cite{mildenhall2020nerfrepresentingscenesneural,Tancik:2020pws}, we expand the standardised target $\widehat u$ using sinusoidal features at frequencies $\pi$ and $2\pi$,
\begin{equation}
    \phi(\widehat u)
    =
    \left(
        \widehat u,\,
        \sin(\pi\widehat u),\,
        \cos(\pi\widehat u),\,
        \sin(2\pi\widehat u),\,
        \cos(2\pi\widehat u)
    \right).
    \label{eq:conditional-observable-features}
\end{equation}
These features are projected to the transformer hidden dimension
by a two-layer MLP, yielding the condition embedding
\begin{equation}
    \bm e_{\mathcal O}(o,g)
    =
    \operatorname{MLP}_{\mathcal O}
    \left(\phi(\widehat u)\right)
    \in\mathbb R^d.
    \label{eq:conditional-observable-embedding}
\end{equation}
Since the target value is encoded directly through continuous-valued features and an MLP, no discretisation of continuous quantities such as $g_s$ and $|W_0|$ is required. This allows the conditional model to accept real-valued targets without imposing a fixed numerical resolution.

The resulting condition token is prepended to the flux-token representations before entering the transformer encoder. The input sequence is therefore
\begin{equation}
    \bm H_{\mathcal O}^{(0)}
    =
    \left(
        \bm e_{\mathcal O}(o,g),
        \bm h_1^{(0)},\ldots,\bm h_{L_g}^{(0)}
    \right).
    \label{eq:conditional-input-sequence}
\end{equation}
Through bidirectional self-attention, each flux coordinate can attend to the condition token, allowing its predicted distribution to depend on the prescribed target value. This follows the in-context conditioning approach of \cite{peebles2023scalablediffusionmodelstransformers}.
The geometry and diffusion-time conditioning through adaLN-Zero
remains unchanged, and the categorical output projection is applied
only to the flux coordinates. The condition token itself is not assigned a prediction target and does not contribute to the denoising loss.

Each conditional denoiser is trained using the masked cross-entropy
objective in Eq.~\eqref{eq:denoising-loss}, with the clean-token
distribution replaced by
$p_{\theta_{\mathcal O}}(x_{0,i}\mid\bm x_t,g,t,o)$. To enable classifier-free guidance, the condition token is omitted independently for each training example with probability $0.1$, by zeroing its representation and excluding it from self-attention. The same network therefore learns to predict the clean flux distribution both with and without the observable condition, while retaining the geometry and diffusion-time conditioning in both cases~\cite{ho2022classifierfreediffusionguidance}.

At sampling time, the conditional and unconditional logits are obtained from two evaluations of the same denoiser, using the same partially masked flux vector, geometry, and diffusion time. The observable condition token is included only in the conditional evaluation. We combine the two predictions using classifier-free guidance,
\begin{equation}
    \bm\ell^{\mathrm{cfg}}
    =
    \bm\ell_{\theta_{\mathcal O}}^{\mathrm{uncond}}
    +
    s_{\mathrm{cfg}}
    \left(
        \bm\ell_{\theta_{\mathcal O}}^{\mathrm{cond}}
        -
        \bm\ell_{\theta_{\mathcal O}}^{\mathrm{uncond}}
    \right),
    \qquad
    s_{\mathrm{cfg}}=2.
    \label{eq:categorical-cfg}
\end{equation}
Here, $s_{\mathrm{cfg}}=0$ corresponds to unconditional generation,
while $s_{\mathrm{cfg}}=1$ recovers the ordinary conditional
prediction. Values $s_{\mathrm{cfg}}>1$ strengthen the influence
of the prescribed target. The guided logits replace
$\bm\ell_i^{(s)}$ in Eq.~\eqref{eq:sampling-probability} during
sequential masked decoding.

The initial-guess generator is unchanged from the unconditional model:
it receives only the generated flux vector and the corresponding
Calabi--Yau geometry, without the additional observable condition.

\subsection{GRPO for diffusion model}\label{app:diffusion-grpo}
We used GRPO as the RL fine-tuning technique for the transformer, as shown in Appendix~\ref{app:transformer_grpo}. GRPO was originally formulated for autoregressive models, whose sequence probabilities admit the factorisation
\begin{equation*}
    \pi_{\theta}(\mathbf{o} \mid q) = \prod_{i=1}^L \pi_{\theta} (o_i \mid \mathbf{o}_{<i}, q).
\end{equation*}
This factorisation provides well-defined token-level importance ratios for the clipped policy objective. However, a masked diffusion model does not have a unique left-to-right factorisation: each prediction is conditioned on an arbitrary set of revealed coordinates, and the same final flux vector can be generated through different decoding orders of coordinates. Applying GRPO directly to one fixed order would therefore optimise the probability of that particular decoding path, rather than an order-independent likelihood for the candidate flux vector generated by the diffusion model.

To address that problem, we use a modified GRPO algorithm. Its central distinction from autoregressive GRPO is how coordinate selection enters the policy: a frozen reference model (pre-trained diffusion model) chooses which masked coordinate to reveal, while the trainable model (that GRPO updates) samples its value.

Let $q$ denote the prompt and $\bm{s}_{j,t}$ the partially filled token vector at step $t$ of sample $j$. Among the remaining masked coordinates $\mathcal{M}(\bm{s}_{j,t})$, the frozen reference model selects
\begin{equation}
    i_{j,t} = \underset{i\in\mathcal M(\boldsymbol{s}_{j,t})}{\arg\max} \max_{v\in\mathcal{V}}
\pi_{\mathrm{ref}}(v\mid\boldsymbol{s}_{j,t},i,q),
\end{equation}
where $\mathcal{V}$ is the flux-token vocabulary. Then rollout samples $v_{j,t}$ are collected from the sampling policy $\pi_{\theta_{\text{old}}}(\cdot \mid \bm{s}_{j,t}, i_{j,t}, q)$ and committed at the selected coordinate $i_{j,t}$. Thus, the selection rule for which coordinate to reveal remains fixed by $\pi_{\text{ref}}$ during GRPO, but the realised order depends on previously sampled values from the sampling policy $\pi_{\theta_{\text{old}}}$, rather than a predetermined sequence for revealing positions such as “$2 \rightarrow 4 \rightarrow 1 \rightarrow 3$”.

For each prompt, we generate a group of $G$ samples $\{\bm{F}_j \}_{j=1}^G$ and calculate their corresponding rewards by evaluating their validity using conditions in Sec.~\ref{sec:conditions}.  For the pre-trained model conditioned on only $(h^{1,2}, h^{1,1})$, the reward $R_j$ is calculated by 
\begin{equation*}
    R_j = \mathbb{I}\big[\bm{F}_j \text{ is a valid solution}\big] \ , 
\end{equation*}
and for the pre-trained model conditioned on an extra physical constraint $\mathcal{O} \in \{ N_{\text{flux}}, |W_0|, g_s \}$,
\begin{equation*}
    R_j = \mathbb{I}\big[\bm{F}_j \text{ is a valid solution}\big] \Big(1+\lambda \exp \big(-\frac{(\mathcal{O}^{\text{achieved}}- \mathcal{O}^{\text{target}})^2}{\alpha} \big) \Big) \ .
\end{equation*}
We then compute their group-relative advantages $A_j$ by
\begin{equation*}
    A_j = \frac{R_j - \mu}{\sigma} \ ,  
\end{equation*}
where $\mu$ and $\sigma$ are the mean and standard deviation of the rewards $\{R_j\}_{j=1}^G$ of the group, respectively. We note that the rewards are at trajectory-level, rather than a separate reward for each token.

Each trajectory records the partial states, selected coordinates, sampled values, and old-policy log-probabilities. During optimisation, the likelihood ratio is
\begin{equation}
    r_{j,t}(\theta)
=
\frac{
\pi_\theta(v_{j,t}\mid\boldsymbol{s}_{j,t},i_{j,t},q)
}{
\pi_{\theta_{\mathrm{old}}}(v_{j,t}\mid\boldsymbol{s}_{j,t},i_{j,t},q)
}.
\end{equation}
Both probabilities refer to the same recorded action and state. Importantly, both policies are evaluated at the same recorded partial sample and coordinate. Since the frozen reference model determines coordinate selection for both policies, the importance ratio only needs to compare their probabilities of the sampled token value. This avoids the inconsistency that would arise if a policy update changed the selected coordinate while the likelihood ratio accounted only for its token value.

The clipped policy contribution for prompt $q$ is
\begin{equation}
    \mathcal S_q(\theta)
=
\frac{1}{G}\sum_{j=1}^{G}\sum_{t=1}^{L_q}
\min\!\left[
r_{j,t}(\theta)A_j,\,
\operatorname{clip}\!\left(r_{j,t}(\theta),1-\epsilon,1+\epsilon\right)A_j
\right],
\end{equation}
where $L_q$ is the number of active flux coordinates. We regularise the policy using the exact categorical KL divergence to the reference, averaged over recorded states and selected coordinates:
\begin{equation}
    \mathcal K_q(\theta)
=
\frac{1}{GL_q}
\sum_{j=1}^{G}\sum_{t=1}^{L_q}
D_{\mathrm{KL}}\!\left[
\pi_\theta(\cdot\mid\boldsymbol{s}_{j,t},i_{j,t},q)
\,\Vert\,
\pi_{\mathrm{ref}}(\cdot\mid\boldsymbol{s}_{j,t},i_{j,t},q)
\right].
\end{equation}
Across geometry prompts $\mathcal Q$, we maximise
\begin{equation}
    \mathcal J(\theta)
=
\sum_{q\in\mathcal Q}
\frac{w_q}{\sum_{q'}w_{q'}}\,\mathcal S_q(\theta)
-
\frac{\beta}{|\mathcal Q|}
\sum_{q\in\mathcal Q}\mathcal K_q(\theta),
\end{equation}
where $w_q$ controls each geometry’s policy weight. Fresh trajectories are generated in each round.

\subsection{Hyperparameters}\label{sec:diffusion-hyperparameters}

\begin{table}[h]
    \centering
    \small
    \begin{tabular}{cc}
        \hline
        \textbf{Architectural parameter} & \textbf{Value} \\
        \hline \hline
        Transformer hidden dimension & 256 \\
        Attention heads & 8 \\
        Transformer blocks & 5 \\
        Feed-forward expansion ratio & 4 \\
        Dropout probability & 0 \\
        \hline
        Total trainable parameters & 8.74M \\
        \hline
    \end{tabular}
    \caption{Hyperparameters for the architecture of the unconditional flux denoiser. The flux denoiser contains five transformer blocks in total.}
    \label{tab:unconditional-architecture-diffusion}
\end{table}

\begin{table}[h]
    \centering
    \small
    \begin{tabular}{ccc}
        \hline
        \textbf{Training parameter} & \textbf{Flux denoiser} & \textbf{Initial-guess generator} \\
        \hline \hline
        Optimiser & AdamW & AdamW \\
        Initial learning rate & $2\times10^{-4}$ & $10^{-4}$ \\
        Adam coefficients $(\beta_1,\beta_2)$
            & $(0.9,0.95)$ & $(0.9,0.95)$ \\
        Weight decay & $10^{-2}$ & $10^{-2}$ \\
        Training batch size & 256 & 256 \\
        \hline
        Maximum epochs & 60 & 40 \\
        Minimum epochs before stopping & 10 & 12 \\
        Early-stopping patience (epochs) & 5 & 4 \\
        Relative improvement threshold & $10^{-3}$ & $10^{-3}$ \\
        Learning-rate reduction factor & 0.5 & 0.5 \\
        Learning-rate plateau patience (epochs) & 2 & 2 \\
        Minimum learning rate & $10^{-5}$ & $5\times10^{-6}$ \\
        \hline
    \end{tabular}
    \caption{Hyperparameters for unconditional flux-diffusion training and subsequent initial-guess generator training.
    Learning-rate reductions and early stopping are controlled by validation loss, and the checkpoint with the lowest validation loss is selected. The diffusion model is frozen during initial-guess generator training.}
    \label{tab:unconditional-training-diffusion}
\end{table}

We list the hyperparameters for the diffusion model architecture used in Sec.~\ref{sec:pre-train_grpo}, \ref{sec:fine-tuning} and \ref{sec:conditional-physical-constraints} in Tab.~\ref{tab:unconditional-architecture-diffusion}, and the hyperparameters for pre-training such a model in Tab.~\ref{tab:unconditional-training-diffusion}. The pre-training stage stops when the validation loss stops improving with the early-stopping patience defined in Tab.~\ref{tab:unconditional-training-diffusion}.

\begin{table}[h]
    \centering
    \small
    \begin{tabular}{cc}
        \hline
        \textbf{Shared GRPO parameter} & \textbf{Value} \\
        \hline \hline
        Maximum rounds & 200 \\
        Sampling temperature & 0.4 \\
        KL coefficient $\beta$ & 0.01 \\
        Learning rate & $2 \times 10^{-5}$ \\
        Weight decay & 0 \\
        Optimiser updates per round & 2 \\
        \hline 
        \textbf{Unconditional GRPO parameter} & \textbf{Value} \\
        \hline 
        Rollout group size per geometry for $h^{1,2}=2,3$ & 16  \\
        Rollout group size per geometry for $h^{1,2}=4,5$ & 256 \\
        \hline  
        \textbf{Conditional GRPO parameter} & \textbf{Value} \\
        \hline  
        Classifier-free guidance scale & 2.0 \\
        Prompts per geometry per round for rollout & 4 \\
        Rollout group size per geometry per prompt for $h^{1,2}=2,3$ & 16 \\
        Rollout group size per geometry per prompt for $h^{1,2}=4,5$ & 64 \\
        Relative policy weights for $h^{1,2}=2,3,4,5$ & $0.1, 0.1, 1, 1$ \\
        \hline
    \end{tabular}
    \caption{Hyperparameters for the GRPO implementation for the diffusion model.}
    \label{tab:diffusion-grpo-hyper-diffusion}
\end{table}

The hyperparameters for GRPO implemented for the diffusion model are summarised in Tab.~\ref{tab:diffusion-grpo-hyper-diffusion}. The stopping condition for unconditional GRPO is when the overall valid rate for unique, novel samples that do not appear in the training set under confidence sampling, averaged over 20 geometries, reaches 90\% or after the 200 total rounds, whichever occurs first. The stopping metric is evaluated every five rounds, using 500 generated samples for each of the 20 geometries with a fixed seed.

\subsection{Scaling over $\hc$}

\begin{figure}
    \centering
    \includegraphics{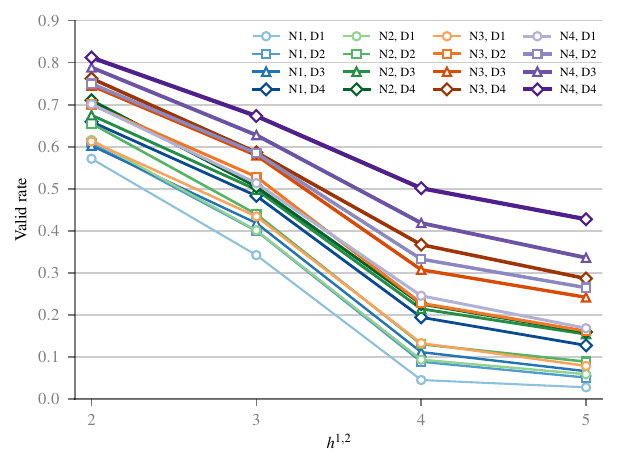}
    \caption{Average valid rate scaling across geometries with same $h^{1,2}$ for the diffusion model. The model sizes in the legends correspond to $N1=2.75$M, $N2=4.40$M, $N3=8.74$M, $N4=20.33$M; The dataset sizes correspond to $D1=10$k, $D2=20$k, $D3=40$k, $D4=80$k per geometry in the training set. }
    \label{fig:per_h12_scaling_validity_diffusion}
\end{figure}

The main motivation for this analysis in addition to the scaling study in Sec.~\ref{sec:scaling_study} is to assess empirically how model capacity and training set size affect performance as $h^{1,2}$ increases. This is particularly relevant for future foundation models, since generative models are expected to be more efficient at vacuum search for larger $h^{1,2}$, where direct numerical searches become increasingly costly. Fig.~\ref{fig:per_h12_scaling_validity_diffusion} shows that the pre-trained diffusion model validity decreases systematically with increasing $h^{1,2}$, while increasing either the model capacity or the amount of training data partially compensates for this drop in performance. Within the range studied, these results therefore provide a practical calibration of the resources needed to maintain a useful pre-training validity as the complexity of the geometries increases. This is particularly relevant for subsequent post-training with GRPO, which relies on the pre-trained model producing a sufficient number of valid samples to provide an effective reward signal. We stress that this study is empirical and restricted to the model sizes, dataset sizes and geometries considered here; it should therefore be interpreted as a guide for future scaling into a foundation model rather than as a universal scaling law. 

\section{Acceptance rates and distributions}\label{app:C}

This appendix supplements the timing comparison of Sec.~\ref{sec:timing} with two diagnostics that the cost per vacuum does not show. The first concerns efficiency: the cost of a vacuum depends not only on the cost of each proposal, but also on how many proposals are needed per vacuum. The second concerns coverage, since methods with similar costs can find very different vacua.

\begin{figure}[t!]
    \centering
    \includegraphics[width=0.92\textwidth]{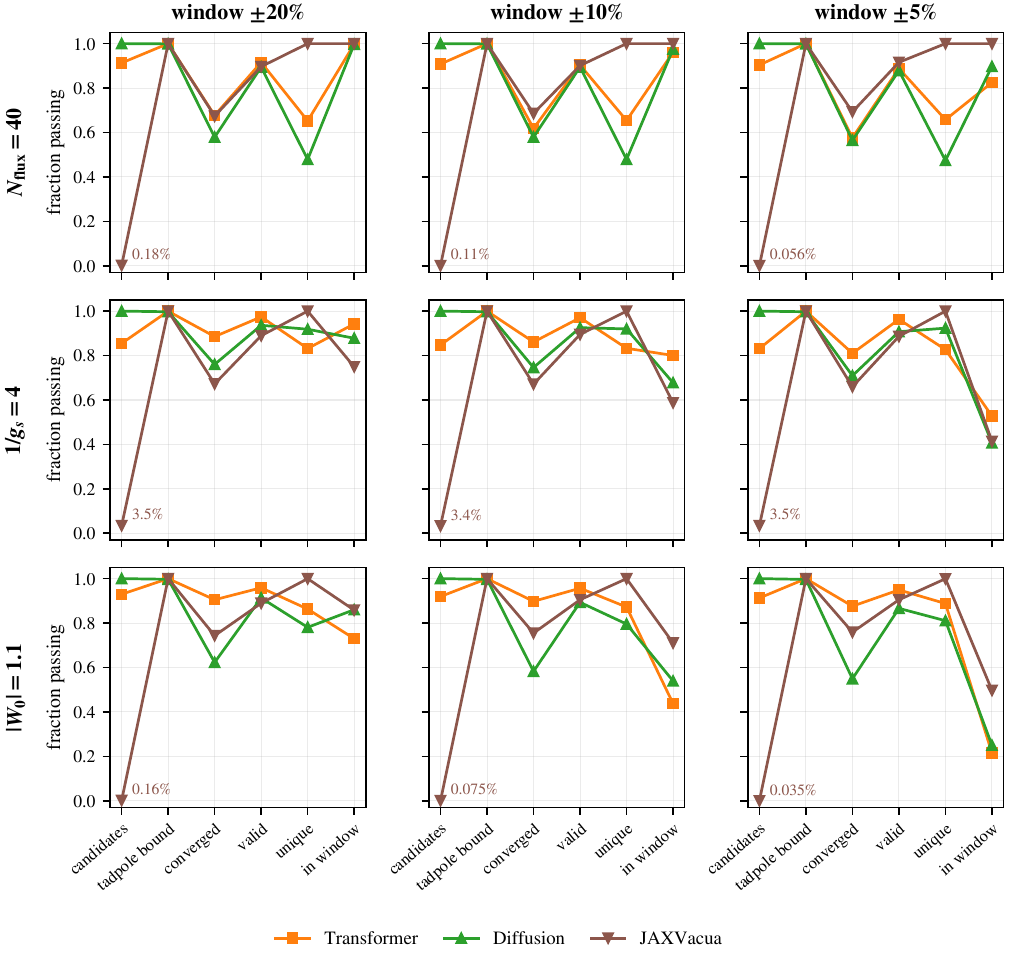}
    \caption{Stage-wise acceptance fractions pooled over the 20 geometries. Each fraction is relative to the number reaching the preceding stage. The first stage converts raw outputs into usable candidates; for \texttt{JAXVacua}, it includes the inexpensive cone, tadpole and provisional target checks. Subsequent stages show tadpole acceptance, solver convergence, validity, uniqueness and agreement with the final target window.}
    \label{fig:rates}
\end{figure}

The conversion of proposals into vacua explains much of this behaviour. In Fig.~\ref{fig:rates}, $83$--$93\%$ of transformer outputs and all diffusion outputs become candidates. Only $0.056$--$0.18\%$ of raw \texttt{JAXVacua} draws survive the initial selection in the $N_{\rm flux}$ experiment, compared with about $3.4\%$ for $1/g_s$ and $0.035$--$0.16\%$ for $|W_0|$. The later solver and validity stages have much closer success rates across methods. Repeated roots reduce the unique yield of the generative models, whereas almost every \texttt{JAXVacua} root retained here is distinct. Its early exact $N_{\rm flux}$ test also means that every distinct root reaching the final window check in that experiment passes it. An initial estimate of $|W_0|$ or $1/g_s$ does not have this property because the moduli move during minimisation.

\begin{figure}[t!]
    \centering
    \includegraphics[width=0.92\textwidth]{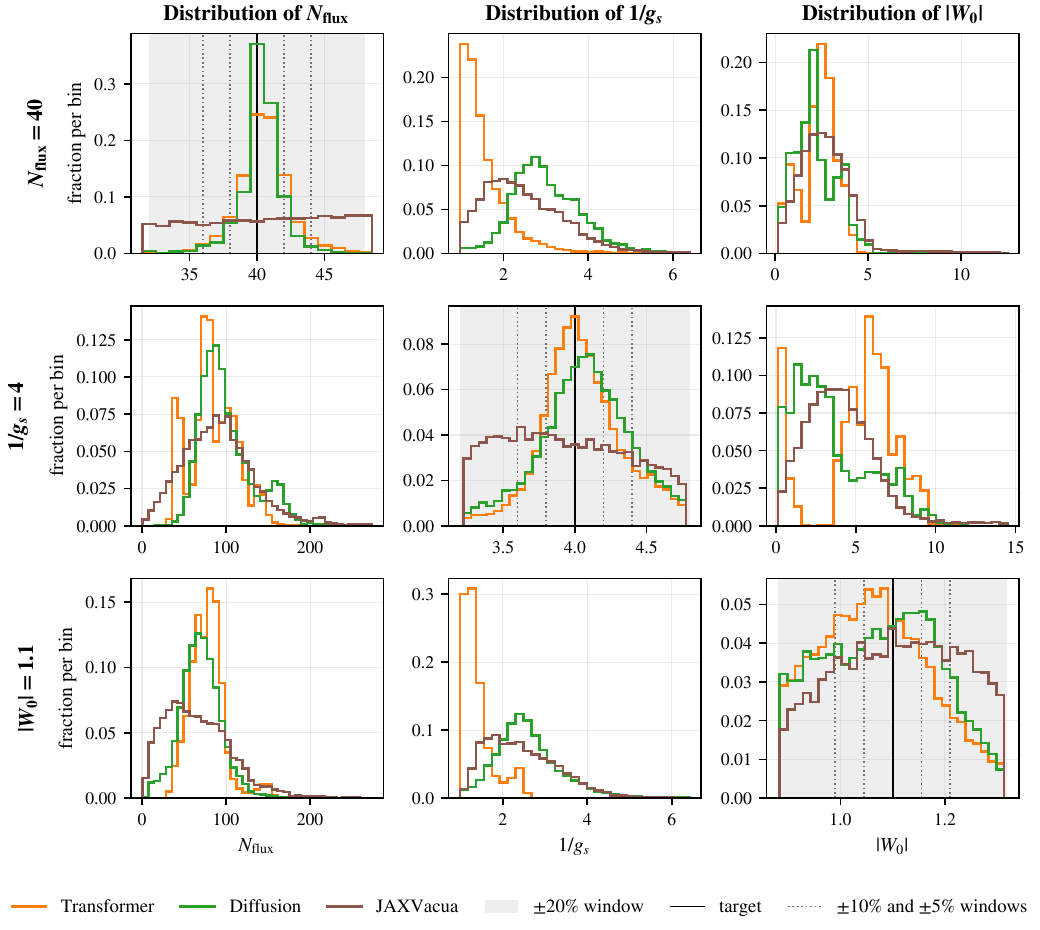}
    \caption{Distributions of all saved accepted roots satisfying the $\pm20\%$ condition, pooled over the 20 geometries and normalised separately for each method. These include later batches processed for the narrower windows and are not limited to the first timing batch. Pooling weights geometries by their respective yields. Rows specify the target experiment and columns the measured observable. In the diagonal panels the grey band marks the $\pm20\%$ window, the solid line the target and the dotted lines the $\pm10\%$ and $\pm5\%$ windows. Off-diagonal ranges are restricted to the pooled $0.5$th--$99.5$th percentiles for legibility.}
    \label{fig:distributions}
\end{figure}

Lastly, Fig.~\ref{fig:distributions} uses all saved roots in the $\pm20\%$ window. Each method is normalised by its own number of roots. In the $N_{\rm flux}$ experiment, $83\%$ of the transformer roots and $90\%$ of the diffusion roots also fall within $\pm5\%$ of the target, compared with $30\%$ for \texttt{JAXVacua}. The corresponding fractions are $57\%$, $48\%$ and $27\%$ for $1/g_s$, but only $35\%$, $33\%$ and $29\%$ for $|W_0|$. The distributions of the unconditioned observables differ as well. For example, in the $N_{\rm flux}$ runs the median $1/g_s$ is $1.39$ for the transformer, $2.92$ for diffusion and $2.33$ for \texttt{JAXVacua}. These histograms describe the roots produced by each search procedure, not an unbiased measure on the landscape. Throughput and coverage should therefore be assessed separately.

\FloatBarrier
\bibliographystyle{JHEP}
\bibliography{bibliography}

@article{Krippendorf:2023idy,
    author = "Krippendorf, Sven and Schachner, Andreas",
    title = "{New non-supersymmetric flux vacua in string theory}",
    eprint = "2308.15525",
    archivePrefix = "arXiv",
    primaryClass = "hep-th",
    reportNumber = "LMU-ASC 30/23",
    doi = "10.1007/JHEP12(2023)145",
    journal = "JHEP",
    volume = "12",
    pages = "145",
    year = "2023"
}

@article{Krippendorf:2021uxu,
    author = "Krippendorf, Sven and Kroepsch, Rene and Syvaeri, Marc",
    title = "{Revealing systematics in phenomenologically viable flux vacua with reinforcement learning}",
    eprint = "2107.04039",
    archivePrefix = "arXiv",
    primaryClass = "hep-th",
    reportNumber = "LMU-ASC 20/21, MPP-2021-108",
    month = "7",
    year = "2021"
}

@article{Demirtas:2021nlu,
    author = {Demirtas, Mehmet and Kim, Manki and McAllister, Liam and Moritz, Jakob and Rios-Tascon, Andres},
    title = {{Small cosmological constants in string theory}},
    journal = {JHEP},
    volume = {12},
    pages = {136},
    year = {2021},
    doi = {10.1007/JHEP12(2021)136},
    eprint = {2107.09064},
    archiveprefix = {arXiv},
    primaryclass = {hep-th},
}

@article{Strominger:1990pd,
    author = {Strominger, Andrew},
    title = {{SPECIAL GEOMETRY}},
    journal = {Commun. Math. Phys.},
    volume = {133},
    pages = {163--180},
    year = {1990},
    doi = {10.1007/BF02096559},
    reportnumber = {UCSB-TH-89-61},
}

@article{Candelas:1990rm,
    author = {Candelas, Philip and De La Ossa, Xenia C. and Green, Paul S. and Parkes, Linda},
    editor = {Yau, Shing-Tung},
    title = {{A Pair of Calabi-Yau manifolds as an exactly soluble superconformal theory}},
    journal = {Nucl. Phys. B},
    volume = {359},
    pages = {21--74},
    year = {1991},
    doi = {10.1016/0550-3213(91)90292-6},
    reportnumber = {UTTG-25-90},
}

@article{Ceresole:1992su,
    author = {Ceresole, Anna and D'Auria, R. and Ferrara, S. and Lerche, W. and Louis, J.},
    title = {{Picard-Fuchs equations and special geometry}},
    journal = {Int. J. Mod. Phys. A},
    volume = {8},
    pages = {79--114},
    year = {1993},
    doi = {10.1142/S0217751X93000047},
    eprint = {hep-th/9204035},
    archiveprefix = {arXiv},
    reportnumber = {CERN-TH-6441-92, UCLA-92-TEP-8, CALT-68-1776, POLFIS-TH-08-92},
}

@article{Hosono:1993qy,
    author = {Hosono, S. and Klemm, A. and Theisen, S. and Yau, Shing-Tung},
    title = {{Mirror symmetry, mirror map and applications to Calabi-Yau hypersurfaces}},
    journal = {Commun. Math. Phys.},
    volume = {167},
    pages = {301--350},
    year = {1995},
    doi = {10.1007/BF02100589},
    eprint = {hep-th/9308122},
    archiveprefix = {arXiv},
    reportnumber = {HUTMP-93-0801, LMU-TPW-93-22},
}

@article{Yip:2025hon,
    author = {Yip, Jacky H. T. and Arnal, Charles and Charton, Fran{\c{c}}ois and Shiu, Gary},
    title = {{Transforming Calabi-Yau Constructions: Generating New Calabi-Yau Manifolds with Transformers}},
    year = {2025},
    month = {7},
    eprint = {2507.03732},
    archiveprefix = {arXiv},
    primaryclass = {hep-th},
}

@article{Eguchi:2005eh,
    author = "Eguchi, Tohru and Tachikawa, Yuji",
    title = "{Distribution of flux vacua around singular points in Calabi-Yau moduli space}",
    eprint = "hep-th/0510061",
    archivePrefix = "arXiv",
    reportNumber = "UT-05-14, NSF-KITP-05-79",
    doi = "10.1088/1126-6708/2006/01/100",
    journal = "JHEP",
    volume = "01",
    pages = "100",
    year = "2006"
}

@article{Candelas:1990pi,
    author = {Candelas, Philip and de la Ossa, Xenia},
    title = {{Moduli Space of {Calabi-Yau} Manifolds}},
    journal = {Nucl. Phys. B},
    volume = {355},
    pages = {455--481},
    year = {1991},
    doi = {10.1016/0550-3213(91)90122-E},
    reportnumber = {UTTG-07-90},
}

@article{Braun:2011av,
    author = {Braun, Andreas P. and Johansson, Niklas and Larfors, Magdalena and Walliser, Nils-Ole},
    title = {{Restrictions on infinite sequences of type IIB vacua}},
    journal = {JHEP},
    volume = {10},
    pages = {091},
    year = {2011},
    doi = {10.1007/JHEP10(2011)091},
    eprint = {1108.1394},
    archiveprefix = {arXiv},
    primaryclass = {hep-th},
    reportnumber = {TUW-11-20, LMU-ASC-35-11},
}

@article{Ashok:2003gk,
    author = {Ashok, Sujay and Douglas, Michael R.},
    title = {{Counting flux vacua}},
    journal = {JHEP},
    volume = {01},
    pages = {060},
    year = {2004},
    doi = {10.1088/1126-6708/2004/01/060},
    eprint = {hep-th/0307049},
    archiveprefix = {arXiv},
    reportnumber = {RUNHETC-2003-21},
}

@article{Acharya:2006zw,
    author = {Acharya, Bobby Samir and Douglas, Michael R},
    title = {{A Finite landscape?}},
    year = {2006},
    month = {6},
    eprint = {hep-th/0606212},
    archiveprefix = {arXiv},
    reportnumber = {IC-2006-42},
}

@article{Douglas:2003um,
    author = {Douglas, Michael R.},
    title = {{The Statistics of string / M theory vacua}},
    journal = {JHEP},
    volume = {05},
    pages = {046},
    year = {2003},
    doi = {10.1088/1126-6708/2003/05/046},
    eprint = {hep-th/0303194},
    archiveprefix = {arXiv},
    reportnumber = {RUNHETC-2003-09},
}

@article{Douglas:2006es,
    author = {Douglas, Michael R. and Kachru, Shamit},
    title = {{Flux compactification}},
    journal = {Rev. Mod. Phys.},
    volume = {79},
    pages = {733--796},
    year = {2007},
    doi = {10.1103/RevModPhys.79.733},
    eprint = {hep-th/0610102},
    archiveprefix = {arXiv},
    reportnumber = {SLAC-PUB-12131},
}

@article{Lu:2009aw,
    author = {Lu, Zhiqin and Douglas, Michael R.},
    title = {{Gauss{\textendash}Bonnet{\textendash}Chern theorem on moduli space}},
    journal = {Math. Ann.},
    volume = {357},
    number = {2},
    pages = {469--511},
    year = {2013},
    doi = {10.1007/s00208-013-0907-4},
    eprint = {0902.3839},
    archiveprefix = {arXiv},
    primaryclass = {math.DG},
}

@article{Cheng:2019mgz,
    author = {Cheng, Miranda C. N. and Moore, Gregory W. and Paquette, Natalie M.},
    title = {{Flux vacua: a voluminous recount}},
    journal = {Commun. Num. Theor. Phys.},
    volume = {16},
    number = {4},
    pages = {761--800},
    year = {2022},
    doi = {10.4310/CNTP.2022.v16.n4.a4},
    eprint = {1909.04666},
    archiveprefix = {arXiv},
    primaryclass = {hep-th},
}

@article{Denef:2004ze,
    author = {Denef, Frederik and Douglas, Michael R.},
    title = {{Distributions of flux vacua}},
    journal = {JHEP},
    volume = {05},
    pages = {072},
    year = {2004},
    doi = {10.1088/1126-6708/2004/05/072},
    eprint = {hep-th/0404116},
    archiveprefix = {arXiv},
}

@article{Kachru:2003aw,
    author = {Kachru, Shamit and Kallosh, Renata and Linde, Andrei D. and Trivedi, Sandip P.},
    title = {{De Sitter vacua in string theory}},
    journal = {Phys. Rev. D},
    volume = {68},
    pages = {046005},
    year = {2003},
    doi = {10.1103/PhysRevD.68.046005},
    eprint = {hep-th/0301240},
    archiveprefix = {arXiv},
    reportnumber = {SLAC-PUB-9630, SU-ITP-03-01, TIFR-TH-03-03},
}

@article{Dasgupta:1999ss,
    author = "Dasgupta, Keshav and Rajesh, Govindan and Sethi, Savdeep",
    title = "{M theory, orientifolds and G - flux}",
    eprint = "hep-th/9908088",
    archivePrefix = "arXiv",
    reportNumber = "IASSNS-HEP-99-75, NSF-ITP-99-095",
    doi = "10.1088/1126-6708/1999/08/023",
    journal = "JHEP",
    volume = "08",
    pages = "023",
    year = "1999"
}

@article{Arnal:2026zyo,
    author = "Arnal, Charles and Yip, Jacky H. T. and Charton, Fran{\c{c}}ois and Shiu, Gary",
    title = "{Generating Special Triangulations with Transformers}",
    eprint = "2606.26660",
    archivePrefix = "arXiv",
    primaryClass = "hep-th",
    month = "6",
    year = "2026"
}

@article{mccabe2024multiplephysicspretrainingphysical,
      title={Multiple Physics Pretraining for Physical Surrogate Models}, 
      author={Michael McCabe and Bruno Régaldo-Saint Blancard and Liam Holden Parker and Ruben Ohana and Miles Cranmer and Alberto Bietti and Michael Eickenberg and Siavash Golkar and Geraud Krawezik and Francois Lanusse and Mariel Pettee and Tiberiu Tesileanu and Kyunghyun Cho and Shirley Ho},
      year={2024},
      eprint={2310.02994},
      archivePrefix={arXiv},
      primaryClass={cs.LG},
      url={https://arxiv.org/abs/2310.02994}, 
}

@article{zeni2024mattergengenerativemodelinorganic,
      title={MatterGen: a generative model for inorganic materials design}, 
      author={Claudio Zeni and Robert Pinsler and Daniel Zügner and Andrew Fowler and Matthew Horton and Xiang Fu and Sasha Shysheya and Jonathan Crabbé and Lixin Sun and Jake Smith and Bichlien Nguyen and Hannes Schulz and Sarah Lewis and Chin-Wei Huang and Ziheng Lu and Yichi Zhou and Han Yang and Hongxia Hao and Jielan Li and Ryota Tomioka and Tian Xie},
      year={2024},
      eprint={2312.03687},
      archivePrefix={arXiv},
      primaryClass={cond-mat.mtrl-sci},
      url={https://arxiv.org/abs/2312.03687}, 
}

@article{Birk:2024knn,
    author = "Birk, Joschka and Hallin, Anna and Kasieczka, Gregor",
    title = "{OmniJet-{\ensuremath{\alpha}}: the first cross-task foundation model for particle physics}",
    eprint = "2403.05618",
    archivePrefix = "arXiv",
    primaryClass = "hep-ph",
    doi = "10.1088/2632-2153/ad66ad",
    journal = "Mach. Learn. Sci. Tech.",
    volume = "5",
    number = "3",
    pages = "035031",
    year = "2024"
}

@article{Golling:2024abg,
    author = "Golling, Tobias and Heinrich, Lukas and Kagan, Michael and Klein, Samuel and Leigh, Matthew and Osadchy, Margarita and Raine, John Andrew",
    title = "{Masked particle modeling on sets: towards self-supervised high energy physics foundation models}",
    eprint = "2401.13537",
    archivePrefix = "arXiv",
    primaryClass = "hep-ph",
    doi = "10.1088/2632-2153/ad64a8",
    journal = "Mach. Learn. Sci. Tech.",
    volume = "5",
    number = "3",
    pages = "035074",
    year = "2024"
}

@article{bodnar2024foundationmodelearth,
      title={A Foundation Model for the Earth System}, 
      author={Cristian Bodnar and Wessel P. Bruinsma and Ana Lucic and Megan Stanley and Anna Vaughan and Johannes Brandstetter and Patrick Garvan and Maik Riechert and Jonathan A. Weyn and Haiyu Dong and Jayesh K. Gupta and Kit Thambiratnam and Alexander T. Archibald and Chun-Chieh Wu and Elizabeth Heider and Max Welling and Richard E. Turner and Paris Perdikaris},
      year={2024},
      eprint={2405.13063},
      archivePrefix={arXiv},
      primaryClass={physics.ao-ph},
      url={https://arxiv.org/abs/2405.13063}, 
}

@article{Giddings:2001yu,
    author = {Giddings, Steven B. and Kachru, Shamit and Polchinski, Joseph},
    title = {{Hierarchies from fluxes in string compactifications}},
    journal = {Phys. Rev. D},
    volume = {66},
    pages = {106006},
    year = {2002},
    doi = {10.1103/PhysRevD.66.106006},
    eprint = {hep-th/0105097},
    archiveprefix = {arXiv},
    reportnumber = {SLAC-PUB-8807, NSF-ITP-01-37, SU-ITP-01-16},
}

@article{Gukov:1999ya,
    author = {Gukov, Sergei and Vafa, Cumrun and Witten, Edward},
    title = {{CFT's from Calabi-Yau four folds}},
    journal = {Nucl. Phys. B},
    volume = {584},
    pages = {69--108},
    year = {2000},
    doi = {10.1016/S0550-3213(00)00373-4},
    eprint = {hep-th/9906070},
    archiveprefix = {arXiv},
    reportnumber = {HUTP-99-A034, IASSNS-HEP-99-52, PUPT-1864},
    note = {[Erratum: Nucl.Phys.B 608, 477--478 (2001)]},
}

@article{Grana:2005jc,
    author = {Grana, Mariana},
    title = {{Flux compactifications in string theory: A Comprehensive review}},
    journal = {Phys. Rept.},
    volume = {423},
    pages = {91--158},
    year = {2006},
    doi = {10.1016/j.physrep.2005.10.008},
    eprint = {hep-th/0509003},
    archiveprefix = {arXiv},
    reportnumber = {LPTENS-05-26, CPHT-RR-049-0805},
}

@article{Dubey:2023dvu,
    author = {Dubey, Abhishek and Krippendorf, Sven and Schachner, Andreas},
    title = {{JAXVacua {\textemdash} a framework for sampling string vacua}},
    journal = {JHEP},
    volume = {12},
    pages = {146},
    year = {2023},
    doi = {10.1007/JHEP12(2023)146},
    eprint = {2306.06160},
    archiveprefix = {arXiv},
    primaryclass = {hep-th},
}

@article{Chauhan:2025rdj,
    author = {Chauhan, Aman and Cicoli, Michele and Krippendorf, Sven and Maharana, Anshuman and Piantadosi, Pellegrino and Schachner, Andreas},
    title = {{Deep observations of the Type IIB flux landscape}},
    journal = {JHEP},
    volume = {07},
    pages = {271},
    year = {2025},
    month = {1},
    doi = {10.1007/JHEP07(2025)271},
    eprint = {2501.03984},
    archiveprefix = {arXiv},
    primaryclass = {hep-th},
}

@article{Chauhan:2026gid,
    author = {Chauhan, Aman and Cicoli, Michele and Krippendorf, Sven and Maharana, Anshuman and Piantadosi, Pellegrino and Schachner, Andreas},
    title = {{Parameter compression in the flux landscape}},
    journal = {JHEP},
    volume = {07},
    pages = {186},
    year = {2026},
    doi = {10.1007/JHEP07(2026)186},
    eprint = {2603.04941},
    archiveprefix = {arXiv},
    primaryclass = {hep-th},
    reportnumber = {LITP-26-05},
}

@article{Demirtas:2019sip,
    author = {Demirtas, Mehmet and Kim, Manki and Mcallister, Liam and Moritz, Jakob},
    title = {{Vacua with Small Flux Superpotential}},
    journal = {Phys. Rev. Lett.},
    volume = {124},
    number = {21},
    pages = {211603},
    year = {2020},
    doi = {10.1103/PhysRevLett.124.211603},
    eprint = {1912.10047},
    archiveprefix = {arXiv},
    primaryclass = {hep-th},
}

@article{Alvarez-Garcia:2020pxd,
    author = {{\'A}lvarez-Garc{\'\i}a, Rafael and Blumenhagen, Ralph and Brinkmann, Max and Schlechter, Lorenz},
    title = {{Small Flux Superpotentials for Type IIB Flux Vacua Close to a Conifold}},
    journal = {Fortsch. Phys.},
    volume = {68},
    pages = {2000088},
    year = {2020},
    doi = {10.1002/prop.202000088},
    eprint = {2009.03325},
    archiveprefix = {arXiv},
    primaryclass = {hep-th},
    reportnumber = {MPP-2020-165},
}

@article{Martinez-Pedrera:2012teo,
    author = {Martinez-Pedrera, Danny and Mehta, Dhagash and Rummel, Markus and Westphal, Alexander},
    title = {{Finding all flux vacua in an explicit example}},
    journal = {JHEP},
    volume = {06},
    pages = {110},
    year = {2013},
    doi = {10.1007/JHEP06(2013)110},
    eprint = {1212.4530},
    archiveprefix = {arXiv},
    primaryclass = {hep-th},
    reportnumber = {DESY-12-246},
}

@article{Plauschinn:2023hjw,
    author = {Plauschinn, Erik and Schlechter, Lorenz},
    title = {{Flux vacua of the mirror octic}},
    journal = {JHEP},
    volume = {01},
    pages = {157},
    year = {2024},
    doi = {10.1007/JHEP01(2024)157},
    eprint = {2310.06040},
    archiveprefix = {arXiv},
    primaryclass = {hep-th},
}

@article{Cole:2019enn,
    author = {Cole, Alex and Schachner, Andreas and Shiu, Gary},
    title = {{Searching the Landscape of Flux Vacua with Genetic Algorithms}},
    journal = {JHEP},
    volume = {11},
    pages = {045},
    year = {2019},
    doi = {10.1007/JHEP11(2019)045},
    eprint = {1907.10072},
    archiveprefix = {arXiv},
    primaryclass = {hep-th},
    reportnumber = {MAD-TH-19-05},
}

@inproceedings{Cole:2021nnt,
    author = {Cole, Alex and Krippendorf, Sven and Schachner, Andreas and Shiu, Gary},
    title = {{Probing the Structure of String Theory Vacua with Genetic Algorithms and Reinforcement Learning}},
    booktitle = {{35th Conference on Neural Information Processing Systems}},
    year = {2021},
    month = {11},
    eprint = {2111.11466},
    archiveprefix = {arXiv},
    primaryclass = {hep-th},
}

@article{Krippendorf:2025mhp,
    author = {Krippendorf, Sven and Liu, Zhimei},
    title = {{Solving inverse problems of Type IIB flux vacua with conditional generative models}},
    journal = {JHEP},
    volume = {07},
    pages = {103},
    year = {2026},
    doi = {10.1007/JHEP07(2026)103},
    eprint = {2506.22551},
    archiveprefix = {arXiv},
    primaryclass = {hep-th},
    url = {https://arxiv.org/abs/2506.22551},
}

@article{Demirtas:2022hqf,
    author = {Demirtas, Mehmet and Rios-Tascon, Andres and McAllister, Liam},
    title = {{CYTools: A Software Package for Analyzing Calabi-Yau Manifolds}},
    year = {2022},
    month = {11},
    eprint = {2211.03823},
    archiveprefix = {arXiv},
    primaryclass = {hep-th},
}

@inproceedings{Vaswani:2017lxt,
    author = {Vaswani, Ashish and Shazeer, Noam and Parmar, Niki and Uszkoreit, Jakob and Jones, Llion and Gomez, Aidan N. and Kaiser, Lukasz and Polosukhin, Illia},
    editor = {I. Guyon and U. Von Luxburg and S. Bengio and H. Wallach and R. Fergus and S. Vishwanathan and R. Garnett},
    title = {{Attention Is All You Need}},
    booktitle = {{31st International Conference on Neural Information Processing Systems}},
    volume = {30},
    pages = {5998--6008},
    publisher = {Curran Associates, Inc.},
    year = {2017},
    month = {6},
    eprint = {1706.03762},
    archiveprefix = {arXiv},
    primaryclass = {cs.CL},
    url = {https://arxiv.org/abs/1706.03762},
}

@article{Demirtas:2020ffz,
    author = "Demirtas, Mehmet and Kim, Manki and McAllister, Liam and Moritz, Jakob",
    title = "{Conifold Vacua with Small Flux Superpotential}",
    eprint = "2009.03312",
    archivePrefix = "arXiv",
    primaryClass = "hep-th",
    doi = "10.1002/prop.202000085",
    journal = "Fortsch. Phys.",
    volume = "68",
    pages = "2000085",
    year = "2020"
}

@article{ho2022classifierfreediffusionguidance,
    author = {Jonathan Ho and Tim Salimans},
    title = {Classifier-Free Diffusion Guidance},
    year = {2022},
    eprint = {2207.12598},
    archiveprefix = {arXiv},
    primaryclass = {cs.LG},
    url = {https://arxiv.org/abs/2207.12598},
}

@article{shao2024deepseekmathpushinglimitsmathematical,
    author = {Zhihong Shao and Peiyi Wang and Qihao Zhu and Runxin Xu and Junxiao Song and Xiao Bi and Haowei Zhang and Mingchuan Zhang and Y. K. Li and Y. Wu and Daya Guo},
    title = {DeepSeekMath: Pushing the Limits of Mathematical Reasoning in Open Language Models},
    year = {2024},
    eprint = {2402.03300},
    archiveprefix = {arXiv},
    primaryclass = {cs.CL},
    url = {https://arxiv.org/abs/2402.03300},
}

@inproceedings{DBLP:conf/nips/AustinJHTB21,
    author = {Jacob Austin and
                  Daniel D. Johnson and
                  Jonathan Ho and
                  Daniel Tarlow and
                  Rianne van den Berg},
    editor = {Marc'Aurelio Ranzato and
                  Alina Beygelzimer and
                  Yann N. Dauphin and
                  Percy Liang and
                  Jennifer Wortman Vaughan},
    title = {Structured Denoising Diffusion Models in Discrete State-Spaces},
    booktitle = {Advances in Neural Information Processing Systems 34: Annual Conference
                  on Neural Information Processing Systems 2021, NeurIPS 2021, December
                  6-14, 2021, virtual},
    volume = {34},
    pages = {17981--17993},
    publisher = {Curran Associates, Inc.},
    year = {2021},
    eprint = {2107.03006},
    archiveprefix = {arXiv},
    primaryclass = {cs.LG},
    url = {https://proceedings.neurips.cc/paper/2021/hash/958c530554f78bcd8e97125b70e6973d-Abstract.html},
    bibsource = {dblp computer science bibliography, https://dblp.org},
}

@article{Ebelt:2023clh,
    author = {Ebelt, Julian and Krippendorf, Sven and Schachner, Andreas},
    title = {{W0{\_}sample = np.random.normal(0,1)?}},
    journal = {Phys. Lett. B},
    volume = {855},
    pages = {138786},
    year = {2024},
    doi = {10.1016/j.physletb.2024.138786},
    eprint = {2307.15749},
    archiveprefix = {arXiv},
    primaryclass = {hep-th},
}

@inproceedings{peebles2023scalablediffusionmodelstransformers,
    author = {William Peebles and Saining Xie},
    title = {Scalable Diffusion Models with Transformers},
    booktitle = {2023 IEEE/CVF International Conference on Computer Vision (ICCV)},
    pages = {4172--4182},
    publisher = {IEEE},
    year = {2023},
    doi = {10.1109/iccv51070.2023.00387},
    eprint = {2212.09748},
    archiveprefix = {arXiv},
    primaryclass = {cs.CV},
    url = {https://arxiv.org/abs/2212.09748},
}

@inproceedings{chang2022maskgitmaskedgenerativeimage,
    author = {Huiwen Chang and Han Zhang and Lu Jiang and Ce Liu and William T. Freeman},
    title = {MaskGIT: Masked Generative Image Transformer},
    booktitle = {2022 IEEE/CVF Conference on Computer Vision and Pattern Recognition (CVPR)},
    pages = {11305--11315},
    publisher = {IEEE},
    year = {2022},
    doi = {10.1109/cvpr52688.2022.01103},
    eprint = {2202.04200},
    archiveprefix = {arXiv},
    primaryclass = {cs.CV},
    url = {https://arxiv.org/abs/2202.04200},
}

@inproceedings{sahoo2024simpleeffectivemaskeddiffusion,
    author = {Subham Sekhar Sahoo and Marianne Arriola and Yair Schiff and Aaron Gokaslan and Edgar Marroquin and Justin T Chiu and Alexander Rush and Volodymyr Kuleshov},
    title = {Simple and Effective Masked Diffusion Language Models},
    booktitle = {Advances in Neural Information Processing Systems 37},
    pages = {130136--130184},
    publisher = {Neural Information Processing Systems Foundation, Inc. (NeurIPS)},
    year = {2024},
    doi = {10.52202/079017-4135},
    eprint = {2406.07524},
    archiveprefix = {arXiv},
    primaryclass = {cs.CL},
    url = {https://arxiv.org/abs/2406.07524},
}

@article{Walden:2025cpf,
    author = {Walden, Moritz and Larfors, Magdalena},
    title = {{Sampling string vacua using generative models}},
    journal = {Mach. Learn. Sci. Tech.},
    volume = {7},
    number = {1},
    pages = {015018},
    publisher = {IOP Publishing},
    year = {2026},
    month = {Jan},
    doi = {10.1088/2632-2153/ae32dc},
    eprint = {2509.16029},
    archiveprefix = {arXiv},
    primaryclass = {hep-th},
    reportnumber = {UUITP-26/25},
    issn = {2632-2153},
    url = {http://dx.doi.org/10.1088/2632-2153/ae32dc},
}

@inproceedings{hoffmann2022trainingcomputeoptimallargelanguage,
    author = {Jordan Hoffmann and Sebastian Borgeaud and Arthur Mensch and Elena Buchatskaya and Trevor Cai and Eliza Rutherford and Diego de Las Casas and Lisa Anne Hendricks and Johannes Welbl and Aidan Clark and Tom Hennigan and Eric Noland and Katie Millican and George van den Driessche and Bogdan Damoc and Aurelia Guy and Simon Osindero and Karen Simonyan and Erich Elsen and Jack W. Rae and Oriol Vinyals and Laurent Sifre},
    title = {Training Compute-Optimal Large Language Models},
    booktitle = {Advances in Neural Information Processing Systems 35},
    pages = {30016--30030},
    publisher = {Neural Information Processing Systems Foundation, Inc. (NeurIPS)},
    year = {2022},
    doi = {10.52202/068431-2176},
    eprint = {2203.15556},
    archiveprefix = {arXiv},
    primaryclass = {cs.CL},
    url = {https://arxiv.org/abs/2203.15556},
}

@article{Kaplan:2020trn,
    author = {Kaplan, Jared and McCandlish, Sam and Henighan, Tom and Brown, Tom B. and Chess, Benjamin and Child, Rewon and Gray, Scott and Radford, Alec and Wu, Jeffrey and Amodei, Dario},
    title = {{Scaling Laws for Neural Language Models}},
    year = {2020},
    month = {1},
    eprint = {2001.08361},
    archiveprefix = {arXiv},
    primaryclass = {cs.LG},
    url = {https://arxiv.org/abs/2001.08361},
}

@article{henighan2020scalinglawsautoregressivegenerative,
    author = {Tom Henighan and Jared Kaplan and Mor Katz and Mark Chen and Christopher Hesse and Jacob Jackson and Heewoo Jun and Tom B. Brown and Prafulla Dhariwal and Scott Gray and Chris Hallacy and Benjamin Mann and Alec Radford and Aditya Ramesh and Nick Ryder and Daniel M. Ziegler and John Schulman and Dario Amodei and Sam McCandlish},
    title = {Scaling Laws for Autoregressive Generative Modeling},
    year = {2020},
    eprint = {2010.14701},
    archiveprefix = {arXiv},
    primaryclass = {cs.LG},
    url = {https://arxiv.org/abs/2010.14701},
}

@inproceedings{li2024scalabilitydiffusionbasedtexttoimagegeneration,
    author = {Hao Li and Yang Zou and Ying Wang and Orchid Majumder and Yusheng Xie and R. Manmatha and Ashwin Swaminathan and Zhuowen Tu and Stefano Ermon and Stefano Soatto},
    title = {On the Scalability of Diffusion-based Text-to-Image Generation},
    booktitle = {2024 IEEE/CVF Conference on Computer Vision and Pattern Recognition (CVPR)},
    pages = {9400--9409},
    publisher = {IEEE},
    year = {2024},
    doi = {10.1109/cvpr52733.2024.00898},
    eprint = {2404.02883},
    archiveprefix = {arXiv},
    primaryclass = {cs.CV},
    url = {https://arxiv.org/abs/2404.02883},
}

@article{Denef:2008wq,
    author = {Denef, Frederik},
    editor = {Bachas, Costas and Baulieu, Laurent and Douglas, Michael and Kiritsis, Elias and Rabinovici, Eliezer and Vanhove, Pierre and Windey, Paul and Cugliandolo, Leticia F.},
    title = {{Lectures on constructing string vacua}},
    journal = {Les Houches},
    volume = {87},
    pages = {483--610},
    year = {2008},
    doi = {10.1016/S0924-8099(08)80029-7},
    eprint = {0803.1194},
    archiveprefix = {arXiv},
    primaryclass = {hep-th},
}

@inproceedings{ICLR2026_880d8999,
    author = {Liang, Zhengyang and He, Hao and Yang, Ceyuan and DAI, Bo},
    editor = {C. Vondrick and B. Hariharan and C. Raffel and L. Pinto and D. Yang and A. Faust},
    title = {Scaling Laws for Diffusion Transformers},
    booktitle = {International Conference on Learning Representations},
    volume = {2026},
    pages = {84488--84521},
    year = {2026},
    eprint = {2410.08184},
    archiveprefix = {arXiv},
    primaryclass = {cs.CV},
    url = {https://proceedings.iclr.cc/paper_files/paper/2026/file/880d8999c07a8efc9bbbeb0c38f50765-Paper-Conference.pdf},
}

@article{yang2026d3lmdiscretednadiffusion,
    author = {Zhao Yang and Hengchang Liu and Chuan Cao and Bing Su},
    title = {D3LM: A Discrete DNA Diffusion Language Model for Bidirectional DNA Understanding and Generation},
    year = {2026},
    eprint = {2603.01780},
    archiveprefix = {arXiv},
    primaryclass = {cs.LG},
    url = {https://arxiv.org/abs/2603.01780},
}

@inproceedings{hoogeboom2021argmaxflowsmultinomialdiffusion,
    author = {Emiel Hoogeboom and Didrik Nielsen and Priyank Jaini and Patrick Forré and Max Welling},
    editor = {M. Ranzato and A. Beygelzimer and Y. Dauphin and P.S. Liang and J. Wortman Vaughan},
    title = {Argmax Flows and Multinomial Diffusion: Learning Categorical Distributions},
    booktitle = {Advances in Neural Information Processing Systems},
    volume = {34},
    pages = {12454--12465},
    publisher = {Curran Associates, Inc.},
    year = {2021},
    eprint = {2102.05379},
    archiveprefix = {arXiv},
    primaryclass = {stat.ML},
    url = {https://arxiv.org/abs/2102.05379},
}

@inproceedings{Devlin:2018mgb,
    author = {Devlin, Jacob and Chang, Ming-Wei and Lee, Kenton and Toutanova, Kristina},
    editor = {Burstein, Jill  and
      Doran, Christy  and
      Solorio, Thamar},
    title = {{BERT: Pre-training of Deep Bidirectional Transformers for Language Understanding}},
    booktitle = {Proceedings of the 2019 Conference of the North {A}merican Chapter of the Association for Computational Linguistics: Human Language Technologies, Volume 1 (Long and Short Papers)},
    pages = {4171--4186},
    publisher = {Association for Computational Linguistics},
    address = {Minneapolis, Minnesota},
    year = {2018},
    month = {10},
    doi = {10.18653/v1/N19-1423},
    eprint = {1810.04805},
    archiveprefix = {arXiv},
    primaryclass = {cs.CL},
    url = {https://aclanthology.org/N19-1423/},
}

@article{Demirtas:2023als,
    author = {Demirtas, Mehmet and Kim, Manki and McAllister, Liam and Moritz, Jakob and Rios-Tascon, Andres},
    title = {{Computational Mirror Symmetry}},
    journal = {JHEP},
    volume = {01},
    pages = {184},
    year = {2024},
    doi = {10.1007/JHEP01(2024)184},
    eprint = {2303.00757},
    archiveprefix = {arXiv},
    primaryclass = {hep-th},
    reportnumber = {MIT-CTP/5528},
    url = {https://arxiv.org/abs/2303.00757},
}

@article{McAllister:2023vgy,
    author = {McAllister, Liam and Quevedo, Fernando},
    title = {{Moduli Stabilization in String Theory}},
    year = {2023},
    month = {10},
    eprint = {2310.20559},
    archiveprefix = {arXiv},
    primaryclass = {hep-th},
    url = {https://arxiv.org/abs/2310.20559},
}

@article{Hosono:1994av,
    author = {Hosono, S. and Klemm, A. and Theisen, S.},
    editor = {Alekseev, Anton and Hietam{\"a}ki, Antero and Huitu, Katri and Morozov, Alexei and Niemi, Antti},
    title = {{Lectures on mirror symmetry}},
    journal = {Lect. Notes Phys.},
    volume = {436},
    pages = {235--280},
    year = {1994},
    doi = {10.1007/3-540-58453-6_13},
    eprint = {hep-th/9403096},
    archiveprefix = {arXiv},
    reportnumber = {HUTMP-94-01, LMU-TPW-94-02},
}

@article{Gopakumar:1998ii,
    author = {Gopakumar, Rajesh and Vafa, Cumrun},
    title = {{M theory and topological strings. 1.}},
    year = {1998},
    month = {9},
    eprint = {hep-th/9809187},
    archiveprefix = {arXiv},
    reportnumber = {HUTP-98-A069},
}

@article{Gopakumar:1998jq,
    author = {Gopakumar, Rajesh and Vafa, Cumrun},
    title = {{M theory and topological strings. 2.}},
    year = {1998},
    month = {12},
    eprint = {hep-th/9812127},
    archiveprefix = {arXiv},
    reportnumber = {HUTP-98-A070},
}

@article{bommasani2022opportunitiesrisksfoundationmodels,
    author = {Rishi Bommasani and Drew A. Hudson and Ehsan Adeli and Russ Altman and Simran Arora and Sydney von Arx and Michael S. Bernstein and Jeannette Bohg and Antoine Bosselut and Emma Brunskill and Erik Brynjolfsson and Shyamal Buch and Dallas Card and Rodrigo Castellon and Niladri Chatterji and Annie Chen and Kathleen Creel and Jared Quincy Davis and Dora Demszky and Chris Donahue and Moussa Doumbouya and Esin Durmus and Stefano Ermon and John Etchemendy and Kawin Ethayarajh and Li Fei-Fei and Chelsea Finn and Trevor Gale and Lauren Gillespie and Karan Goel and Noah Goodman and Shelby Grossman and Neel Guha and Tatsunori Hashimoto and Peter Henderson and John Hewitt and Daniel E. Ho and Jenny Hong and Kyle Hsu and Jing Huang and Thomas Icard and Saahil Jain and Dan Jurafsky and Pratyusha Kalluri and Siddharth Karamcheti and Geoff Keeling and Fereshte Khani and Omar Khattab and Pang Wei Koh and Mark Krass and Ranjay Krishna and Rohith Kuditipudi and Ananya Kumar and Faisal Ladhak and Mina Lee and Tony Lee and Jure Leskovec and Isabelle Levent and Xiang Lisa Li and Xuechen Li and Tengyu Ma and Ali Malik and Christopher D. Manning and Suvir Mirchandani and Eric Mitchell and Zanele Munyikwa and Suraj Nair and Avanika Narayan and Deepak Narayanan and Ben Newman and Allen Nie and Juan Carlos Niebles and Hamed Nilforoshan and Julian Nyarko and Giray Ogut and Laurel Orr and Isabel Papadimitriou and Joon Sung Park and Chris Piech and Eva Portelance and Christopher Potts and Aditi Raghunathan and Rob Reich and Hongyu Ren and Frieda Rong and Yusuf Roohani and Camilo Ruiz and Jack Ryan and Christopher Ré and Dorsa Sadigh and Shiori Sagawa and Keshav Santhanam and Andy Shih and Krishnan Srinivasan and Alex Tamkin and Rohan Taori and Armin W. Thomas and Florian Tramèr and Rose E. Wang and William Wang and Bohan Wu and Jiajun Wu and Yuhuai Wu and Sang Michael Xie and Michihiro Yasunaga and Jiaxuan You and Matei Zaharia and Michael Zhang and Tianyi Zhang and Xikun Zhang and Yuhui Zhang and Lucia Zheng and Kaitlyn Zhou and Percy Liang},
    title = {On the Opportunities and Risks of Foundation Models},
    year = {2022},
    eprint = {2108.07258},
    archiveprefix = {arXiv},
    primaryclass = {cs.LG},
    url = {https://arxiv.org/abs/2108.07258},
}

@article{Bousso:2000xa,
    author = {Bousso, Raphael and Polchinski, Joseph},
    title = {{Quantization of four form fluxes and dynamical neutralization of the cosmological constant}},
    journal = {JHEP},
    volume = {06},
    pages = {006},
    year = {2000},
    doi = {10.1088/1126-6708/2000/06/006},
    eprint = {hep-th/0004134},
    archiveprefix = {arXiv},
    reportnumber = {SU-ITP-00-12, NSF-ITP-00-40},
}

@incollection{Susskind:2003kw,
    author = {Susskind, Leonard},
    editor = {Carr, Bernard J.},
    title = {{The Anthropic landscape of string theory}},
    booktitle = {Universe or Multiverse?},
    pages = {247--266},
    publisher = {Cambridge University Press},
    year = {2007},
    month = {2},
    doi = {10.1017/CBO9781107050990.018},
    eprint = {hep-th/0302219},
    archiveprefix = {arXiv},
    isbn = {9780521848411},
    url = {https://www.cambridge.org/core/books/universe-or-multiverse/38972284ED1AECC0A692CF429DF57D53},
}

@incollection{mildenhall2020nerfrepresentingscenesneural,
    author = {Ben Mildenhall and Pratul P. Srinivasan and Matthew Tancik and Jonathan T. Barron and Ravi Ramamoorthi and Ren Ng},
    title = {NeRF: Representing Scenes as Neural Radiance Fields for View Synthesis},
    booktitle = {Lecture Notes in Computer Science},
    pages = {405--421},
    publisher = {Springer International Publishing},
    year = {2020},
    doi = {10.1007/978-3-030-58452-8_24},
    eprint = {2003.08934},
    archiveprefix = {arXiv},
    primaryclass = {cs.CV},
    isbn = {9783030584511},
    url = {https://arxiv.org/abs/2003.08934},
}

@inproceedings{Tancik:2020pws,
    author = {Tancik, Matthew and Srinivasan, Pratul P. and Mildenhall, Ben and Fridovich-Keil, Sara and Raghavan, Nithin and Singhal, Utkarsh and Ramamoorthi, Ravi and Barron, Jonathan T. and Ng, Ren},
    editor = {H. Larochelle and M. Ranzato and R. Hadsell and M.F. Balcan and H. Lin},
    title = {{Fourier Features Let Networks Learn High Frequency Functions in Low Dimensional Domains}},
    booktitle = {Advances in Neural Information Processing Systems},
    volume = {33},
    pages = {7537--7547},
    publisher = {Curran Associates, Inc.},
    year = {2020},
    month = {6},
    eprint = {2006.10739},
    archiveprefix = {arXiv},
    primaryclass = {cs.CV},
    url = {https://proceedings.neurips.cc/paper_files/paper/2020/file/55053683268957697aa39fba6f231c68-Paper.pdf},
}

@article{jaegle2022perceiveriogeneralarchitecture,
    author = {Andrew Jaegle and Sebastian Borgeaud and Jean-Baptiste Alayrac and Carl Doersch and Catalin Ionescu and David Ding and Skanda Koppula and Daniel Zoran and Andrew Brock and Evan Shelhamer and Olivier Hénaff and Matthew M. Botvinick and Andrew Zisserman and Oriol Vinyals and Joāo Carreira},
    title = {Perceiver IO: A General Architecture for Structured Inputs \& Outputs},
    year = {2022},
    eprint = {2107.14795},
    archiveprefix = {arXiv},
    primaryclass = {cs.LG},
    url = {https://arxiv.org/abs/2107.14795},
}

@article{McAllister:2025qwq,
    author = "McAllister, Liam and Schachner, Andreas",
    title = "{TASI Lectures on de Sitter Vacua}",
    eprint = "2512.17095",
    archivePrefix = "arXiv",
    primaryClass = "hep-th",
    month = "12",
    year = "2025"
}

@article{ba2016layernormalization,
      author={Jimmy Lei Ba and Jamie Ryan Kiros and Geoffrey E. Hinton},
      title={Layer Normalization}, 
      year={2016},
      eprint={1607.06450},
      archivePrefix={arXiv},
      primaryClass={stat.ML},
      url={https://arxiv.org/abs/1607.06450}, 
}

@article{JMLR:v15:srivastava14a,
  author  = {Nitish Srivastava and Geoffrey Hinton and Alex Krizhevsky and Ilya Sutskever and Ruslan Salakhutdinov},
  title   = {Dropout: A Simple Way to Prevent Neural Networks from Overfitting},
  journal = {Journal of Machine Learning Research},
  year    = {2014},
  volume  = {15},
  number  = {56},
  pages   = {1929--1958},
  url     = {http://jmlr.org/papers/v15/srivastava14a.html}
}

@article{Cole:2018emh,
    author = "Cole, Alex and Shiu, Gary",
    title = "{Topological Data Analysis for the String Landscape}",
    eprint = "1812.06960",
    archivePrefix = "arXiv",
    primaryClass = "hep-th",
    reportNumber = "MAD-TH-18-09",
    doi = "10.1007/JHEP03(2019)054",
    journal = "JHEP",
    volume = "03",
    pages = "054",
    year = "2019"
}

@article{charton2025int2intframeworkmathematicstransformers,
      title={Int2Int: a framework for mathematics with transformers}, 
      author={François Charton},
      year={2025},
      eprint={2502.17513},
      archivePrefix={arXiv},
      primaryClass={cs.LG},
      url={https://arxiv.org/abs/2502.17513}, 
}
\end{document}